\documentclass[review,sort&compress]{elsarticle}

\usepackage{lineno,hyperref,fullpage}
\usepackage{afterpage}
\usepackage{enumitem}
\modulolinenumbers[5]

\journal{}

\usepackage{nomencl}
\usepackage{amsmath}
\usepackage{amsfonts}
\usepackage{multirow}
\usepackage{subcaption}

\usepackage{floatrow}
\makenomenclature
\usepackage{xpatch}
\xpatchcmd{\thenomenclature}{\section*{\nomname}
}{}{\typeout{Success}}{\typeout{Failure}}
\makenomenclature
\RequirePackage{ifthen}
\printnomenclature[0.8cm]

\usepackage{soul}
\usepackage{xcolor}
\usepackage{bm}
\usepackage{url}

\graphicspath{}

\hypersetup{nolinks=true}

\let\oldequation\equation
\let\oldendequation\endequation
\renewenvironment{equation}
  {\linenomathNonumbers\oldequation}
  {\oldendequation\endlinenomath}

\begin{document}

\begin{frontmatter}

    \title{Combining direct and indirect sparse data for learning generalizable turbulence models}

  \author[cas,ucas]{Xin-Lei Zhang}

  \author[sim]{Heng Xiao}
  \ead{heng.xiao@simtech.uni-stuttgart.de}

  \author[norce]{Xiaodong Luo}
  
  \author[cas,ucas]{Guowei He}
  \ead{hgw@lnm.imech.ac.cn}
  \address[cas]{The State Key Laboratory of Nonlinear Mechanics, Institute of Mechanics, Chinese Academy of Sciences, Beijing 100190, China}
  \address[ucas]{School of Engineering Sciences, University of Chinese Academy of Sciences, Beijing 100049, China}
  \address[sim]{Stuttgart Center for Simulation Science (SC SimTech), University of Stuttgart, 70569 Stuttgart, Germany}
  \address[norce]{Norwegian Research Centre (NORCE), 5008 Bergen, Norway}

    \begin{abstract}
    Learning turbulence models from observation data is of significant interest in discovering a unified model for a broad range of practical flow applications.
    Either the direct observation of Reynolds stress or the indirect observation of velocity has been used to improve the predictive capacity of turbulence models.
    In this work, we propose combining the direct and indirect sparse data to train neural network-based turbulence models.
    The backpropagation technique and the observation augmentation approach are used to train turbulence models with different observation data in a unified ensemble-based framework.
    These two types of observation data can explore synergy to constrain the model training in different observation spaces, which enables learning generalizable models from very sparse data.
    The present method is tested in secondary flows in a square duct and separated flows over periodic hills.
    Both cases demonstrate that combining direct and indirect observations is able to improve the generalizability of the learned model in similar flow configurations, compared to using only indirect data.
    The ensemble-based method can serve as a practical tool for model learning from different types of observations due to its non-intrusive and derivative-free nature.
    \end{abstract}

  \begin{keyword}
     ensemble Kalman method \sep turbulence modeling \sep direct data \sep indirect data
  \end{keyword}
\end{frontmatter}

\section{Introduction}

Data-driven methods are emerging to improve the predictive capability of turbulence models by incorporating available observation data~\cite{duraisamy2019turbulence, xiao2019quantification, duraisamy2021perspectives}.
Such approaches can use flexible model representations with great expressive powers, e.g., neural networks, to approximate complex mapping from mean velocity to Reynolds stresses.
Therefore, it is promising to build generalizable turbulence models based on these methods, given various data sources from a broad range of flow applications.

Different observation data, such as Reynolds stress, velocity, and lift force, have been used to learn turbulence models for the Reynolds-Averaged Navier-Stokes (RANS) equations~\cite{singh2017machine, wang2017physics, schmelzer2020discovery}.
These observation data can be categorized into direct and indirect observations based on their relationship to the Reynolds stress.
The direct observation is associated with the Reynolds stress directly, such as Reynolds shear stress.
In contrast, the indirect observation is related to physical quantities propagated from the Reynolds stress via the RANS equations, such as mean velocity.
Specialized training strategies need to be developed due to the different relationships between these observations and model parameters.
For instance, training with direct observations can use analytical gradients that are readily available based on differentiable model functions, e.g., neural networks.
In contrast, training with indirect observations requires estimating the gradient of the indirect quantities to the model parameters, which typically resorts to solving certain adjoint equations since the RANS equations are involved in the training process.
In the following, we provide detailed discussions on the training strategies for each type of observation data.

Learning from direct observations constructs turbulence models in \textit{a priori} sense.
That is, the models are trained by minimizing the differences between model predictions and the training data in Reynolds stresses.
Such a training strategy has been demonstrated to improve the prediction of both Reynolds stress and velocity in various flow applications~\cite{ling2016reynolds}.
However, the trained models would lack robustness and have poor generalizability as observed in Ref.~\cite{zhang_ensemble-based_2022}.
This deficiency can be caused by several reasons.
First, learning with direct observations often requires full-field Reynolds stress data, while in practical applications, available data are often sparse from experimental measurements.
Learning from the sparse data would suffer from the issue of ill-posedness.
That is, different model functions can achieve good agreement with the sparse data but may lead to significant discrepancies at unobserved locations.
Second, the model form is often intrinsically inadequate to represent the Reynolds stress, which would cause the issue of robustness when coupling the learned model with the RANS solver.
For instance, the commonly used turbulence models such as the $k$--$\omega$ model~\cite{wilcox2006turbulence} are under the equilibrium assumption, which is inadequate to represent high-fidelity data.
The data-model incompatibility would lead to nonphysical model outputs, e.g., negative eddy viscosity, and further divergence issues of computational solvers.
Finally, the inconsistency between training and prediction environments can lead to large prediction discrepancies in posterior tests~\cite{duraisamy2021perspectives}.
On the one hand, small training errors can lead to large discrepancies in the velocity prediction even in a simple plane channel case~\cite{wu2019reynolds, brener2021conditioning} due to the intrinsic ill-conditioning of the RANS operator.
On the other hand, the model is learned statically in the prior training but repeatedly invoked in posterior tests.
The error accumulation would lead to severe prediction discrepancies in the posterior tests.
Given these limitations, the learning strategy with indirect data is recently proposed to reduce the data requirements and simultaneously alleviate the inconsistency issues.

Learning from indirect observations such as mean velocity is to train turbulence models in \textit{a posteriori} sense or model--consistent manner~\cite{duraisamy2021perspectives} by involving the RANS equations in the training process.
That is, in every training step, the model is re-evaluated to obtain the Reynolds stress fields, which are further propagated to the indirect quantities by solving the RANS equations.
The model parameters are thus optimized by minimizing the discrepancies between model predictions and the indirect data.
By doing so, the data requirement can be reduced, and using partial indirect observations is usually sufficient to learn a model with good predictions of velocity fields~\cite{zhang_ensemble-based_2022}.
Nevertheless, learning from indirect data can also be ill-posed, particularly in the scenario of sparse data.
Different Reynolds stress fields are able to provide good velocity predictions at observed positions.
Particularly in the region with strong pressure gradients or weak strain rates, the Reynolds stress can deviate much from the truth but have limited effects on the velocity, since the velocity is insensitive to the Reynolds stress.
Therefore, learning with only indirect observations is also challenging to provide generalizable turbulence models, particularly in the scenario of sparse data.

In view of the limitations of learning from either direct or indirect observations, combining these two types of observations is promising to take synergy effects and thus enhance the generalizability of the learned model.
Waschkowski et al.~\cite{waschkowski2022multi} made such attempts to combine different types of observation data with genetic programming.
They demonstrated that combining the direct data of turbulent kinetic energy and the indirect data of velocity can improve the predictions of both quantities for flows over periodic hills.
Genetic programming is a global optimization method, which is able to find the global optimal minimum but would tend to be less efficient than the gradient-based local optimization method, e.g., the adjoint method~\cite{darwish2018aerodynamic, simanowitsch2022comparison}.
However, the adjoint method requires extra efforts in developing the adjoint solver, which has become a barrier for the model-consistent training~\cite{duraisamy2021perspectives}.

More recently, Zhang et al.~\cite{zhang_ensemble-based_2022} leveraged an ensemble Kalman method to learn turbulence models from indirect observation data, thereby circumventing the difficulty of adjoint-based model learning. 
The ensemble method demonstrated excellent training efficiency compared to the adjoint-based method, since the ensemble-based approximated gradient and Hessian information are used to achieve a second-order optimization~\cite{zhang_ensemble-based_2022}. The same ensemble learning method has been successfully utilized for constitutive modeling in subsurface flows, where it is used to learn the relative permeability curve for reservoir rocks based on saturation data~\cite{zhou2023inference}.
In this work, we propose using the ensemble-based learning framework to combine the direct and indirect data for turbulence modeling.
Specifically, the back-propagation and the ensemble-based gradient are integrated to combine the Reynolds stress and velocity data based on the regularized ensemble Kalman method~\cite{zhang2020regularized}.
Moreover, the observation augmentation combines the direct data of turbulent kinetic energy (TKE) and the indirect data of velocity, since the model gradient for TKE is not readily available.
The present ensemble method, on the one hand, combines direct and indirect observations to allow the learning of generalizable models from very sparse data.
On the other hand, available analytical gradients and ensemble-based approximated gradients are integrated into a unified framework to train neural-network-based turbulence models.
This is in stark contrast to the earlier work~\cite{zhang_ensemble-based_2022}, where dense indirect data are used to learn turbulence models with the ensemble method, and the readily available analytical gradient from neural networks is not used.
The present work highlights the synergy effects of direct and indirect observation data for turbulence modeling.

The remainder of the paper is organized as follows.
The neural network-based turbulence closure and the proposed training algorithm to combine the direct and indirect observation data are elaborated in Section~\ref{sec:method}.
The case setup for testing the performance of the proposed training method is detailed in Section~\ref{sec:case_setup}.
The numerical results are presented and analyzed in Section~\ref{sec:results}.
Finally, the paper is concluded in Section~\ref{sec:conclusion}.

\section{Methodology}
\label{sec:method}

This work aims to learn generalizable turbulence models from direct and indirect sparse data.
To achieve this goal, the Reynolds stress representation should have sufficient expressive powers to represent various flow characteristics.
Moreover, the training algorithm should be flexible to incorporate different sparse observation data.
In the following, we present the Reynolds stress representation and training strategy, respectively.

\subsection{Reynolds stress representation}

We use the nonlinear eddy viscosity model as the baseline model to represent the Reynolds stress due to two aspects.
On the one hand, the nonlinear eddy viscosity model is in a more general form than the linear eddy viscosity model, since it uses an integrity library of tensor bases instead of using only the first tensor basis as the linear eddy viscosity model.
On the other hand, it is computationally efficient compared to the Reynolds stress transport model, where additional transport equations are required to solve.

In the nonlinear eddy viscosity model, the Reynolds stress\footnote{Here we followed Pope's convention~\cite{pope2001turbulent} of defining Reynolds stress as the covariance of the velocity fluctuations i.e., $\tau_{ij} = \left<u'_i u'_j\right>$.
We note that in the turbulence modeling literature (e.g.,~\cite{wilcox2006turbulence}) it is more common to refer to $- \left<u'_i u'_j\right>$ as the Reynolds stress because of its role in the RANS momentum equations.}~$\boldsymbol{\tau}$
is first normalized by its trace $2k$ and then decomposed into a deviatoric part~$\mathbf{b}$ and a spherical part as
\begin{equation}
    \boldsymbol{\tau} = 2k(\mathbf{b} + \frac{1}{3} \mathbf{I}) \text{,}
    \label{eq:tau}
\end{equation}
where $k$ is the turbulent kinetic energy (TKE) and $\mathbf{I}$ is the identity tensor of second rank.
The two quantities~$\mathbf{b}$ and $k$ both need to be modeled to close the RANS equations.
The modeling strategies for each term are discussed below.

The normalized deviatoric part of Reynolds stress $\mathbf{b}$, which is also referred to as Reynolds stress anisotropy, can be represented in the form of tensor bases based on the Cayley-Hamilton theorem as~\cite{pope1975more} 
\begin{equation}
\begin{aligned}
    &\mathbf{b}=\sum_{i=1}^{10} g^{(i)} \mathbf{T}^{(i)} \\
    &g^{(i)}=g^{(i)}\left(\theta_{1}, \ldots, \theta_{5}\right) \text{,}
    \label{eq:b}
\end{aligned}
\end{equation}
where $\mathbf{T}$ and $\boldsymbol{\theta}$ are the tensor basis and scalar invariant of the input tensors, and $\boldsymbol{g}$ is the coefficient function to be learned.
The functional mapping from the scalar invariant~$\boldsymbol{\theta}$ to the coefficients~$\boldsymbol{g}$ can be represented in analytical forms~\cite{shih1993realizable, weatheritt2016novel} or based on neural networks~\cite{ling2016reynolds}.
There are $10$ independent tensors that provide the most general form in the Reynolds stress anisotropy.
The tensor bases are related to the strain rate~$\mathbf{S}$ and rotation rate~$\mathbf{W}$ normalized with the turbulence time scale~$\tau_s$ as $\mathbf{S}=\frac{1}{2} \tau_s(\nabla \mathbf{u}+\nabla \mathbf{u}^\top)$ and $\mathbf{W}=\frac{1}{2} \tau_s(\nabla \mathbf{u}-\nabla \mathbf{u}^\top)$, where $\mathbf{u}$ represents velocity vectors.
The time scale can be estimated by $\tau_s = 1/\omega$, where $\omega$ is the specific dissipation rate.
The first four tensors and the first two scalar invariants are written as
\begin{equation}
\begin{aligned}
    \mathbf{T}^{(1)} &= \mathbf{S}, \qquad \mathbf{T}^{(2)} = \mathbf{S W} - \mathbf{W S}, \\ 
    \mathbf{T}^{(3)} &= \mathbf{S}^2 - \frac{1}{3}\{\mathbf{S}^2\} \mathbf{I}, \qquad \mathbf{T}^{(4)} = \mathbf{W}^2-\frac{1}{3}\{\mathbf{W}^2\}\mathbf{I}, \\
    \theta_1 &= \{\mathbf{S}^2\} , \qquad \theta_2 = \{\mathbf{W}^2\} \text{,}
\end{aligned}
\end{equation}
where the curly bracket~$\{ \bullet \}$ indicates the matrix trace.
For two-dimensional flows, only two scalar invariants are nonzero, and the first three tensor bases are linearly independent.
Further for incompressible flows, the third tensor basis can be incorporated into the pressure term in the RANS equation, leaving only two tensor functions and two scalar invariants~\cite{strofer2021end}.

Apart from the normalized deviatoric part~$\mathbf{b}$, the turbulent kinetic energy~$k$ is also needed to estimate the Reynolds stress based on Eq.~\eqref{eq:tau}.
It is obtained by solving the TKE transport equation as
\begin{equation}
\begin{aligned}
    \frac{D k}{D t} &= \mathcal{P} -\mathcal{E} + \frac{\partial}{\partial x_j} \left[\left(\nu+\sigma_{k} \nu_{t} \right) \frac{\partial k}{\partial x_j} \right] \\
    \mathcal{P} &= {-2k}b_{ij} \frac{\partial u_i}{ \partial x_j} \text{,}
\end{aligned}
\end{equation}
where~$\mathcal{P}$ is the turbulence production, $\mathcal{E}$ is the turbulence dissipation, $\nu$ is the molecular viscosity, $\nu_t$ is turbulent eddy viscosity, and $\sigma_k$ is a model parameter.
The TKE transport equation can be derived rigorously, but the dissipation term~$\mathcal{E}$ often needs to be estimated with \textit{ad hoc} models.
Here we take the $k$--$\omega$ shear stress transport (SST) model~\cite{menter1994two} as an example.
The dissipation term is formulated as $\mathcal{E} = \beta^{*} \omega k$, where $\beta^*$ is a model constant. 
The specific dissipation rate $\omega$ is estimated by solving the transport equation, i.e.,
    \begin{equation}
    \frac{D \omega}{D t} = \frac{\alpha}{\nu_{t}} \mathcal{P} - \beta \omega^{2} + \frac{\partial }{\partial x_j} \left[ \left( \nu+\sigma_{\omega} \nu_{t} \right) \frac{\partial \omega}{\partial x_{j}} \right] + Q_\text{SST} \text{.}
    \label{eq:omega}
    \end{equation}
The term~$Q_\text{SST}$ and the model coefficients~$\alpha$,~$\beta$,~$\sigma_{\omega}$ are used to blend the $k$--$\varepsilon$ model and $k$--$\omega$ model.
The detailed formulations of the blending functions are presented in~\ref{sec:sst}.

\begin{figure}
    \centering
    \subfloat[classic TBNN]{\includegraphics[width=0.4\textwidth]{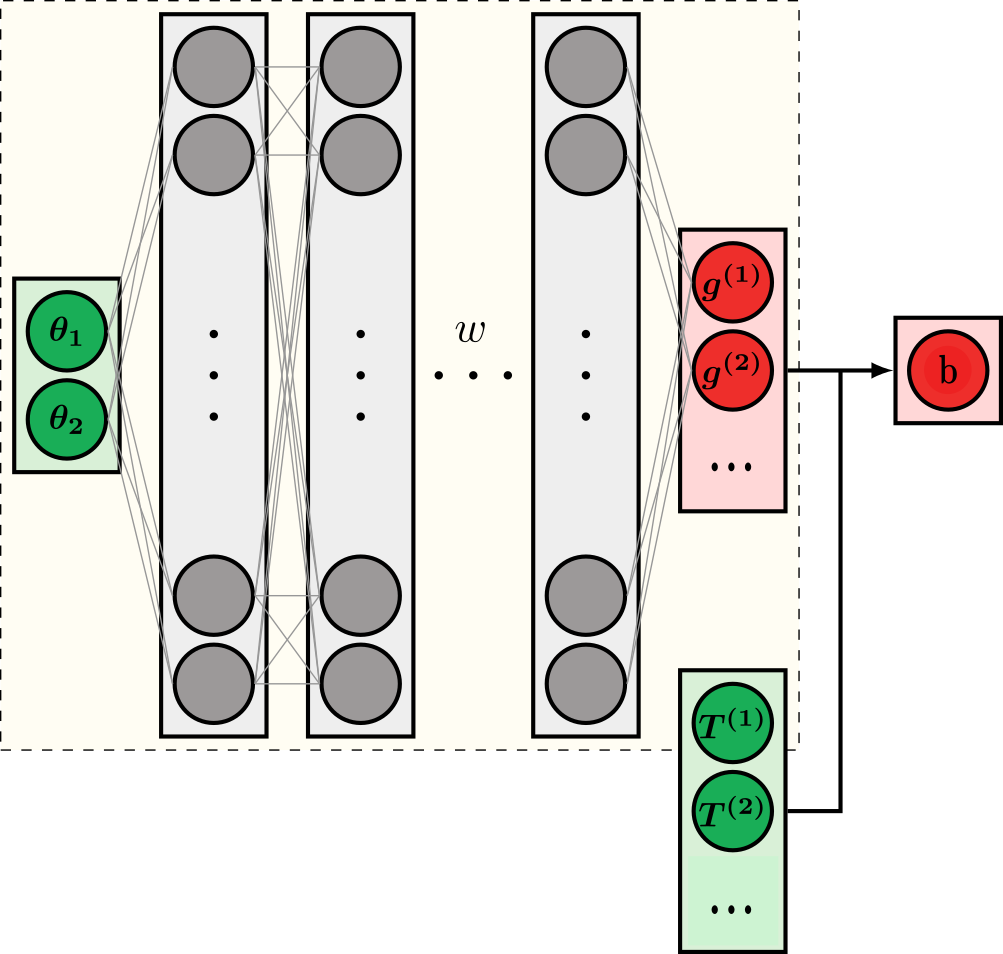}}
    \hspace{2cm}
    \subfloat[TBNN with correction for TKE production]{\includegraphics[width=0.4\textwidth]{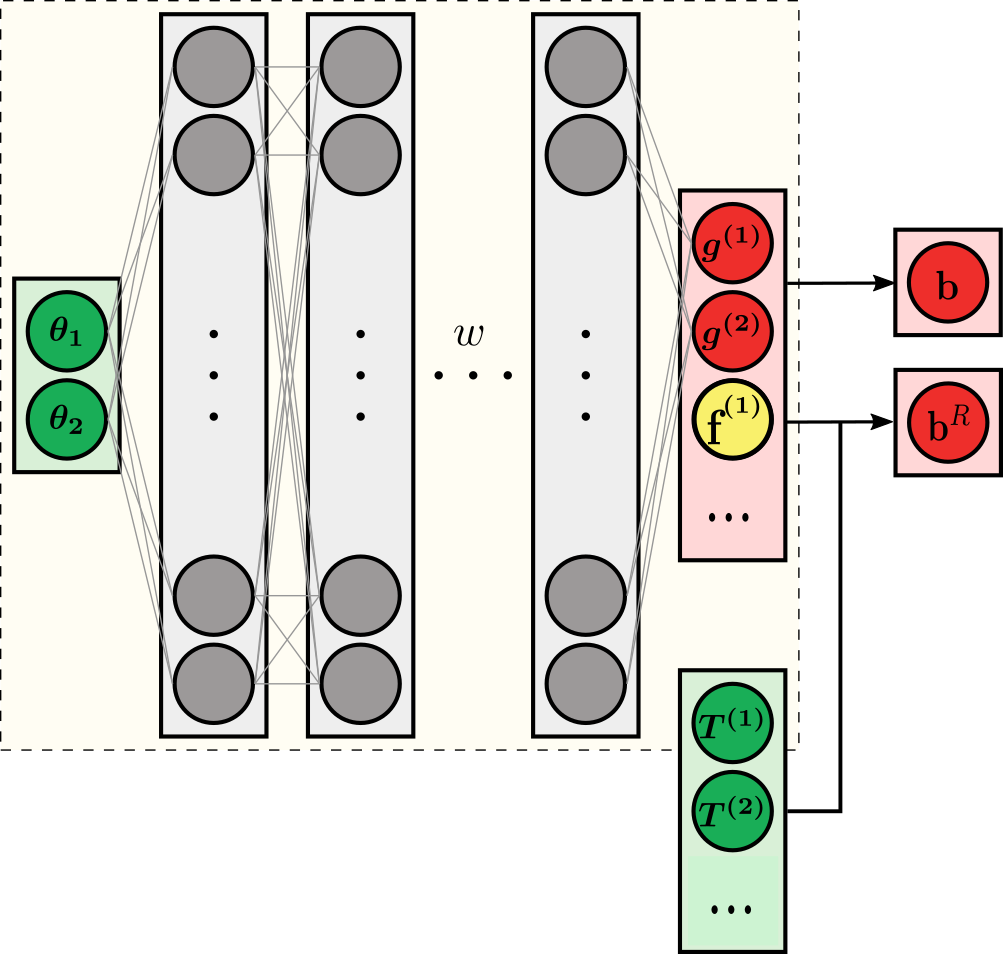}}
    \caption{Schematic of the tensor basis neural network (TBNN) architectures. The classic tensor basis neural network combines only the tensor bases and $g$ functions to form the Reynolds stress anisotropy. 
    The present work uses a modified tensor basis neural network
    where the outputs of the network are used to form both the Reynolds stress anisotropy~$\mathbf{b}$ and a corrective tensor~$\mathbf{b}^R$ in the TKE production.}
    \label{fig:scheme}
\end{figure}

In the transport equation of turbulent kinetic energy, the dissipation term is modeled in an \textit{ad hoc} form, which poses difficulties in recovering the TKE field.
To remedy this deficiency, the production term can be augmented with additional correction~$R$ as~\cite{schmelzer2020discovery} 
\begin{equation}
        \tilde{\mathcal{P}} = \mathcal{P} + R \text{,} \quad
        \text{with} \quad R = {-2k}b_{ij}^R \frac{\partial u_i}{\partial x_j} \text{,}
        \label{eq:TKE_corr}
\end{equation}
where $b_{ij}^R$ is the correction tensor to compute the additional production $R$.
Such a correction is equivalent to adding a multiplication on the turbulent production as in the work of Singh et al.~\cite{singh2017machine}.
Further, the tensor bases~$\mathbf{T}$ are used to construct the normalized Reynolds stress anisotropy~$\mathbf{b}$ based on Eq.~\eqref{eq:b} and the correction term~$\mathbf{b}^R$ as follows:
\begin{equation}
\begin{aligned}
    & \mathbf{b}^{R}=\sum_{i=1}^{10} f^{(i)} \mathbf{T}^{(i)} \\
    & \text{with} \quad f^{(i)}=f^{(i)}\left(\theta_{1}, \ldots, \theta_{5}\right) \text{.}
\end{aligned}
\end{equation}
In this work we use the neural network to represent the functions of $\boldsymbol{g}$ and $\boldsymbol{f}$ due to its flexibility in expressing complicated functional relationships.
Thus, the functions of $\boldsymbol{g}$ and $\boldsymbol{f}$ are regarded as the output of neural networks.
This is in contrast to the classic architecture~\cite{ling2016reynolds} where only $\boldsymbol{g}$ functions are considered the output of the neural network.
The schematic of the neural network architecture is shown in Figure~\ref{fig:scheme} in comparison to the classic tensor basis neural network.

\subsection{Training strategy for combining direct and indirect observation}
The neural network-based turbulence model is trained with both direct and indirect observations in this work. 
The direct observations are measured quantities in a straight connection with the Reynolds stress, e.g., the Reynolds stress anisotropy~$\mathbf{b}$ and TKE~$k$.
In contrast, the indirect observations are the quantities propagated from the Reynolds stress via the RANS equation, e.g., the mean velocity.
These data can be very sparse from experimental measurements, particularly for flows with high Reynolds numbers.
Learning turbulence models from sparse data is challenging to provide generalizable models since such an inverse problem of inferring model parameters from data is highly ill-posed.
Combining the direct and indirect observations can alleviate the ill-posedness and enable the learned model to have better generalizability by constraining the model training in different observation spaces.

Specialized training strategies are required to incorporate direct and indirect observations, since these quantities are usually estimated in different ways.
Specifically, the direct quantities~$\mathbf{b}$ can be obtained via evaluating neural networks as $$\mathbf{b} = \mathcal{N}[\boldsymbol{w}],$$
where $\mathcal{N}$ represents the operator of neural network.
The direct quantities~$k$ and indirect quantities~$\mathbf{u}$ are estimated by solving the TKE transport equation and the RANS equations.
It can be formulated as~$$\mathbf{u}, k = \mathcal{H}[\boldsymbol{w}],$$ where $\mathcal{H}$ represents the model operator mapping from the weights to the velocity and TKE at observation locations.
Accordingly, the gradients of these quantities with respect to the weights, i.e., $\partial \mathcal{N}[\boldsymbol{w}] / \partial \boldsymbol{w}$ and $\partial \mathcal{H}[\boldsymbol{w}] / \partial \boldsymbol{w}$, have to be estimated to train the neural network-based model in different manners. 
Specifically, the gradient of the Reynolds stress anisotropy $\mathbf{b}$ (direct quantity) to the weight can be obtained from the back-propagation algorithm.
In contrast, the gradient of the TKE (direct quantity) and the velocity (indirect quantity) need to be estimated based on the adjoint method~\cite{singh2016using} or ensemble-based approximation~\cite{strofer2021ensemble}.
In view of such differences in estimating the model gradients, two scenarios can be distinguished based on the available observation data.
In Scenario I, the training data are the direct observation of $\mathbf{b}$ and the indirect observation of $\mathbf{u}$.
In Scenario II, the training data are the direct observation of $k$ and indirect observation of $\mathbf{u}$.
We employ the ensemble Kalman method to learn turbulence models based on the used data in a unified framework. 
The training methods for the two scenarios are summarized in Table~\ref{tab:schemes}.
Note that multiple data sources, e.g., $k$, $\mathbf{b}$, and $\mathbf{u}$, can also be combined within the framework by using the regularized ensemble method and observation augmentation simultaneously.

In Scenario I, where both direct observation~$\mathbf{b}$ and indirect observation~$\mathbf{u}$ are available, we adopt the regularized ensemble Kalman method~\cite{zhang2020regularized,zhang2022assessment} to train neural network-based models.
In such a scenario, the analytical gradient~$\partial \mathbf{b} / \partial \boldsymbol{w}$ is available based on the back-propagation technique.
The method can integrate the analytical gradient and ensemble-based gradient to train the model with direct and indirect observations.
The objective for the regularized ensemble Kalman method is to minimize the cost function as
\begin{equation}
    J = \parallel \boldsymbol{w}^{i+1}_j - \boldsymbol{w}^{i}_j \parallel^2_\mathsf{P} + \parallel \mathsf{y}_j^{u} - \mathcal{H}[\boldsymbol{w}^{i+1}_j] \parallel^2_\mathsf{R} + \parallel \mathsf{y}_j^{b} - \mathcal{N}[\boldsymbol{w}^{i+1}_j] \parallel^2_\mathsf{Q} \text{,}
    \label{eq:cost}
\end{equation}
where $\parallel \cdot \parallel_A^2$ represents the weighted Euclidean norm, which is defined as $\parallel \bm{v} \parallel_A^2=\bm{v}^\top A^{-1} \bm{v}$, $\mathsf{P}$ is the model error covariance indicating the uncertainties on the weights, $\mathsf{R}$ is the error covariance matrix of indirect data, $\mathsf{Q}$ is the error covariance matrix of direct data, $\mathsf{y}$ indicates the training data, the superscripts ${u}$ and ${b}$ represent observed quantities, and the subscripts $i$ and $j$ are the indexes of iteration and sample, respectively.
The observation errors are independent and identically distributed samples from a Gaussian distribution,
and the model error covariance is estimated based on
\begin{equation}
    \mathsf{P} = \frac{1}{N-1} \Delta \mathsf{W} \Delta \mathsf{W}^\top \text{,}
\end{equation}
where $\Delta \mathsf{W} = \{\boldsymbol{w}_j - \overline{\mathsf{W}} \}_{j=1}^N$, 
$\mathsf{W}=\{\boldsymbol{w}_j\}_{j=1}^N$, $\overline{\mathsf{W}}$ is the sample mean, and $N$ is number of samples.
The first term in Eq.~\eqref{eq:cost} is to regularize the difference of the neural network weights between adjacent iterations, the second term aims to train the model based on the indirect data~$\mathsf{y}^\text{u}$, and the final term penalizes the difference between the model evaluation and the direct data~$\mathsf{y}^\text{b}$.
The three terms in Eq.~\eqref{eq:cost} are weighted by associated error covariances, respectively.
Specifically, the first term is weighted by the model error covariance~$\mathsf{P}$, the second term is weighted by the observation error covariance of indirect data~$\mathsf{R}$, and the third term is weighted by the observation error covariance of direct data~$\mathsf{Q}$.
The observation data error typically includes measurement errors and process errors~\cite{xiao_quantifying_2016,dennis2006estimating,luo2021accounting}.
The measurement error can be originated from experimental conditions and is not considered for the DNS data.
In contrast, the process error stems from the intrinsic discrepancies between the RANS model and observation data, e.g., due to ensemble averaging.
Such error correlations among different data points are challenging to estimate in practice and often neglected, as in this work.
This can result in inferior training results with dense observation data due to significant observation error correlations.
Moreover, the correlated observation errors lead to non-diagonal error covariance matrices, which would consume a large amount of memory for storing them in the presence of huge datasets.
Therefore, the present ensemble Kalman method is suitable for the sparse-data scenario, where data points are placed spatially far from each other to have negligible error correlations.
Nevertheless, it is noted that the issues caused by neglecting observation error correlations are not particular for ensemble Kalman methods, and similar algorithms, e.g., the four-dimensional variational (4DVar) method~\cite{mons2016reconstruction,chandramouli20204d}, suffer the same problems.

The update scheme of the regularized ensemble Kalman method can be derived by minimizing the cost function in Eq.~\eqref{eq:cost}.
It can be formulated as two steps:
\begin{equation}
\begin{aligned}
    \tilde{\boldsymbol{w}}_j &= \boldsymbol{w}_j^{i} - \mathsf{P}(\mathcal{N}'[\boldsymbol{w}_j^i])^\top \mathsf{Q}^{-1}(\mathcal{N}[\boldsymbol{w}_j^{i}]-\mathsf{y}_j^\text{b})  \\
    \boldsymbol{w}_j^{i+1} &= \tilde{\boldsymbol{w}}_j + \mathsf{K}(\mathsf{y}_j^\text{u} - \mathsf{H}\tilde{\boldsymbol{w}}_j) \\
    \text{with} \quad \mathsf{K} &= \mathsf{PH}^{\top}(\mathsf{HPH}^\top + \mathsf{R})^{-1}
    \text{,}
\end{aligned}
\end{equation}
where $\mathsf{H}$ is the local gradient of the observation operator $\mathcal{H}$ to the parameter $\boldsymbol{w}$, and $\mathcal{N'}[\boldsymbol{w}]$ is the gradient of the neural network outputs to the weight.
In the first step, the regularized ensemble Kalman method performs a first-order gradient descent to minimize the difference between the model prediction and direct data~$\mathsf{y}^\text{b}$.
Herein, the gradient of the neural network outputs to the weights $\mathcal{N'}[\boldsymbol{w}]=\partial \mathbf{b} / \partial \boldsymbol{w}$ can be obtained from the back-propagation algorithm.
It is noted that the gradient can also be approximated with the sample statistics, i.e., $\mathcal{N'}[\boldsymbol{w}]=\Delta \mathcal{N}[\mathsf{W}] (\Delta \mathsf{W})^{-1}$, if the analytical gradient is not available.
The second step performs a second-order optimization to reduce the discrepancy in the prediction of indirect quantities~$\boldsymbol{u}$.
The local gradient~$\mathsf{H}$ is estimated with the samples as $\mathsf{H} = \Delta \mathcal{H}[\mathsf{W}] (\Delta \mathsf{W})^{-1}$.
The pseudo-inverse~$(\Delta \mathsf{W})^{-1}$ is computed based on the singular value decomposition.
Further, the ensemble-based gradient~$\mathsf{H}$ is pre-multiplied by the model error covariance matrix~$\mathsf{P}$, which is a common method for preconditioning of the ensemble-based optimization~\cite{chen2009efficient,schillings2017analysis}.
In addition, the Hessian information can be also estimated with the ensemble of samples~\cite{luo2015iterative, zhang_ensemble-based_2022}.
The merits of the present ensemble method are two-fold.
On the one hand, the method inherits the advantage of the ensemble method in learning models from indirect observations without requiring solving adjoint equations. 
On the other hand, the method allows regularizing the training process with the extra direct observation~$\mathbf{b}$ based on the back-propagation technique.

In Scenario II, where we have only TKE as direct observations, the associated gradient~$\partial k / \partial \boldsymbol{w}$ is not readily available since the TKE is estimated by solving the transport equations.
Typically, the gradient can be obtained by solving adjoint equations, but it requires extra effort in the code redevelopment.
Here we leverage the ensemble method to incorporate the two data sources of the TKE and the velocity through observation augmentation.
Specifically, we augment the observation vector and formulate the cost function as
\begin{equation}
    \begin{aligned}
    J &= \parallel \boldsymbol{w}_j^{i+1} - \boldsymbol{w}_j^{i} \parallel^2_\mathsf{P} + \parallel \mathsf{y}_j^\text{a} - \mathcal{H}^\text{a}[\boldsymbol{w}_j^{i+1}] \parallel^2_{\mathsf{R}^\text{a}} \\
    \mathsf{y}^\text{a} &= [\mathsf{y}^{u}, \mathsf{y}^{k}] \text{,}
    \end{aligned}
    \label{eq:aug_cost}
\end{equation}
where the superscript~$\text{a}$ indicates the augmented quantities.
Further, the ensemble Kalman method can be used to train the neural network by minimizing the cost function~\eqref{eq:aug_cost}.
The update scheme is formulated as
\begin{equation}
\begin{aligned}
    \boldsymbol{w}_j^{i+1} &=  \boldsymbol{w}_j^{i} + \mathsf{K}^\text{a} (\mathsf{y}_j^\text{a} - \mathsf{H}^\text{a}[\boldsymbol{w}_j^i]) \\
    \text{with} \quad \mathsf{K}^\text{a} &= \mathsf{PH}^{\text{a},\top}(\mathsf{H}^{\text{a}} \mathsf{P} \mathsf{H}^{\text{a} \top} + \mathsf{R}^\text{a})^{-1} \text{.}
\end{aligned}
\end{equation}
As such, the two data sources are incorporated by the ensemble method without requiring analytical gradients.

\begin{table}[!htb]
    \centering
    \begin{tabular}{ccc}
    \hline
         Scenarios & Scenario I & Scenario II \\ \hline
         direct data & $\mathbf{b}$ & $k$ \\
         indirect data & $\mathbf{u}$ & $\mathbf{u}$ \\
         training method & \shortstack{Ensemble Kalman method \\
         with regularization} & \shortstack{Ensemble Kalman method \\ with observation augmentation} \\ \\
        cost function &   \shortstack[r]{
        \begin{math}
        \begin{aligned}[c] 
        J = & \parallel \boldsymbol{w}^{i+1} - \boldsymbol{w}^{i} \parallel^2_\mathsf{P} + \parallel \mathsf{y}^{\boldsymbol{u}} - \mathcal{H}[\boldsymbol{w}^{i+1}] \parallel^2_\mathsf{R} \\
        & + \parallel \mathsf{y}^{\boldsymbol{b}} - \mathcal{N}[\boldsymbol{w}^{i+1}] \parallel^2_\mathsf{Q}
        \end{aligned}
        \end{math}
        }
        & 
        \shortstack[r]{
        \begin{math}
        \begin{aligned}[c] 
        \mathsf{y}^\text{a} &= [\mathsf{y}^{\boldsymbol{u}}, \mathsf{y}^k] \\
         J &= \parallel \boldsymbol{w}^{i+1} - \boldsymbol{w}^{i} \parallel^2_\mathsf{P} + \parallel \mathsf{y}^\text{a} - \mathcal{H}^\text{a}[\boldsymbol{w}^{i+1}] \parallel^2_{\mathsf{R}^\text{a}}
        \end{aligned}
        \end{math}
        }
         \\ \\
         update scheme & \shortstack[r]{
         \begin{math}
         \begin{aligned}[c] 
         \tilde{\boldsymbol{w}} &= \boldsymbol{w}^{i} - \mathsf{P}(\mathcal{N'}[\boldsymbol{w}^i])^{\top}\mathsf{Q}^{-1}(\mathcal{N}[\boldsymbol{w}^i]-\mathsf{y}^{\boldsymbol{b}}) \text{;} \\
          \boldsymbol{w}^{i+1} &= \tilde{\boldsymbol{w}} + \mathsf{K}(\mathsf{y}_j - \mathsf{H}\tilde{\boldsymbol{w}}) 
          \end{aligned}
          \end{math}
          } & 
          \shortstack{
          \\
          $\boldsymbol{w}^{i+1} = \boldsymbol{w}^i + \mathsf{K}^\text{a} (\mathsf{y}^\text{a} - \mathsf{H}^\text{a}[\boldsymbol{w}^i])$} \\
    \hline
    \end{tabular}
    \caption{Summaries of the ensemble-based training strategies for combining direct and indirect data sources.
    The direct data include the turbulent kinetic energy $k$ and the Reynolds stress anisotropy~$\mathbf{b}$, which are combined with the indirect data of velocity~$\mathbf{u}$ in two different update schemes.
    The two scenarios are compared in terms of the cost functions, training methods, and update schemes.
    }
    \label{tab:schemes}
\end{table}

The ensemble-based method is able to learn turbulence models from both direct and indirect data sources, which has the following three practical benefits.
First, the ensemble Kalman method is non-intrusive and thus can be applied to black-box systems without the need for code re-developments.
This feature promotes its ease of implementation and practical use for different problems compared to the adjoint method.
Second, it is flexible to incorporate different data sources~\cite{zhang2021assimilation, zhang2022acoustic,schneider2022ensemble}, including integral data, point-wise measurements, or statistic data~\cite{wang2018spanwise,WuICA}.
Finally, the method performs a second-order optimization based on the low-rank approximated gradient and Hessian information.
As such, it could be more efficient than the first-order adjoint-based optimization~\cite{zhang_ensemble-based_2022}.

We note that hybrid gradients~\cite{oliver2022hybrid} can also be used to train the turbulence model by combining the analytical and ensemble-based gradients.
Such methods decompose the full gradient into two parts as $$\frac{\partial \mathcal{H}[\boldsymbol{w}]}{\partial \boldsymbol{w}} =  \frac{\partial \mathcal{H}[\boldsymbol{w}]}{\partial \boldsymbol{\tau}} \frac{\partial \boldsymbol{\tau}}{\partial \boldsymbol{w}}.$$
The gradient $\partial \mathcal{H}[\boldsymbol{w}] / \partial \boldsymbol{\tau}$ is estimated with the ensemble samples as $\partial \mathcal{H}[\boldsymbol{w}]/ \partial \boldsymbol{\tau} = \Delta \mathcal{H}[\mathsf{W}] (\Delta \boldsymbol{\tau})^{-1}$, and the gradient $\partial \boldsymbol{\tau} / \partial \boldsymbol{w}$ is obtained through the back-propagation algorithm.
We test the method in a square duct case, and the results are presented in~\ref{sec:hybrid}.
Generally, the hybrid method can also provide a turbulence model with improved predictions of the velocity and Reynolds stress compared to the baseline model.
However, the predictions are slightly inferior to the ensemble method, likely due to the fact that the pseudo inversion of the ill-conditioned matrix~$\Delta \boldsymbol{\tau}$ is involved in computing $\partial \mathcal{H}[\boldsymbol{w}] / \partial \boldsymbol{\tau}$.
In contrast, the ensemble method can avoid the pseudo inversion after pre-multiplying the model error covariance~$\mathsf{P}$ as shown in~\ref{sec:hybrid}. 
We note that the physics-informed neural networks (PINNs)~\cite{raissi2019physics} are increasingly used to solve partial differential equations, including the RANS equations~\cite{eivazi2022physics}.
It is of great interest to apply this framework for training neural-network-based turbulence models with various data sources since the model sensitivity in terms of different physical quantities can be readily available based on the auto-differentiation technique.

\subsection{Comparison to other learning methods in algorithmic details}
\label{sec:comparison}

Various data-driven approaches, such as symbolic regression~\cite{schmelzer2020discovery} and genetic programming~\cite{waschkowski2022multi}, have been proposed to combine different observation data for learning turbulence models.
These methods have been demonstrated to improve the predictive accuracy of the learned models in different physical quantities, including velocity and turbulent kinetic energy.
However, they differ from the present method in certain algorithmic aspects, such as cost-function formulation and data usage. 
In the following, we provide an algorithmic comparison among these methods and the present ensemble method.

The symbolic regression approach utilizes the Reynolds stress and the velocity data to learn turbulence models in a symbolic form.
The two data sources are firstly used to build a library of candidate models~$\bm{C}$ in the form of tensor bases and scalar invariants.
Specifically, the velocity data is used to construct the tensor bases and corresponding scalar invariants. 
Then the velocity data and the Reynolds stress data are combined to estimate the time scale by solving the transport equation of turbulence frequency~$\omega$, which is used to nondimensionalize the tensor bases.
After building the model library~$\bm{C}$, the sparse regression technique is applied to select appropriate candidate models based on the Reynolds stress data.
Finally, the model coefficients~$\bm{w}$ are determined based on the ridge regression~\cite{bishop2006pattern} by minimizing the model discrepancies from the Reynolds stress data.
We note that the symbolic regression method uses indirect velocity data to generate candidate model libraries instead of as a training objective.

The genetic programming method~\cite{waschkowski2022multi} combines different data sources through the evolutionary algorithm.
It is achieved by evaluating two cost functions, one corresponding to the Reynolds stress data and the other to the velocity data, to guide the model learning.
The balance between these cost functions is determined based on the Pareto domination, which can identify optimal solutions that are not dominated by other solutions.
Further, the evolutionary algorithm is used to minimize the cost functions by searching for the Pareto front, i.e., the set of non-dominated solutions to the optimization problem.
The method regards both the indirect velocity data and the direct Reynolds stress data as the training objective, which is in contrast to the sparse regression method~\cite{schmelzer2020discovery} that uses the direct data as the training objective and the indirect data for generating model candidates.

The proposed method in this work is extended from the ensemble Kalman method~\cite{zhang_ensemble-based_2022} to combine both direct data and indirect data for turbulence modeling.
The ensemble Kalman method typically uses only indirect velocity data to learn turbulence models, while the present method integrates the ensemble method and the back-propagation technique to train the model with these different types of observation data.
It is achieved by either augmenting the observation or leveraging the neural network gradient depending on the used direct data, as illustrated in Table~\ref{tab:schemes}.
This method regards both direct and indirect data as the training objectives, which is similar to the genetic programming method~\cite{waschkowski2022multi}.
However, the balance between different observation data is determined by the prescribed observation error in the present method. In contrast, genetic programming uses Pareto domination to balance the cost function associated with different observation data. 
Another difference between the two methods is that the ensemble method is a derivative-free local optimization method, which implicitly uses the information of approximate gradient and Hessian to identify local minima, while genetic programming is a global optimization method that can find global minima, but at a relatively high computational cost~\cite{simanowitsch2022comparison}.
A comparison among these different methods in their algorithmic details is provided in Table~\ref{tab:sum_methods}.

The novelty of the present approach lies in integrating different gradient information to train models with direct and indirect data in a unified ensemble-based framework.
The backpropagation algorithm and the ensemble method are able to evaluate the gradient information for direct data and indirect data, respectively.
However, these gradients are often used individually for neural network training.
The proposed framework provides an alternative strategy to take advantage of both gradient information to improve model training. 
The models can be trained with direct data based on the analytical gradient of the neural network and with indirect data based on the ensemble-based gradient and Hessian simultaneously.
By doing so, the direct and indirect data are combined effectively and take synergy effects to enhance the predictive ability of the learned model.

\begin{table}[!htb]
    \centering
    \begin{tabular}{c|c|c|c}
    \hline
    method  & symbolic regression~\cite{schmelzer2020discovery} & genetic programming~\cite{waschkowski2022multi} & present \\
    cost function  & \shortstack{$\bm{C} = f(\bm{\tau}, \bm{u})$ \\ $J = \|\bm{C w} - \mathsf{y}^{\boldsymbol{\tau}} \| + \gamma \| \bm{w} \|$} &
    \shortstack{ $J_1 = \| \bm{\tau(w)} - \mathsf{y}^{\boldsymbol{\tau}} \|$ \\  $ J_2 = \| \bm{ u(\tau(w))} - \mathsf{y}^{\boldsymbol{u}} \|$ } &
    \shortstack{$J=\| \bm{\tau(w)} - \mathsf{y}^{\boldsymbol{\tau} }\|_\mathsf{Q}$ \\ $+ \| \bm{u(\tau(w))}-\mathsf{y}^{\boldsymbol{u}} \|_\mathsf{R}$ } \\
    update algorithm & regression analysis & evolutionary algorithm & \shortstack{ensemble Kalman update \& \\ back propagation}  \\
    usage of indirect data & candidate model generation & cost evaluation & \shortstack{ensemble gradient/hessian \\ approximation} \\
    usage of direct data   & model selection \& inference & cost evaluation & analytic gradient evaluation \\
    data balance           & -- & Pareto dominance  & prescribed data noise \\
    \hline
    \end{tabular}
    \caption{Comparison of the present method with other learning approaches, including the symbolic regression and genetic programming, to combine different data sources in algorithmic details}
    \label{tab:sum_methods}
\end{table}

\subsection{Procedures}

\begin{figure}[!htb]
    \centering
    \includegraphics[width=0.8\textwidth]{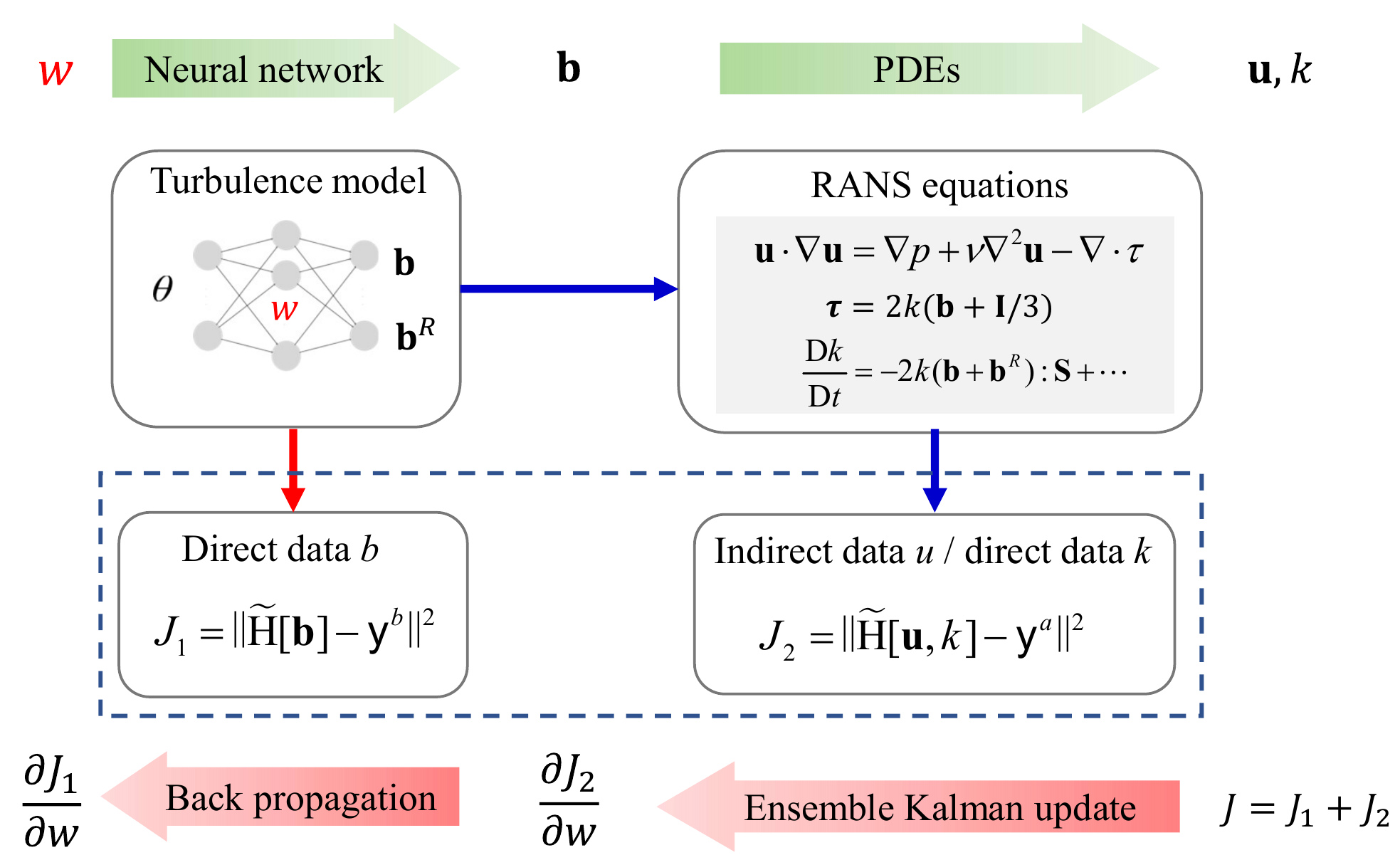}
    \caption{Schematic of the ensemble-based learning framework to combine direct and indirect observations. 
    The weights of the neural network are propagated to the Reynolds stress anisotropy and further to the velocity and turbulent kinetic energy through solving the RANS equations. 
    The gradient is estimated to optimize the weights in different ways based on the data sources.
    The back-propagation is used to train the neural network with the direct data in the Reynolds stress anisotropy~$\mathbf{b}$.
    The ensemble method is used to estimate the gradient with the direct data of turbulent kinetic energy and the indirect data of velocity.
    The symbol~$\tilde{\text{H}}$ indicates the observation operator that extracts the predictions at observed locations.
    }
    \label{fig:schematic}
\end{figure}

The present method trains the neural network-based turbulence model with direct and indirect observations in a unified ensemble-based framework.
The procedure of the training algorithm is illustrated as follows.
\begin{itemize}
    \item Pre-training: The neural network is pre-trained to be a linear eddy viscosity model, i.e., $g^{(1)}=-0.09, g^{(2)-(10)}=0$~\cite{strofer2021end}.
    The conventional weight initialization would lead to nonphysical coefficients such as negative eddy viscosity.
    For this reason, pretraining is required to address this issue and also improve the efficiency of model learning.
    
    \item Sampling: We draw samples of $\boldsymbol{w}$ from the normal distribution $\mathcal{N}(w^0, \sigma^2)$.
    The pretrained weight~$w^0$ is used as the mean of the distribution.
    For each realization, we can obtain the tensor coefficient $\boldsymbol{g}$ and the gradient~$\partial \mathbf{b} / \partial \boldsymbol{w}$ with the back-propagation algorithm.

    \item Feature extraction: The computed flow field and turbulence time scale are used to estimate the scalar invariants and the tensor bases.
    The time scale can be estimated from the specific turbulence dissipation~$\omega$, which is obtained by solving the corresponding transport equation.
    The scalar invariants are then adopted as the input features of the neural network, while the tensor bases are combined with the outputs of the neural network to reconstruct the Reynolds stress and the correction term~$R$.
    The input features of the neural network are scaled into the range of $0$ to $1$ with the maximum and minimum values.
    Note that the min-max normalization would lead to feature clustering in the scenario of having the singularity due to local extreme velocity gradients such as shock waves.
    In such cases, the local normalization~\cite{ling2015evaluation,wang2017physics,wu2019physics} can be used to avoid the feature clustering, e.g., $\hat{\theta} = \theta / ( | \theta | + | \theta^* |)$, where $\theta^*$ is local normalization factors.

    \item Propagation: Each realization of the weights~$\boldsymbol{w}$ is used to propagate the input features to the $\boldsymbol{g}$ and  $\boldsymbol{f}$ functions.
    The $\boldsymbol{g}$ function is further used to construct the Reynolds stress anisotropy~$\mathbf{b}$, while the $\boldsymbol{f}$ function is used to construct the correction term~$\mathbf{b}^R$ in the TKE production~\eqref{eq:TKE_corr}.
    The turbulent kinetic energy is obtained by solving the TKE transport equation, where the production term is estimated with the reconstructed Reynolds stress anisotropy~$\mathbf{b}$ and the additional correction term~$\mathbf{b}^R$.
    Further, we can obtain the velocity prediction by solving the RANS equation for each constructed Reynolds stress.

    \item Regularization: The neural network is trained based on the direct observation~$\mathbf{b}$ with a regularization step.
    The update scheme is written as
    $\tilde{\boldsymbol{w}} = \boldsymbol{w}^{i} - \mathsf{P}(\mathcal{N'}[w])^\top \mathsf{Q}^{-1} (\mathcal{N}[\boldsymbol{w}^i]-\mathsf{y}^{\boldsymbol{b}})$.
    Note that if the direct observation~$\mathbf{b}$ is not available, this step will be passed as~$\tilde{\boldsymbol{w}}=\boldsymbol{w}^{i}$.

    \item Kalman update: The ensemble Kalman method with observation augmentation is used to learn turbulence models from the direct observation of turbulent kinetic energy and the indirect observation of velocity.
    The observation is augmented to include both data sources, i.e.,~$\mathsf{y}^\text{a}=[\mathsf{y}^u, \mathsf{y}^k]$
    The update scheme is expressed as
    $\boldsymbol{w}^{i+1} = \tilde{\boldsymbol{w}} + \mathsf{K}^\text{a}(\mathsf{y}^\text{a} - \mathsf{H}^\text{a}[\tilde{\boldsymbol{w}}])$.
\end{itemize}
The iteration is stopped until the data misfit reduces below the noise level based on the discrepancy principle~\cite{ernst2015analysis}.
The schematic of the ensemble-based method is presented in Figure~\ref{fig:schematic}. 

\section{Test cases}
\label{sec:case_setup}

We demonstrate the capability of the present ensemble method in two flow cases, i.e., flows in a duct and flows over periodic hills.
They are representative cases for secondary flows and separated flows, respectively, both of which are challenging for linear eddy viscosity models.
The learned model is tested in similar configurations to evaluate the generalizability of the model as summarized in Table~\ref{tab:setup}.
Specifically, in the square duct flow case, the learned model is generalized to different ducts with varying aspect ratios.
In the periodic hill case, the learned model is generalized to the convergent-divergent channel and the curved back-forward step. 
In the following, the case setups are illustrated in detail.
In this work, both the direct data and indirect data have severe sparsity, which necessitates combining different data sources for learning generalizable turbulence models.

\begin{table}[!htb]
    \centering
    \begin{tabular}{ccccc}
    \hline
        training case &  direct data & indirect data & \shortstack{number of \\ data points} & test cases \\
        \hline
        \\
        \shortstack{Square duct with $z/y=1$ \\ (synthetic truth)} & $\mathbf{b}$ & $\mathbf{u}$ & 3 & \shortstack{Ducts with varying aspect ratios\\ ($z/y=1.2, 1.5, 1.8, 2.0$)}  \\ \\
        \shortstack{Periodic hills \\ (LES~\cite{breuer2009flow})} 
        &  $k$ & $\mathbf{u}$ & 16 & \shortstack{Convergent-divergent \\ channel~\cite{laval2011direct} \\ Curved back-forward \\ step~\cite{bentaleb2012large}}
        \\
    \hline
    \end{tabular}
    \caption{Summary of the case setup for flows in a square duct and flows over periodic hills.}
    \label{tab:setup}
\end{table}

\subsection{Flows in a square duct}

The square duct case is used to demonstrate the ability of the present ensemble method to recover underlying model functions based on direct and indirect sparse data.
In this case, the training data is from the RANS prediction with the Shih's quadratic model~\cite{shih1993realizable}, which is a widely used and validated turbulence model for predicting secondary flows~\cite{mompean1996predicting,lien2004simulation}.
By using this synthetic data, we ensure the compatibility between the model and the data.
In other words, there exist underlying models in the form of tensor bases to represent the training data.
Moreover, the analytical model form can provide the ground truth of tensor coefficients to evaluate the accuracy of the learned model function.
The $k$--$\varepsilon$ model~\cite{launder1974application} is used to estimate the turbulence time scale in this case.
In the Shih's quadratic model, the $\boldsymbol{g}$ function of the scalar invariant~$\boldsymbol{\theta}$ is written as
\begin{equation}
\begin{aligned}
  & g^{(1)}\left(\theta_{1}, \theta_{2}\right)=\frac{-2 / 3}{1.25+\sqrt{2 \theta_{1}}+0.9 \sqrt{-2 \theta_{2}}} \text{,} \\
  & g^{(2)}\left(\theta_{1} , \theta_{2}\right)=\frac{7.5}{1000+\left(\sqrt{2 \theta_{1}}\right)^{3}} \\
  & g^{(3)}\left(\theta_{1} , \theta_{2}\right)=\frac{1.5}{1000+\left(\sqrt{2 \theta_{1}}\right)^{3}} \\
  & g^{(4)}\left(\theta_{1}, \theta_{2}\right)=\frac{-9.5}{1000+\left(\sqrt{2 \theta_{1}}\right)^{3}} \text{.}
\end{aligned}
\end{equation}
The flow in a square duct is fully developed, and only one cell is used in the stream-wise direction.
Moreover, one-quarter of the domain is employed as the computation domain based on the symmetry property.

We regard the velocity as the indirect data and the Reynolds stress anisotropy as the direct data.
Only three data points are used to train the model, which highlights the ability of the method to learn reasonably good turbulence models from very sparse data. 
The data points are positioned around the corner of the computational domain as shown in Figure~\ref{fig:case_obs}.
The point at the left-top corner is discarded because of the symmetry.
The scalar invariants for the square duct case are in the range of $[0, 8]$.
It can be seen from Figure~\ref{fig:case_obs}(d) that the selected three points include two points with large scalar invariants and one point with small scalar invariants.
We assume that the training data have a relative standard deviation of $10^{-3}$ in this case.
Details results of the parameter study in terms of data noise, data positions, data amounts, the standard deviation of prior sampling, and the neural network architecture are presented in~\ref{sec:sensitivity}.

As for the neural network architecture, two scalar invariants~$\theta_1$ and $\theta_2$ are used as input features, and four tensor coefficients~$g^{(1-4)}$ are used in the output layer.
The model errors in the TKE transport equation are not involved in this case, since the used data are from the prediction of Shih's model.
For this reason, the correction function of $\boldsymbol{f}$ in the TKE equation is not used in this synthetic case. 
Two hidden layers with $5$ neurons per layer are employed based on our sensitivity study of the neural network architectures.
The regularized ensemble-based method is used to learn the neural network-based model from data of $\mathbf{b}$ and $\mathbf{u}$ by integrating the ensemble method and the back-propagation algorithm.
Specifically, the analytical gradient~${\partial \mathbf{b}}/ {\partial \boldsymbol{w}}$ from the back-propagation algorithm is first used to achieve good data fits in the Reynolds stress anisotropy~$\mathbf{b}$.
Further, the ensemble method is used to train the model based on velocity data~$\mathbf{u}$.
In order to highlight the superiority of combining the direct and indirect data, we compare the training result with that using only indirect data.
The results of using only direct data are omitted since it has been noted~\cite{duraisamy2021perspectives} that training with direct data is prone to provide inaccurate models that make poor velocity predictions in posterior tests due to the inconsistency between the training and prediction environments.
Moreover, it has been observed~\cite{zhang_ensemble-based_2022} that training with indirect data can provide more robust and accurate models than using direct data.
Therefore, we compare the results of combining direct and indirect data with those of using only indirect data in this work.

\subsection{Flows over periodic hills}

The flow over periodic hills is one canonical separated flow case, which is widely used for assessments of turbulence models.
The Reynolds number based on the bulk velocity and the height of the crest~$H$ is $10595$ in this case.
The large eddy simulation (LES) results~\cite{breuer2009flow} are used as the training data.
The nonlinear eddy viscosity model is inadequate to represent the LES data due to the weak equilibrium assumption.
As such, we demonstrate the capability of the proposed training method in the scenario of having inadequate model representation.
We use the $k$--$\omega$ SST model~\cite{menter1994two} as the baseline model.
The periodic boundary condition is imposed at the inlet and outlet of the computational domain, and the no-slip condition is imposed at the walls.

\begin{figure}[!htb]
    \centering
    \subfloat[square duct: $\theta_1$]{\includegraphics[width=0.2\textwidth]{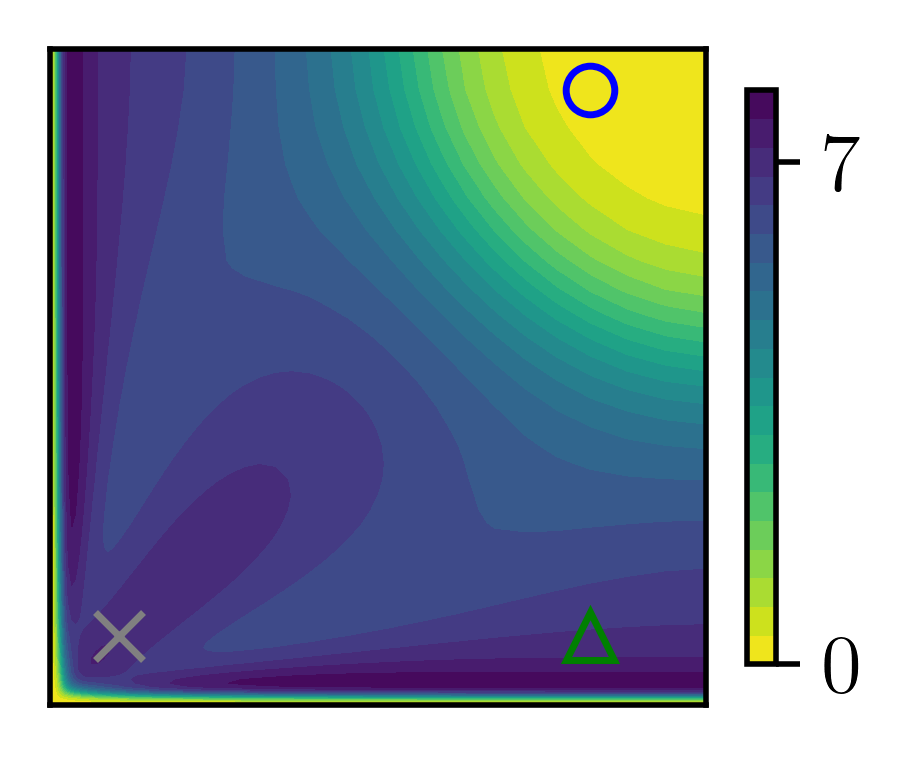}}
    \subfloat[periodic hills: $\theta_1$]{\includegraphics[width=0.35\textwidth]{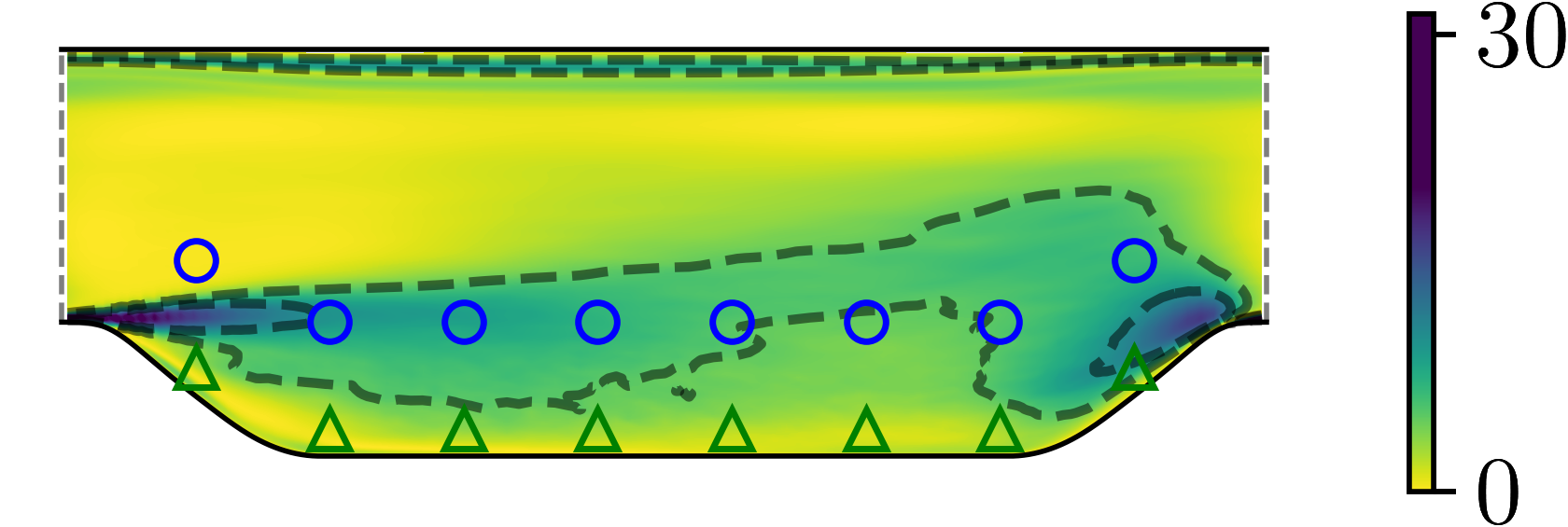}}
    \subfloat[periodic hills: $\theta_2$]{\includegraphics[width=0.35\textwidth]{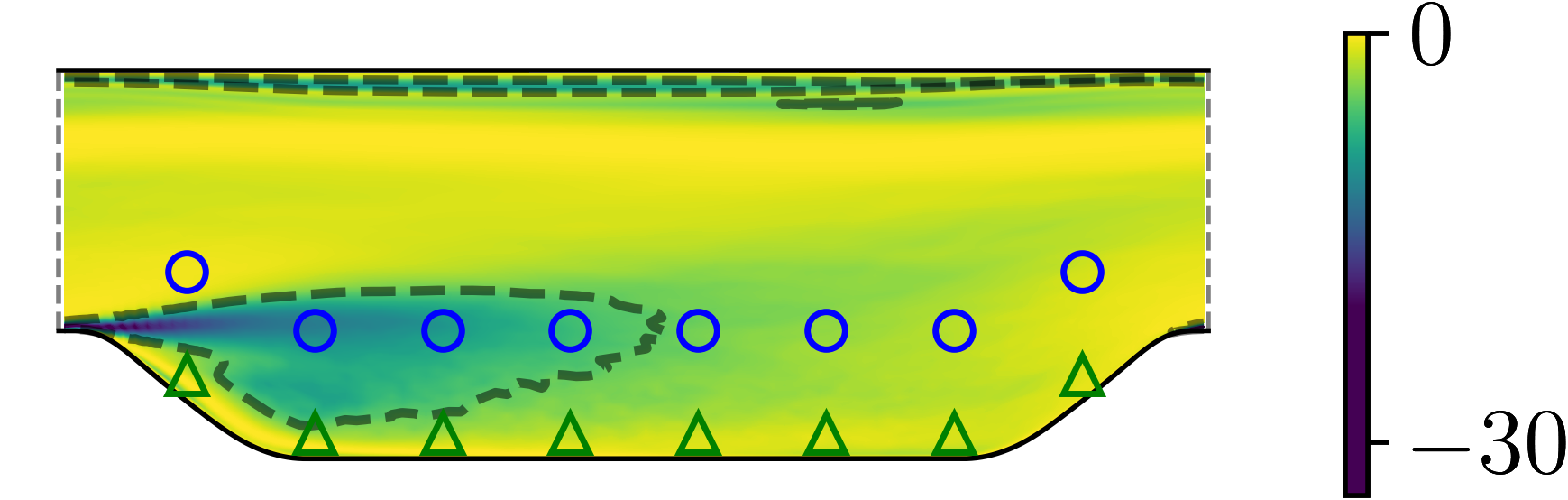}} \\
    \subfloat[square duct: feature distributions]{\includegraphics[width=0.3\textwidth]{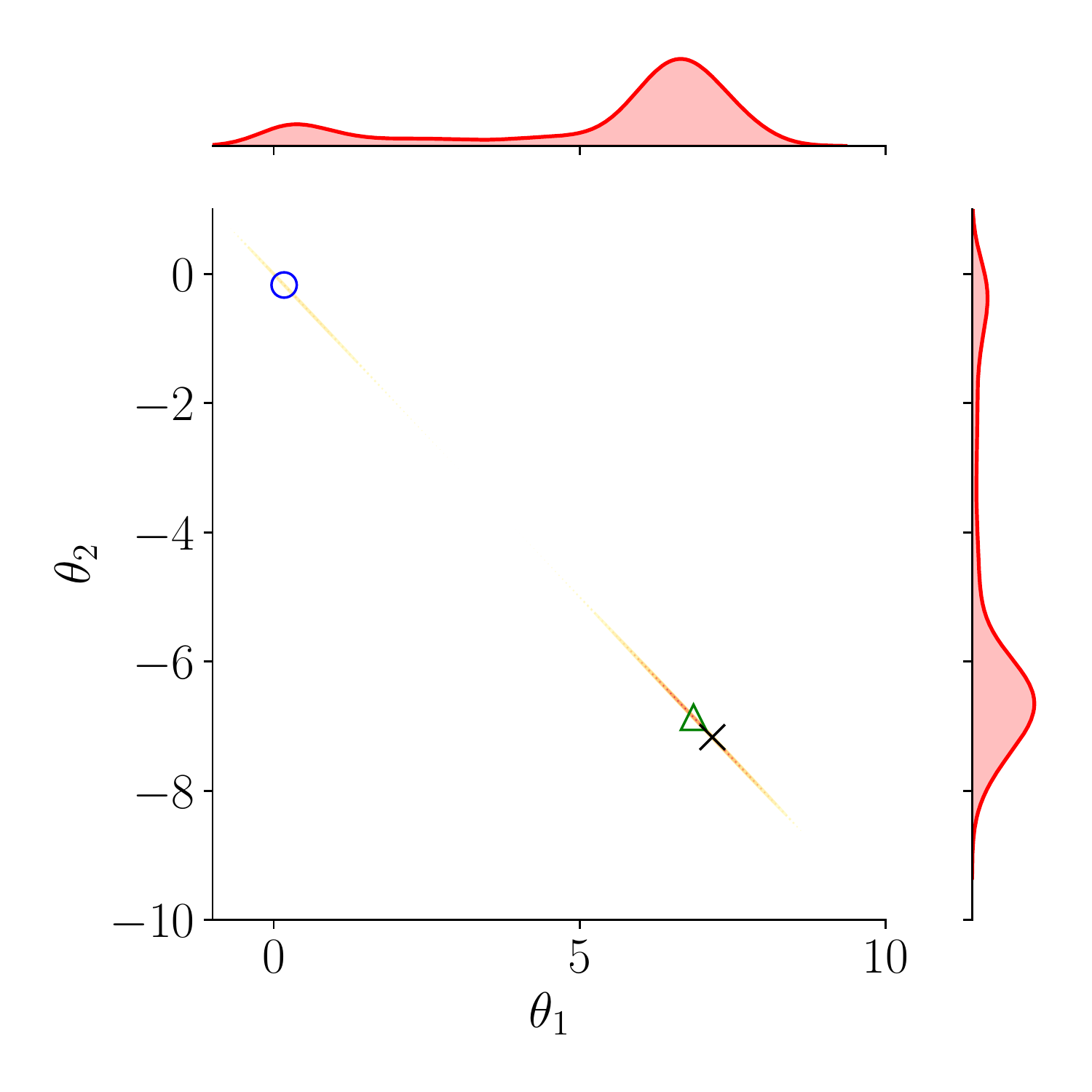}}
    \hspace{2cm}
    \subfloat[periodic hills: feature distributions]{\includegraphics[width=0.3\textwidth]{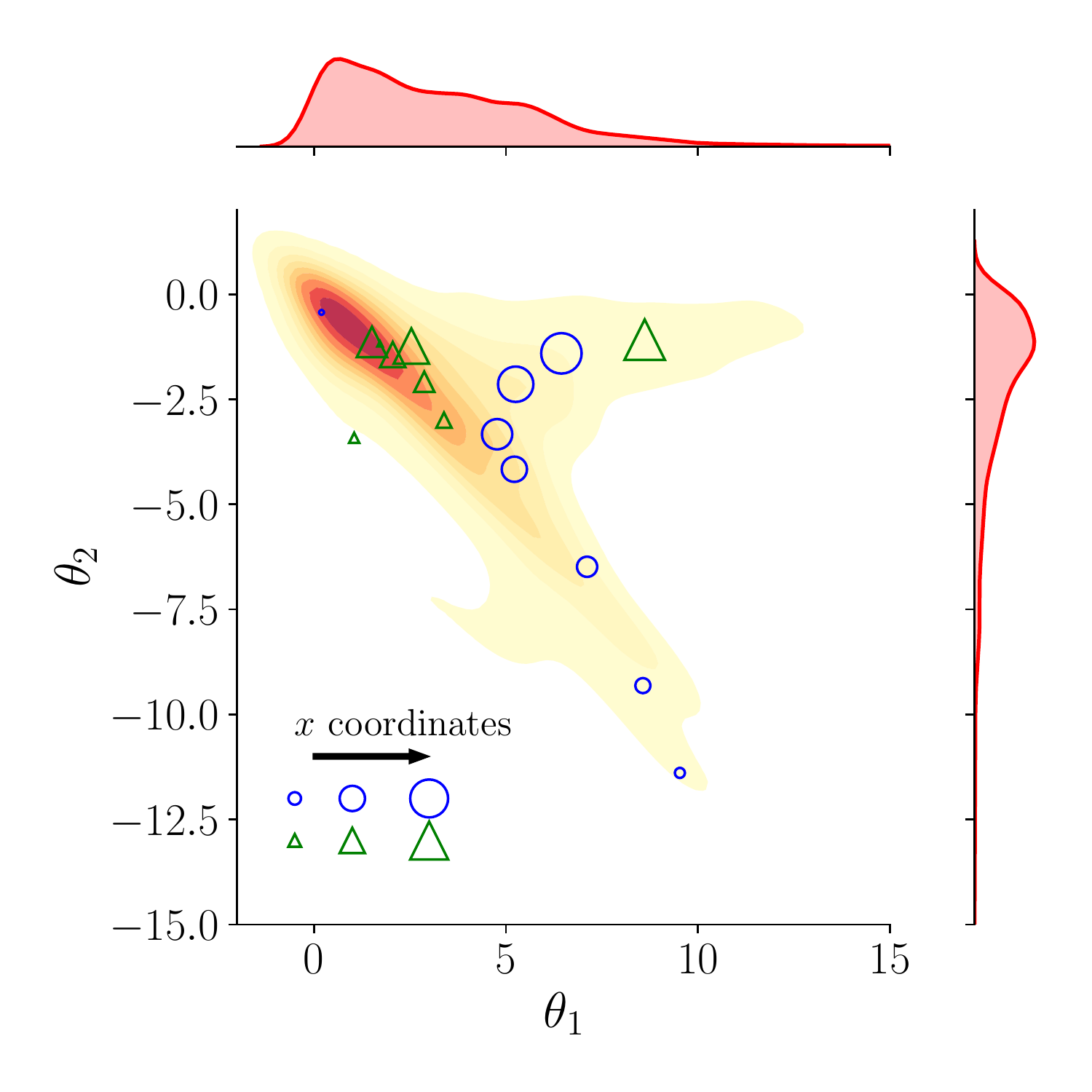}}
    \caption{Contour plots of scalar invariants in the square duct case and the periodic hill case. The data points are indicated with symbols. Panels (d) and (e) show the distribution of the data point in the space of $\theta_1$--$\theta_2$ in the two cases, respectively.}
    \label{fig:case_obs}
\end{figure}

In the periodic hill case, the turbulent kinetic energy is taken as the direct data, and the velocity is used as the indirect data.
The case with only indirect data of velocity is also provided for comparison.
We use very sparse observations (only $16$ data points) to improve the prediction of the flow field as indicated in Figure~\ref{fig:case_obs}.
The relative standard deviation of both direct and indirect data is set as $0.01$ for this case.
The observation data are positioned along two lines: one is $0.2H$ away from the bottom wall, and the another is $H$ away from the bottom wall.
It can be seen from Figure~\ref{fig:case_obs}(g) that the observation data near the wall have relatively small magnitudes in scalar invariants.
The magnitude of scalar invariants is mostly less than $5$, except for one point with $\theta_1$ around $10$.
This point is located at the windward side of hills, and the large magnitude in scalar invariants is due to flow accelerations by the channel contraction.
In contrast, the observation data at the distance of $H$ away from the wall have scalar invariants in a wider range of $[0, 10]$.
One point has a magnitude around zero for both scalar invariants, which is the leftmost point in the computational domain where the velocity gradient is small.

The trained model is further tested on two different configurations, i.e., curved back-forward step (hereafter referred to as CBFS) and convergent-divergent channel (hereafter referred to as ConvDiv).
Both test cases have similar Reynolds numbers as the training case.
The CBFS case has the Reynolds number of $13700$, and the ConvDiv case has the Reynolds number of $12600$.
The high-fidelity data are available from the LES~\cite{bentaleb2012large} for the CBFS case and the DNS~\cite{marquillie2008direct} for the ConVDiv case.
These data are collected by McConkey et al.~\cite{mcconkey2021curated} and adopted in this work.

In the periodic hill cases, the ensemble method with augmented observation is employed to combine the direct data of TKE and the indirect data of velocity.
The gradient of the TKE to the neural network weights is typically obtained by solving the adjoint equations, which pose difficulties in code re-development.
Hence, we use the ensemble method to incorporate the two data sources simultaneously by augmenting the indirect velocity data with the direct data~$k$, which corresponds to Scenario II as illustrated in Table~\ref{tab:schemes}.
Moreover, in this case, the LES results are used as the training data, while the baseline model, i.e., the nonlinear eddy viscosity model, is inadequate to represent the LES data since it is under weak equilibrium assumptions.
For this reason, the additional correction in the TKE transport equation has to be used to remedy this data-model incompatibility.
The output layer of neural network includes $g^{(1-2)}$ and $f^{(1)}$, and the input features are the first two scalar invariants~$\theta_1$ and $\theta_2$.
The first tensor basis has been identified as most relevant to the correction term in TKE transport equation~\cite{schmelzer2020discovery}. 
Hence we only learn $f^{(1)}$ function in this case.
We apply the neural network with ten layers and 10 neurons per layer based on our previous sensitivity studies~\cite{zhang_ensemble-based_2022}.
The coefficients of $\boldsymbol{g}$ and $\boldsymbol{f}$ are combined with tensor basis~$\mathbf{T}$ to reconstruct the Reynolds stress anisotropy~$\mathbf{b}$ and corrective term~$\mathbf{b}^R$ in the TKE transport equation, respectively.
The TKE fields are obtained by solving the TKE transport equation with the correction terms.
The velocity field is propagated from the reconstructed Reynolds stress fields by solving the RANS equations.
Further the propagated TKE and velocity are coupled with training data based on the ensemble method to update the weights of the neural network.

We compare with the results of Schmelzer et al.~\cite{schmelzer2020discovery}, which are considered a benchmark, to demonstrate the capability of the present method.
Note that the training method and data used in the work of Schmelzer et al.~\cite{schmelzer2020discovery} are different from the present work as illustrated in Section~\ref{sec:comparison}.
Specifically, Schmelzer et al.~\cite{schmelzer2020discovery} use the sparse regression to learn turbulence models, while we use the ensemble Kalman method to achieve this goal.
As to training data, they use the full fields of velocity, TKE, and Reynolds stresses from high-fidelity simulations.
Moreover, the indirect data of velocity is used in their framework to estimate the reference value for specific dissipation rate instead of as training data.
On the contrary, the present work uses very sparse data at $16$ measurement points in both direct data of TKE and indirect data of velocity.
To highlight the data sparsity used in this work, we compare the data sources among different works, including learning from direct data~$\boldsymbol{b}$~\cite{ling2016reynolds}, learning with sparse regression~\cite{schmelzer2020discovery}, learning from indirect data~$\boldsymbol{u}$ with the ensemble method~\cite{zhang_ensemble-based_2022}, learning from indirect data $\boldsymbol{u}$ and direct data $k$ with genetic programming, and the present work.
The data used in these works are summarized in Table~\ref{tab:data_sparsity}, which clearly shows the high data sparsity of the present work.

\begin{table}[!htb]
    \centering
    \begin{tabular}{cccccc}
    \hline
        \shortstack{data \\ sources} & present & Ling et al. &  Schmelzer et al. &
        Zhang et al. &
        Waschkowski et al. \\
        \hline
        $\bm{b}$     & --     & full        & full         & --   & --    \\ \\
        $k$          & 16 points  & --          & full         & --   & \shortstack{6 profiles\\(960 points)} \\ \\
        $\mathbf{u}$ & 16 points & --          & full         & \shortstack{4 profiles\\(600 points)}     & \shortstack{6 profiles\\(960 points)} \\
    \hline
    \end{tabular}
    \caption{Summary of data sparsity in different works, including learning from direct data~\cite{ling2016reynolds}, learning with sparse regression~\cite{schmelzer2020discovery}, learning with ensemble method~\cite{zhang_ensemble-based_2022}, learning with genetic programming~\cite{waschkowski2022multi}, and the present work.
    The full field indicates the number of data points larger than $10000$.}
    \label{tab:data_sparsity}
\end{table}

The open-source code OpenFOAM is used in this work to solve the RANS equations.
Specifically, the built-in solver~\textit{simpleFOAM} is applied to solve the RANS equation based on the given Reynolds stress fields.
The DAFI code~\cite{strofer2021dafi} is used to implement the ensemble-based training algorithm.
The TensorFlow is employed to construct the tensor basis neural network for evaluating the Reynolds stress anisotropy.
The rectified linear unit (ReLU) activation function is used for the hidden layers, and the linear activation function is used for the output layer.

\section{Results}
\label{sec:results}

\subsection{Flows in a square duct}
\begin{figure}[!htb]
    \centering
    \begin{tabular}{ccccc}
        & $u_y$ & $\tau_{xy}$ & $\tau_{yz}$ & $\tau_{yy}-\tau_{zz}$ \\
        \rotatebox[origin=c]{90}{ground truth} & 
        \raisebox{-.5\height}{\includegraphics[scale=0.5]{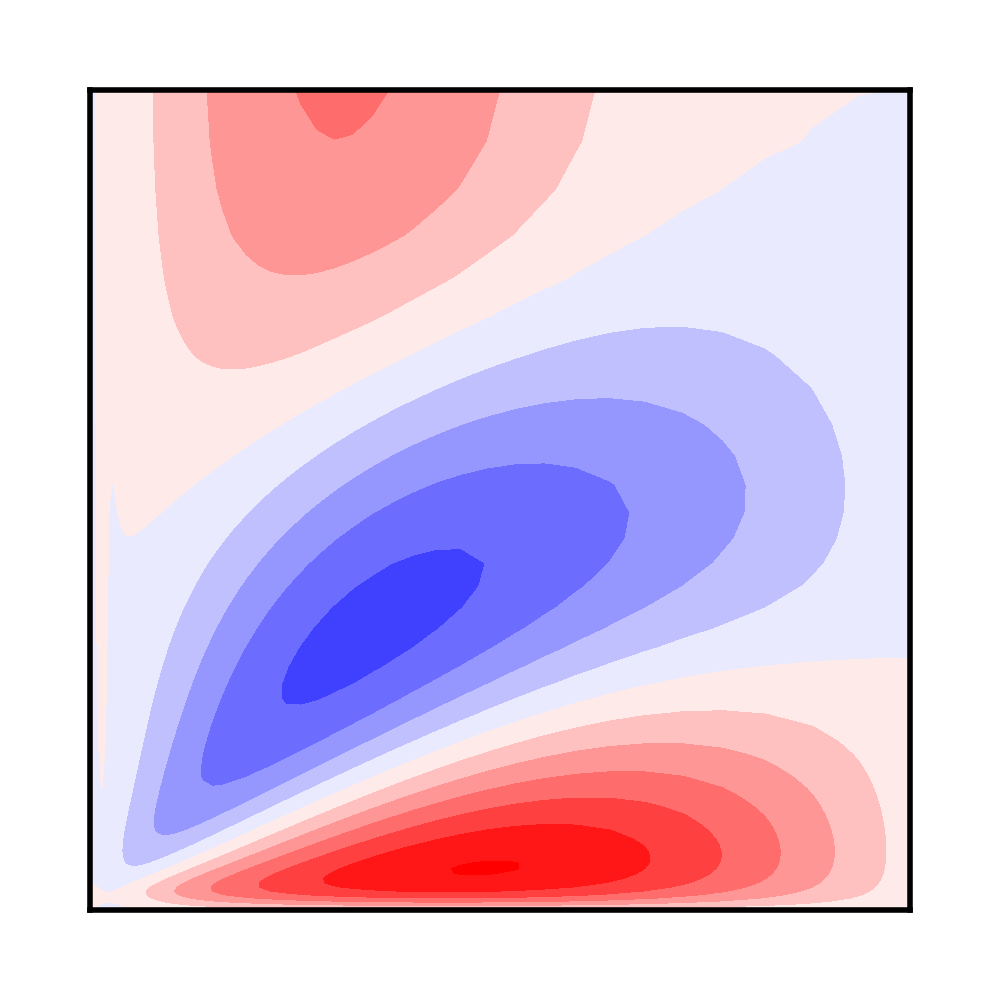}} &
        \raisebox{-.5\height}{\includegraphics[scale=0.5]{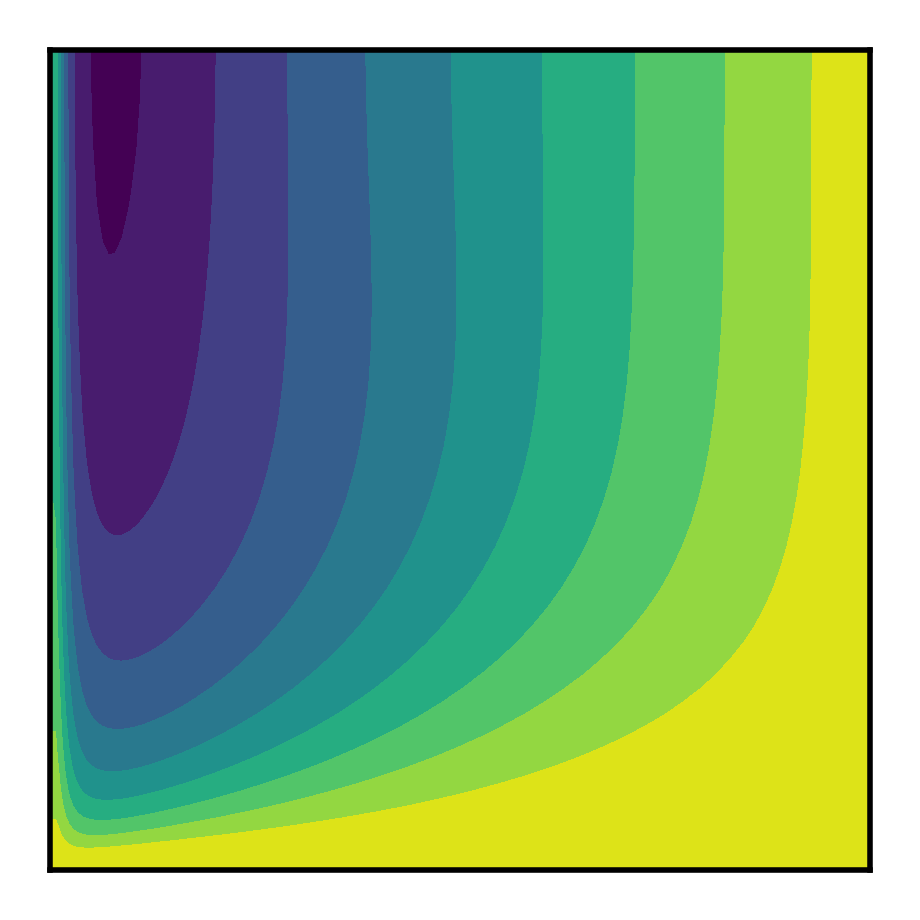}} &
        \raisebox{-.5\height}{\includegraphics[scale=0.5]{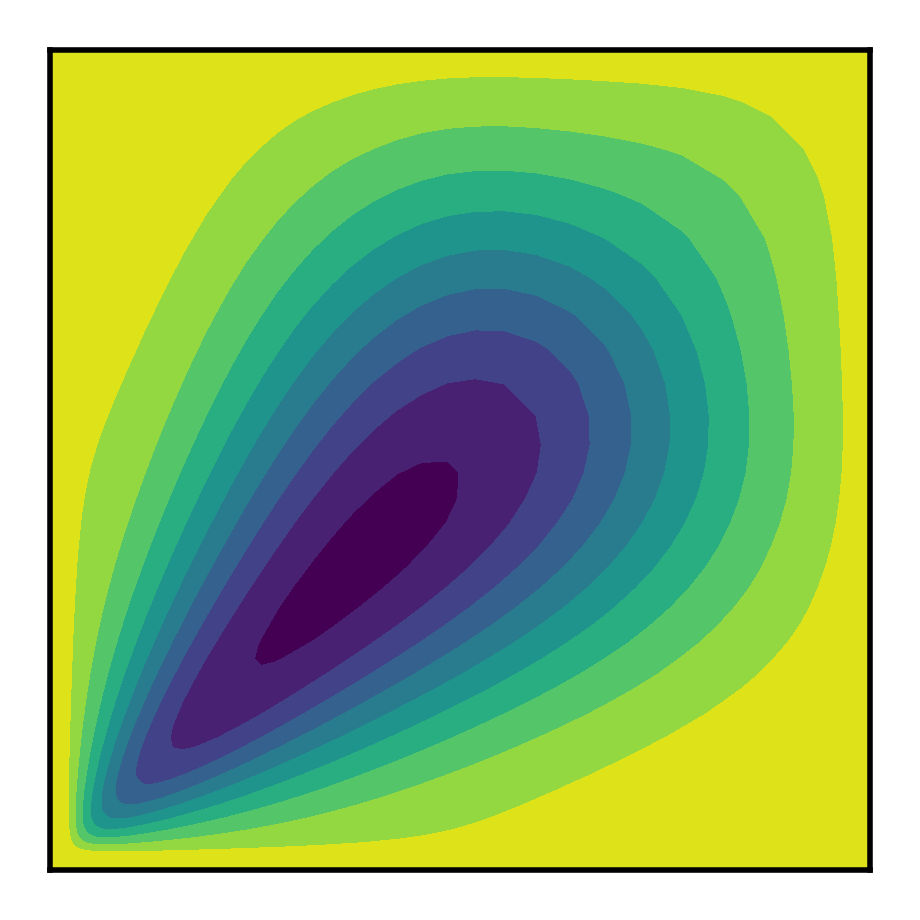}} &
        \raisebox{-.5\height}{\includegraphics[scale=0.5]{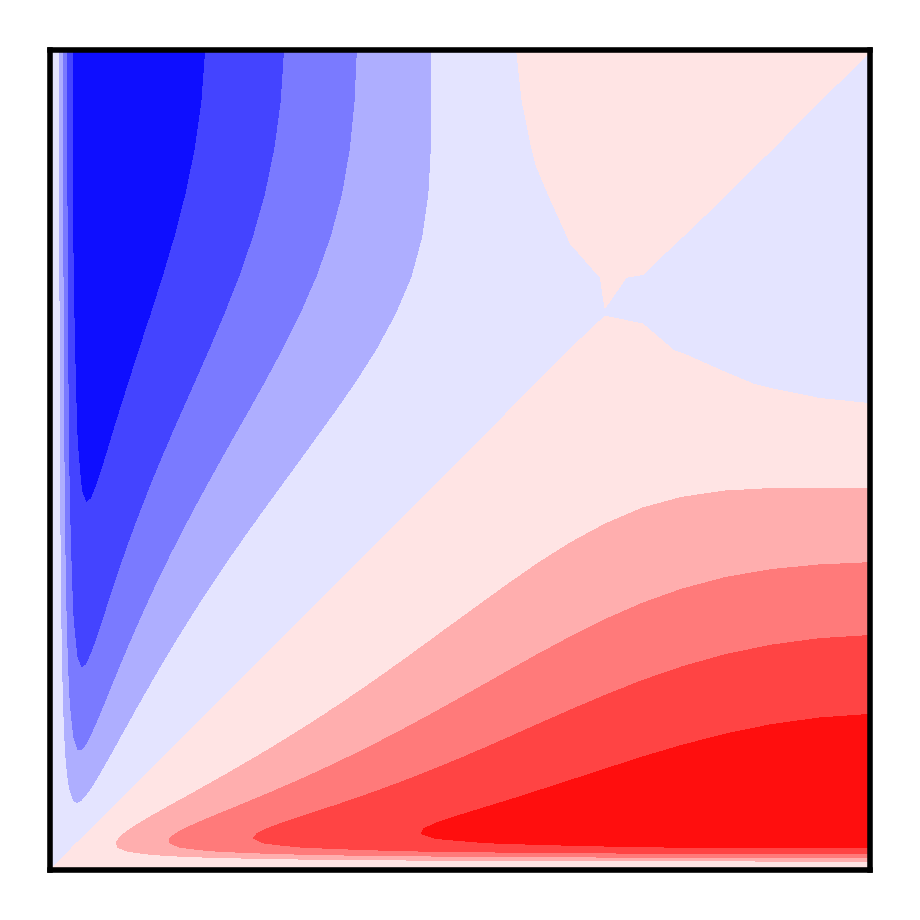}}
        \\
        \rotatebox[origin=c]{90}{indirect} & 
        \raisebox{-.5\height}{\includegraphics[scale=0.5]{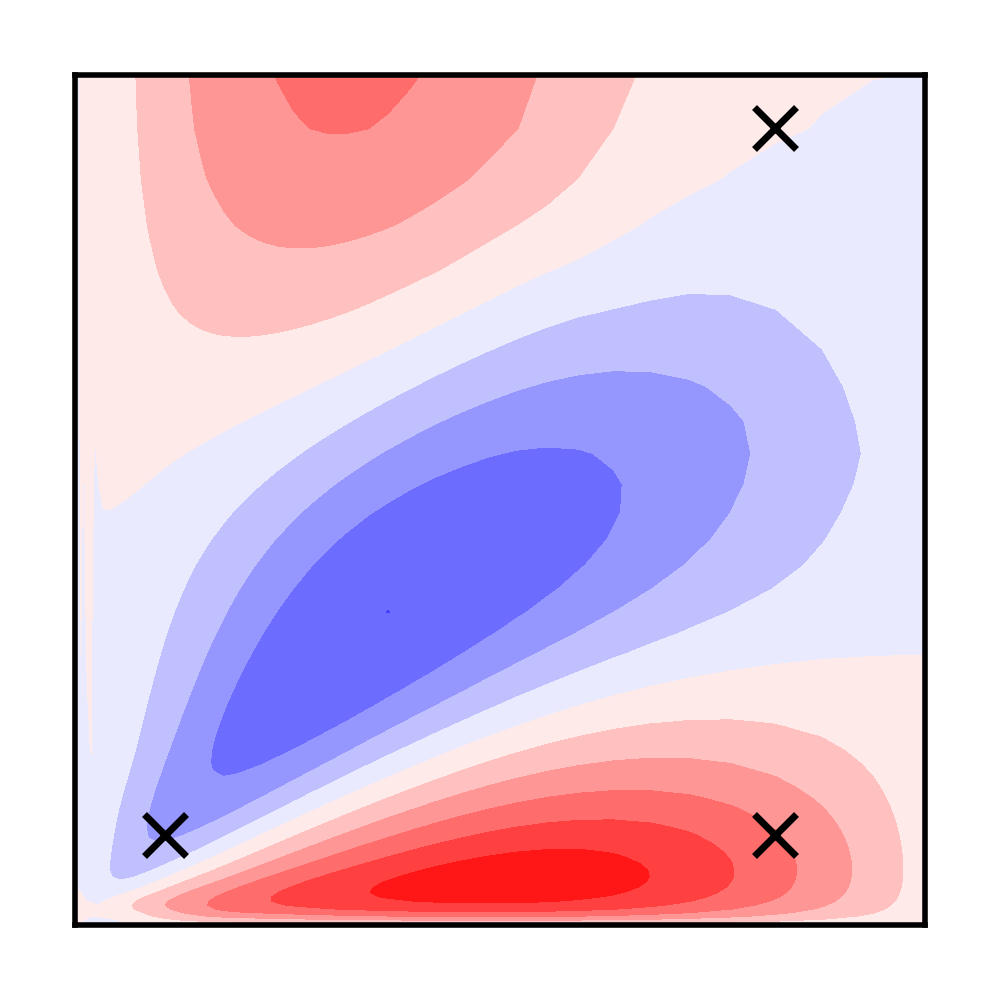}} &
        \raisebox{-.5\height}{\includegraphics[scale=0.5]{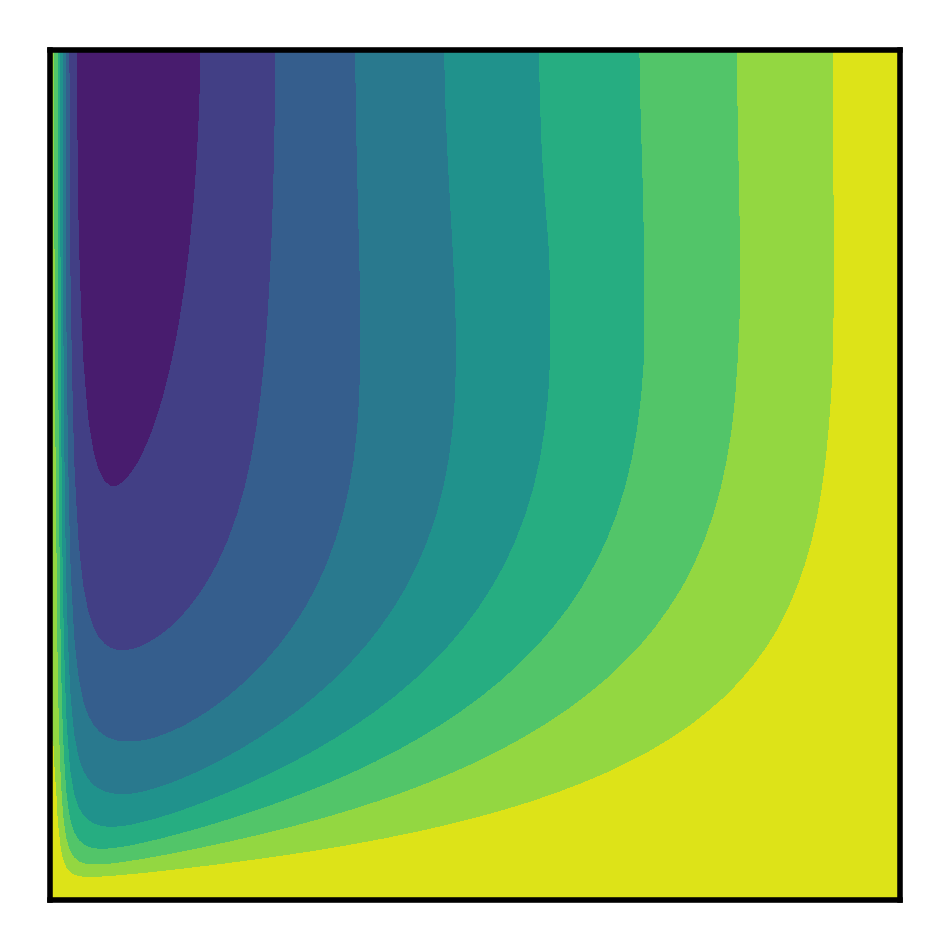}} &
        \raisebox{-.5\height}{\includegraphics[scale=0.5]{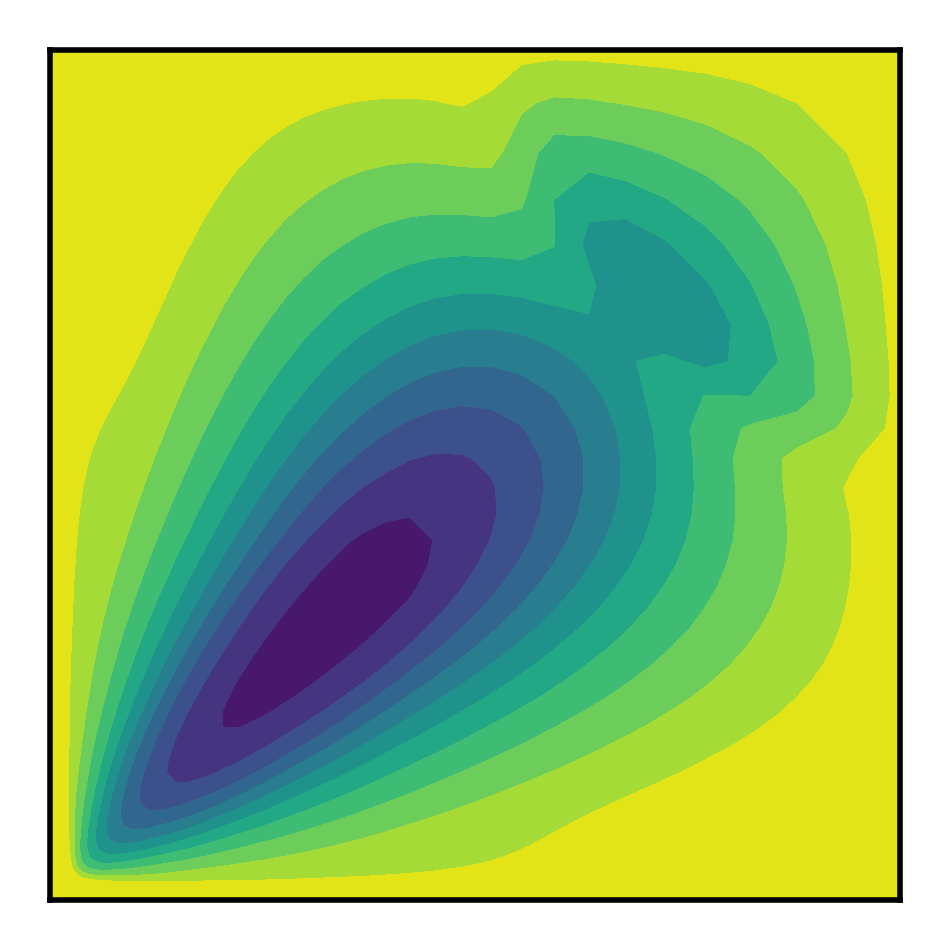}} &
        \raisebox{-.5\height}{\includegraphics[scale=0.5]{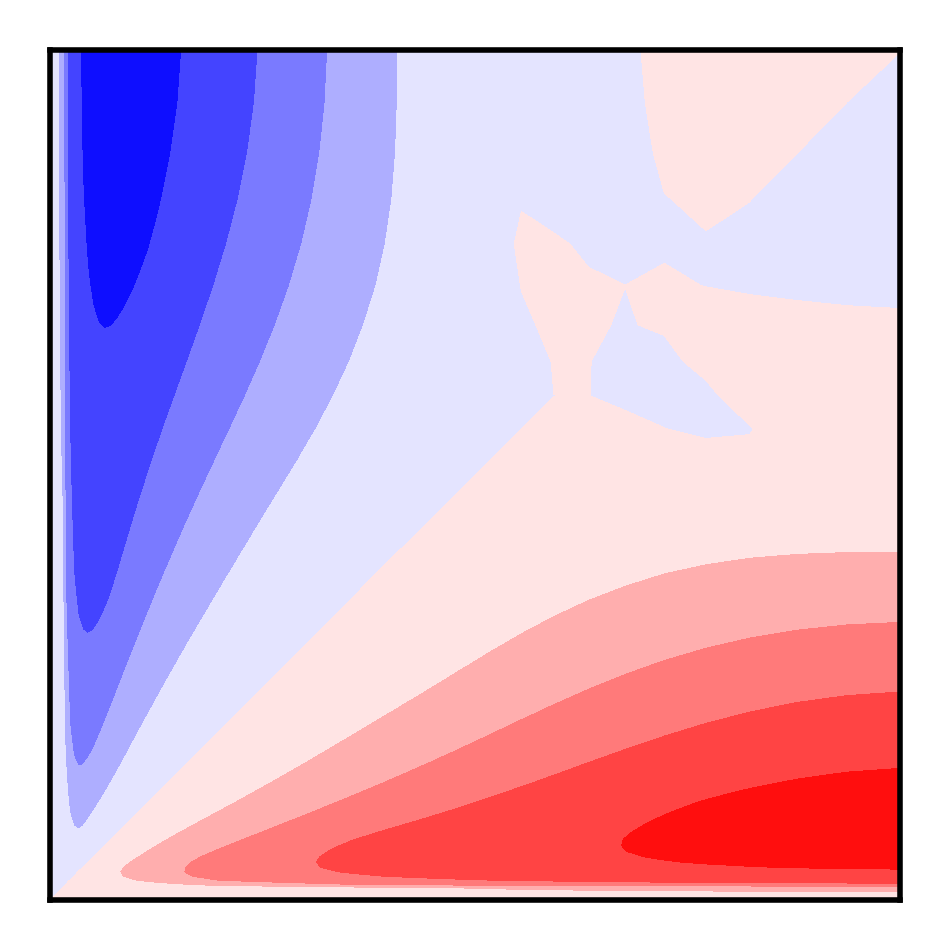}}
        \\
        \rotatebox[origin=c]{90}{direct + indirect} &
        \raisebox{-.5\height}{\includegraphics[scale=0.5]{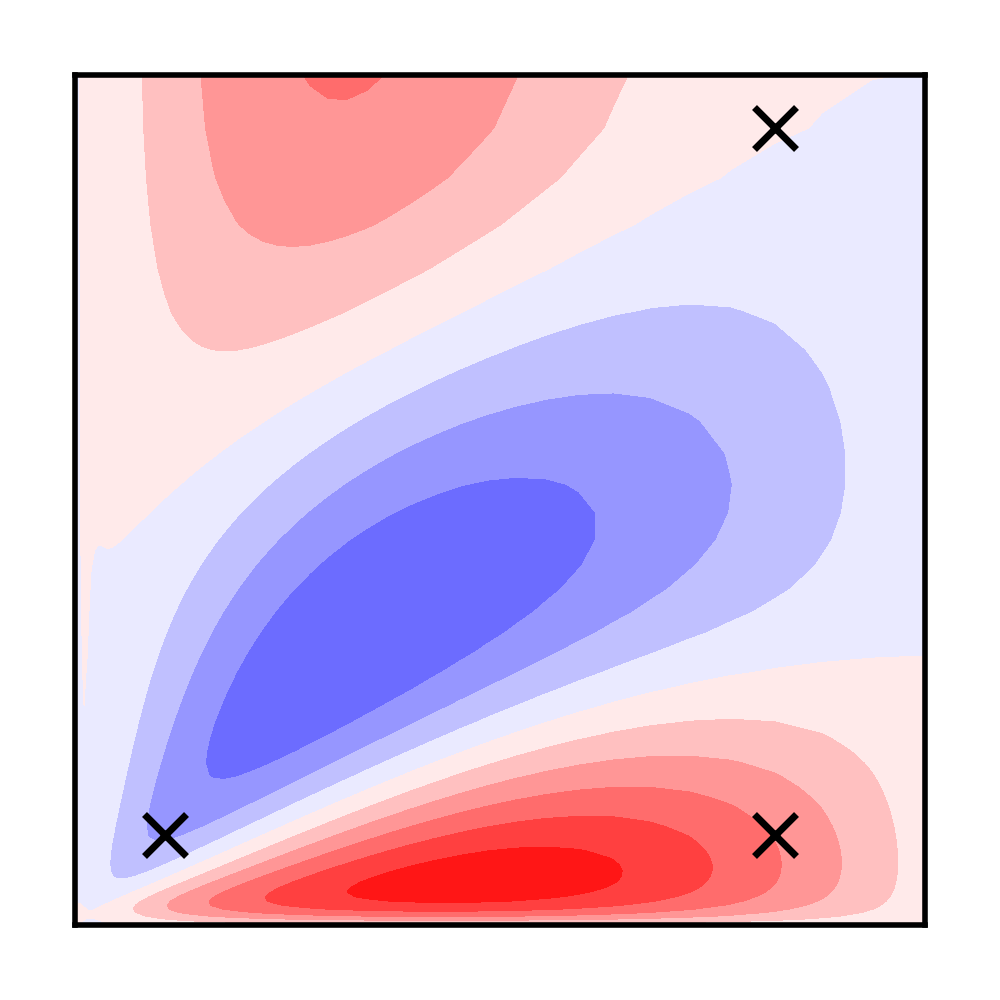}} &
        \raisebox{-.5\height}{\includegraphics[scale=0.5]{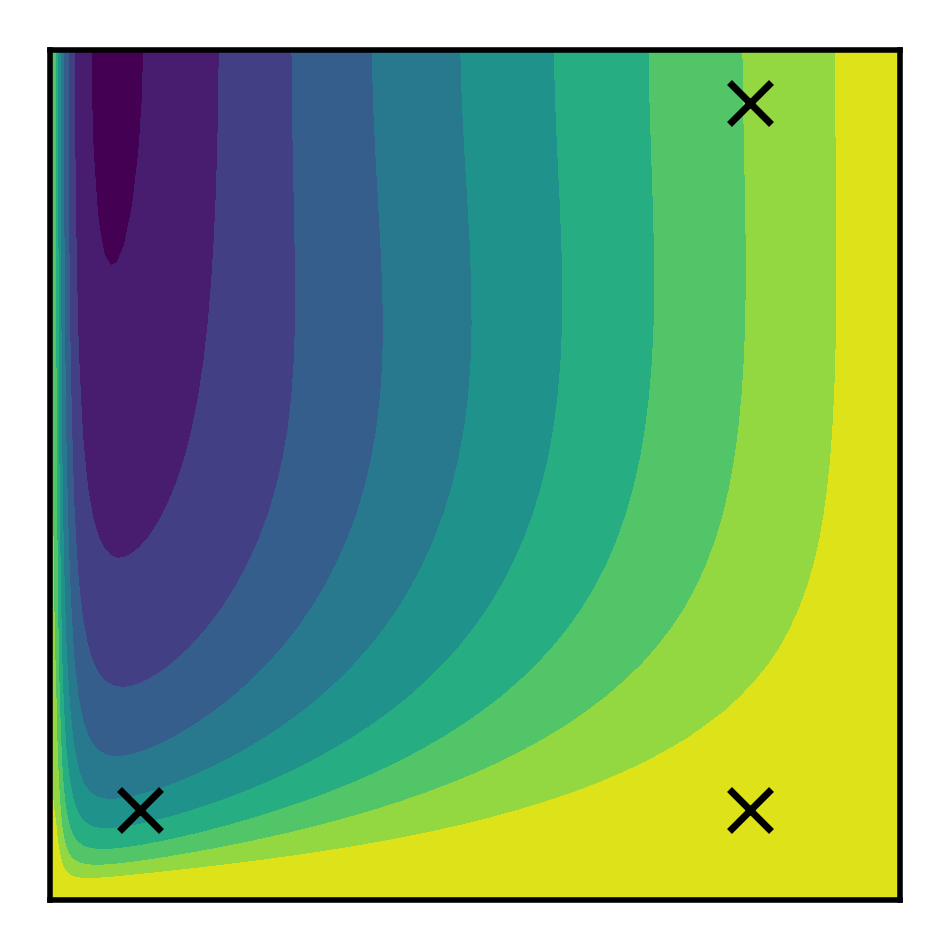}} &
        \raisebox{-.5\height}{\includegraphics[scale=0.5]{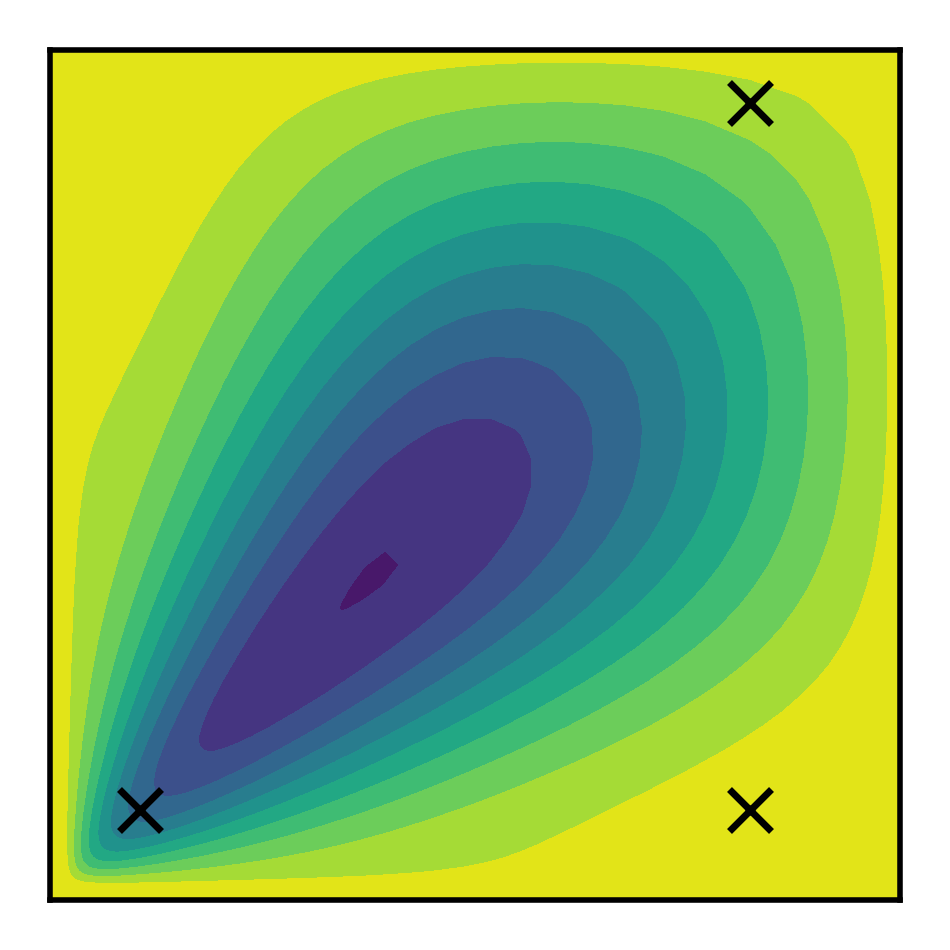}} &
        \raisebox{-.5\height}{\includegraphics[scale=0.5]{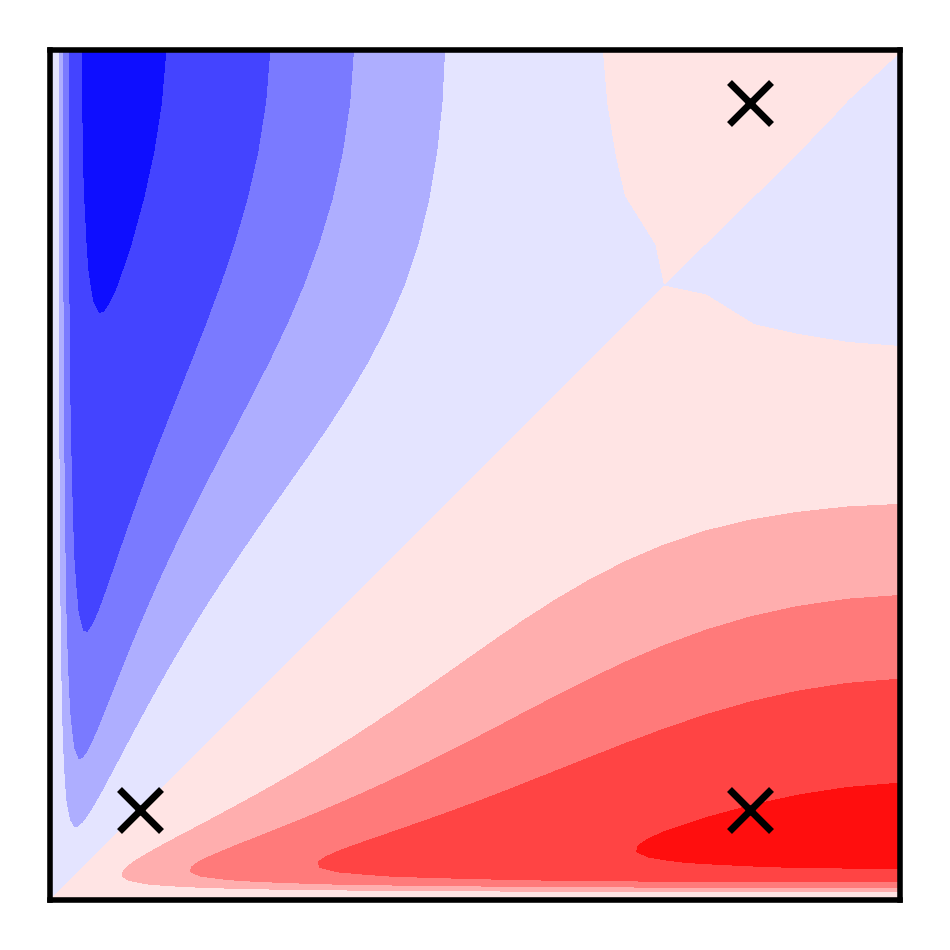}}
        \\
        & \raisebox{-.5\height}{\includegraphics[scale=0.7]{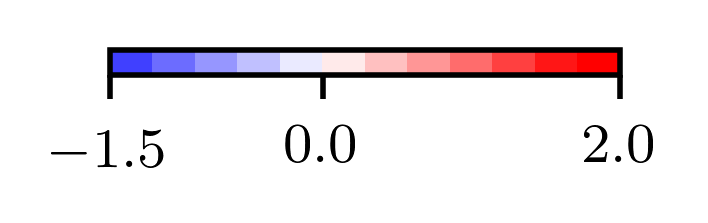}} 
        & \raisebox{-.5\height}{\includegraphics[scale=0.7]{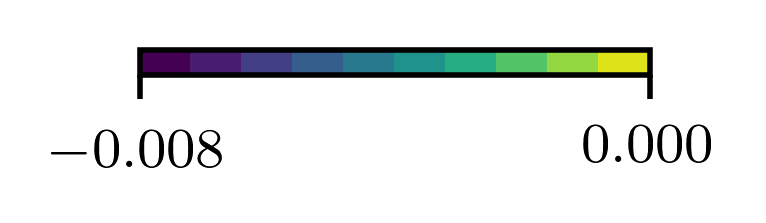}} 
        & \raisebox{-.5\height}{\includegraphics[scale=0.7]{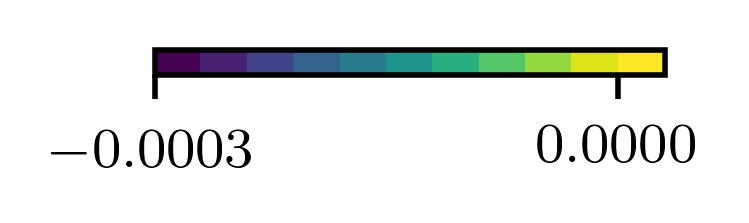}}
        &\raisebox{-.5\height}{\includegraphics[scale=0.7]{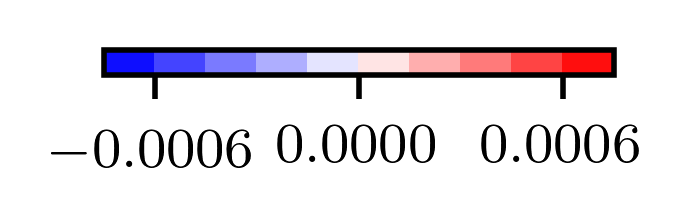}}
    \end{tabular}
    \caption{ 
     Plots of velocity~$u_y$, Reynolds shear stresses $\tau_{xy}$ and $\tau_{yz}$, and normal stresses imbalance $\tau_{yy}-\tau_{zz}$ predicted from the learned models with only indirect data (center row) and with the direct and indirect data (bottom row), compared with the ground truth (top row), for the square duct case. 
     }
    \label{fig:sd_results}
\end{figure}

Combining the direct and indirect data can achieve good predictions of both Reynolds stress and velocity for flows in a square duct.
It can be seen from Figure~\ref{fig:sd_results}, where the positions of three data points are indicated with crosses to highlight the data sparsity.
However, using only the indirect velocity data is not able to recover well the Reynolds stress.
Around the duct center, there exist noticeable discrepancies between the model output and the ground truth.
That is because the magnitude of the Reynolds stress is small near the duct center, leading to the insensitivity of Reynolds stress to the velocity.
The learned model can provide different Reynolds stress fields that make satisfactory velocity predictions at the observed location near the duct center.
Moreover, only one data point of velocity is placed near the duct center, which poses additional challenges in recovering the associated Reynolds stress.
Combining the indirect and direct observation can improve the prediction of the Reynolds stress as shown in the last row of Figure~\ref{fig:sd_results}.
By adding Reynolds stress data, the method can take synergy effects between the velocity and Reynolds stress data, thereby improving the reconstruction of the Reynolds stress field.
The results highlight the necessity of direct data to reconstruct well the Reynolds stress field, particularly at the positions where the Reynolds stress is insensitive to the velocity.

\begin{figure}[!htb]
    \centering
    \includegraphics[width=0.7\textwidth]{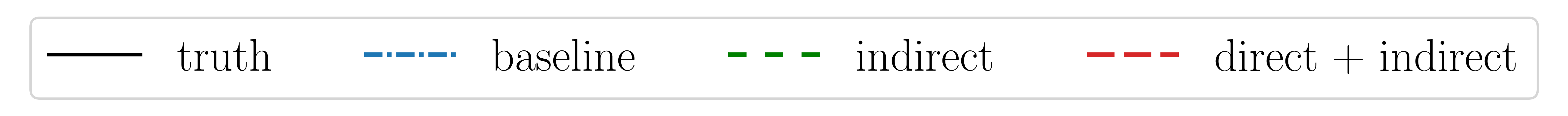}
    \subfloat[$u_y$]{\includegraphics[width=0.3\textwidth]{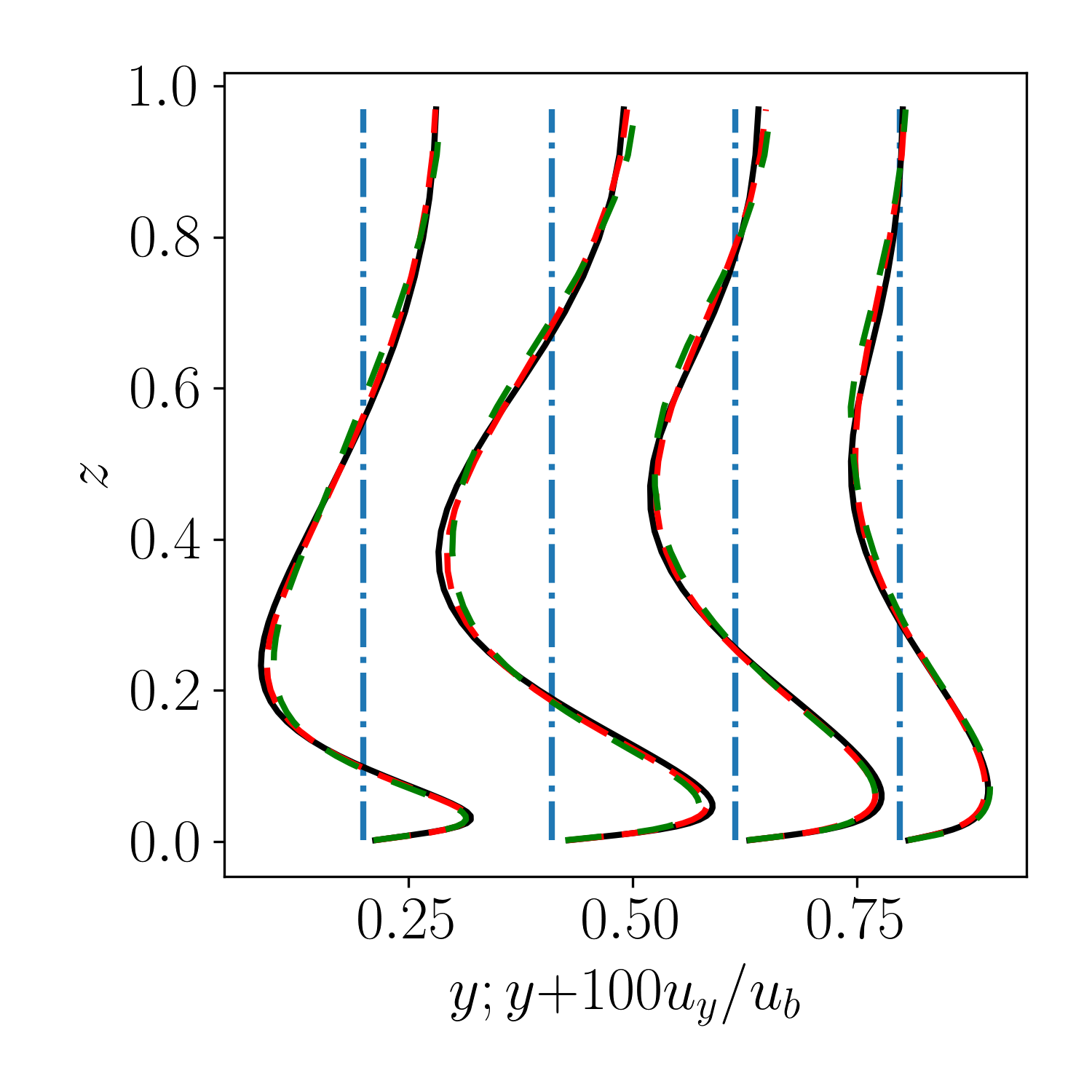}}
    \subfloat[$\tau_{yz}$]{\includegraphics[width=0.3\textwidth]{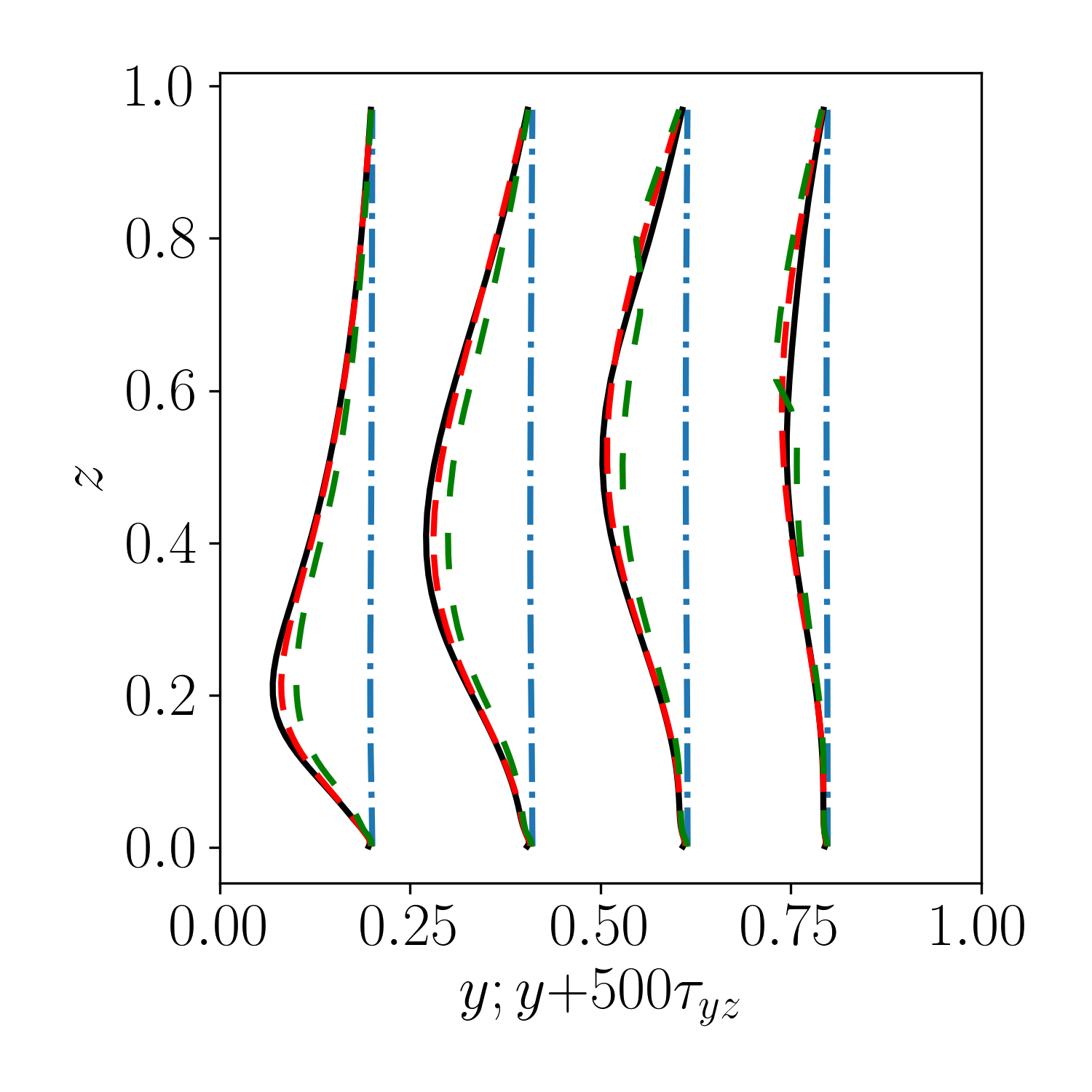}}
    \subfloat[$\tau_{yy}-\tau_{zz}$]{\includegraphics[width=0.3\textwidth]{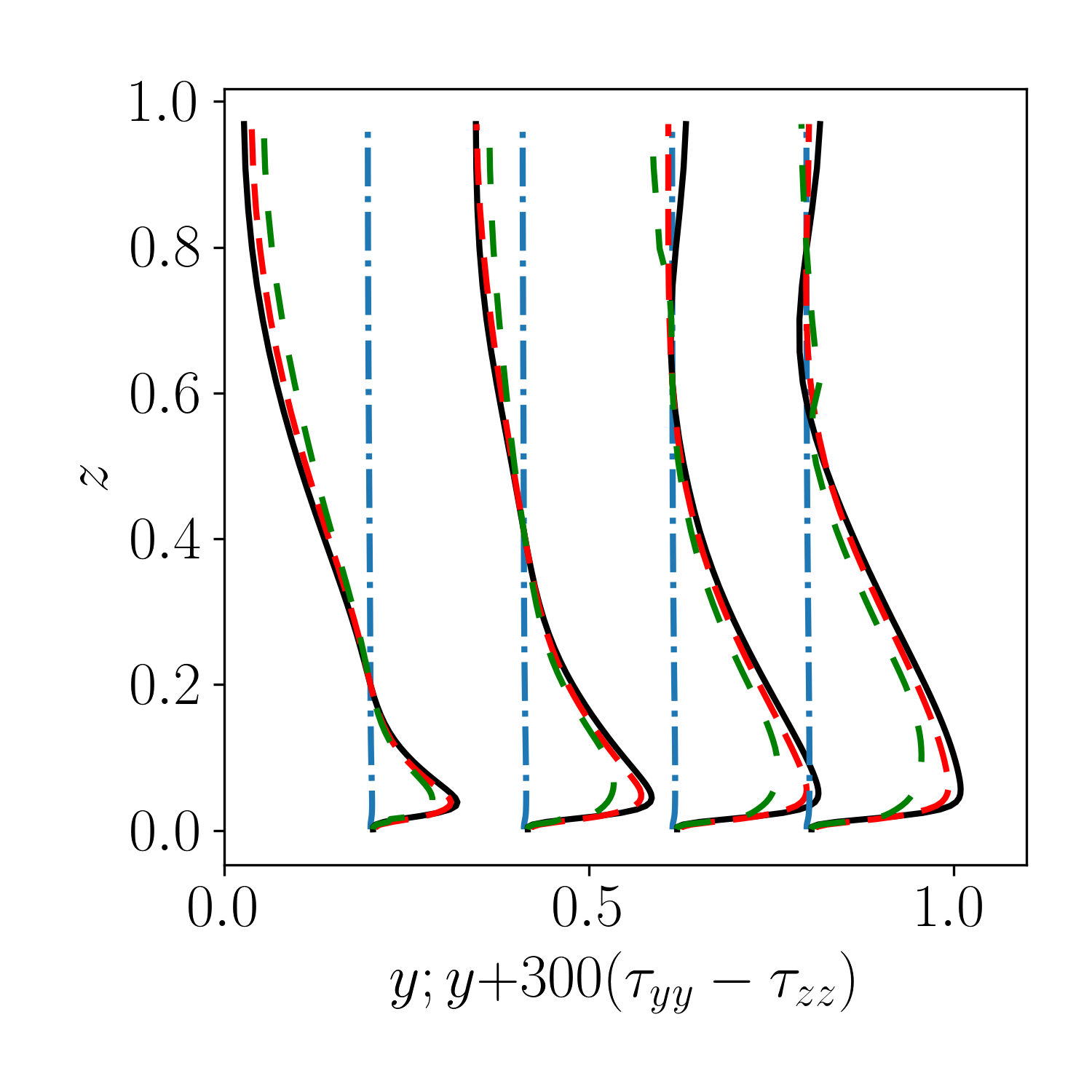}}
    \caption{Comparison of in-plane velocity and Reynolds stress along profiles among the truth, the model learned from indirect data, and the model learned from indirect and direct data for the square duct case.}
    \label{fig:prof_sd}
\end{figure}

The profiles of the predicted velocity and Reynolds stresses are plotted in Figure~\ref{fig:prof_sd}, which clearly shows that combining direct and indirect data can achieve better agreement with the synthetic truth compared to only using the indirect data.
Specifically, combining the direct and indirect data can make better predictions in the Reynolds stress than using only indirect data.
For the Reynolds shear stress~$\tau_{yz}$ and the imbalance of the Reynolds normal stresses~$\tau_{yy}-\tau_{zz}$, using only the indirect data underestimates the magnitude, while combining direct and indirect data can noticeably improve the model prediction and achieve a good agreement with the data.
The two methods produce very similar velocity predictions since both use the velocity data to train the models.
The improved Reynolds stress prediction occurs mainly in regions where the Reynolds stress is insensitive to the velocity, e.g., near the duct center from the plots in Figure~\ref{fig:sd_results}.
Hence, combining the direct and indirect data leads to improved Reynolds stress prediction and similar velocity prediction compared to using only indirect data.

\begin{figure}[!htb]
    \centering
    \subfloat[contours of invariants ]{
    \begin{tabular}{ccccl} 
        & indirect & indirect $+$ direct & truth & \\
        \rotatebox[origin=c]{90}{$\theta_1$} & 
        \raisebox{-.5\height}{\includegraphics[scale=0.45]{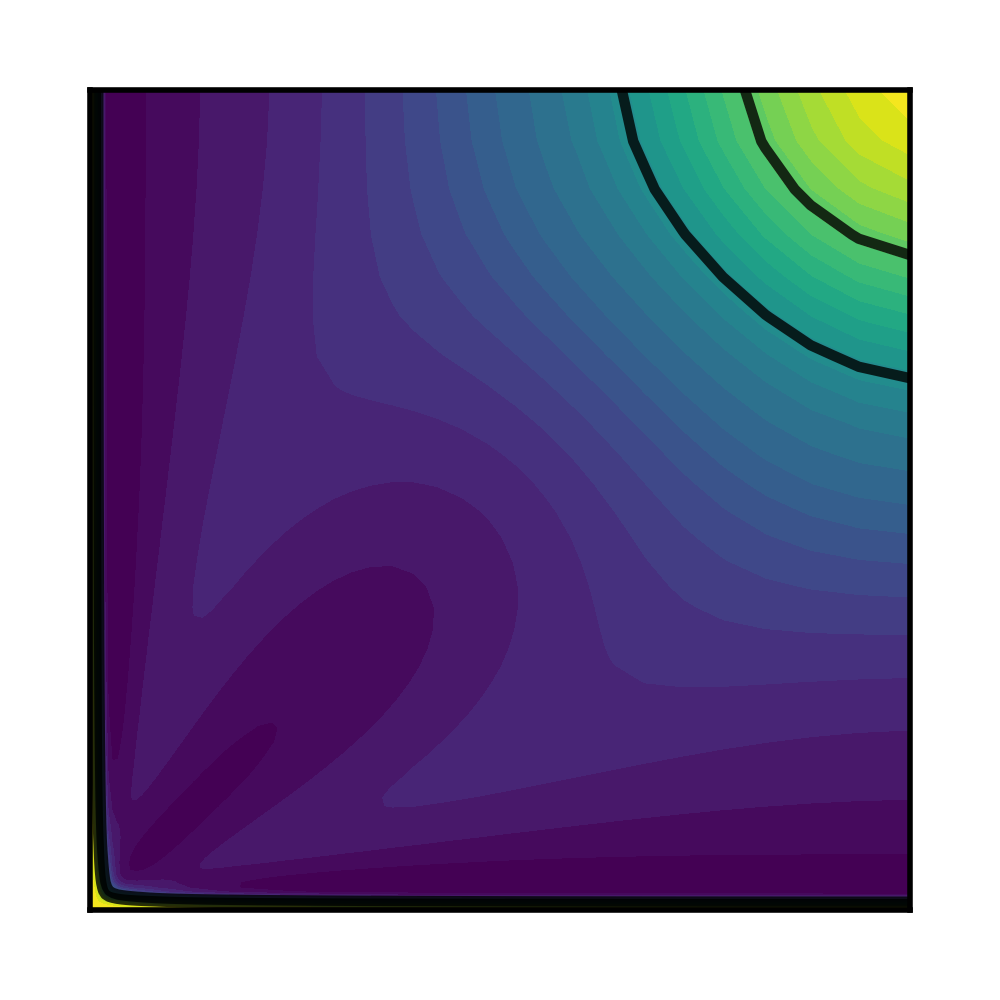}} &
        \raisebox{-.5\height}{\includegraphics[scale=0.45]{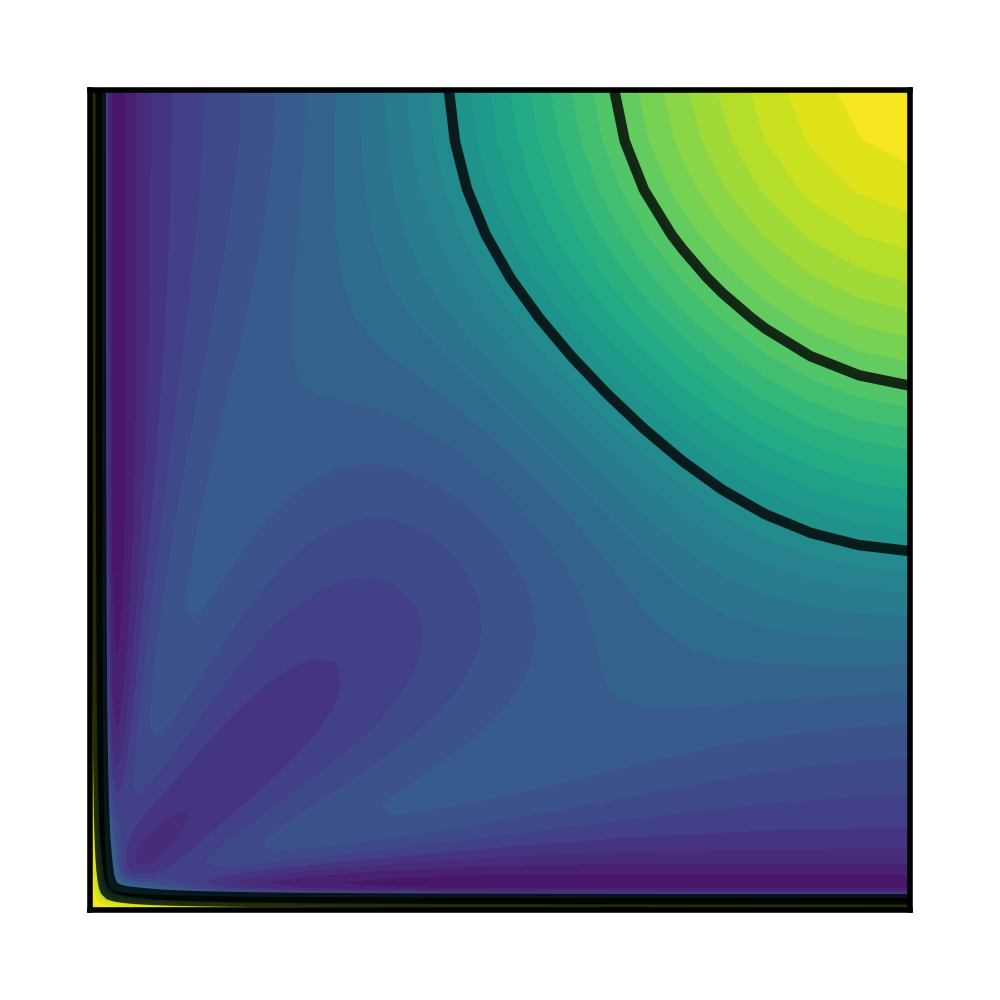}} &
        \raisebox{-.5\height}{\includegraphics[scale=0.45]{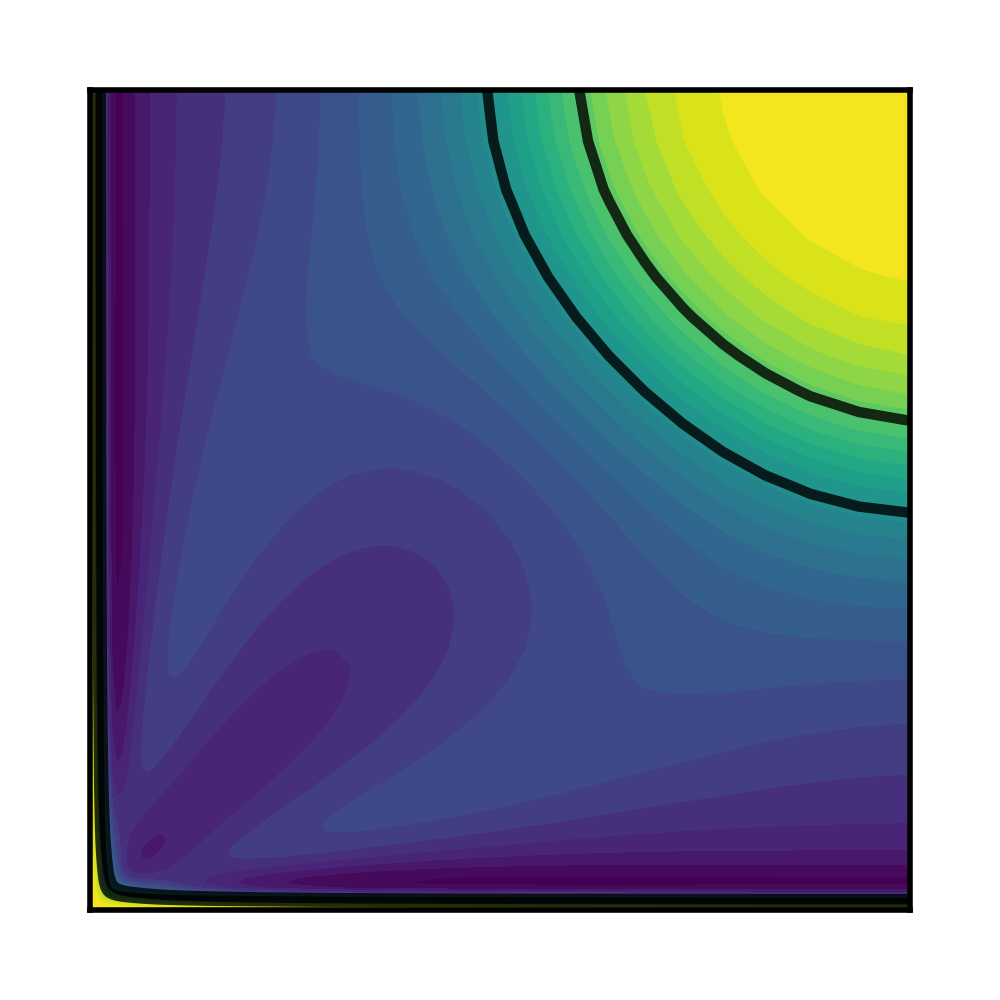}} & 
        \raisebox{-.5\height}{\includegraphics[scale=0.6]{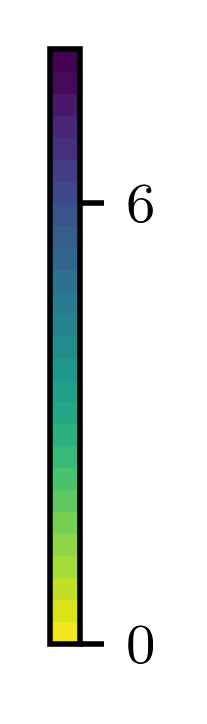}}
        \\
        \rotatebox[origin=c]{90}{$|\theta_1|-|\theta_2|$} & 
        \raisebox{-.5\height}{\includegraphics[scale=0.45]{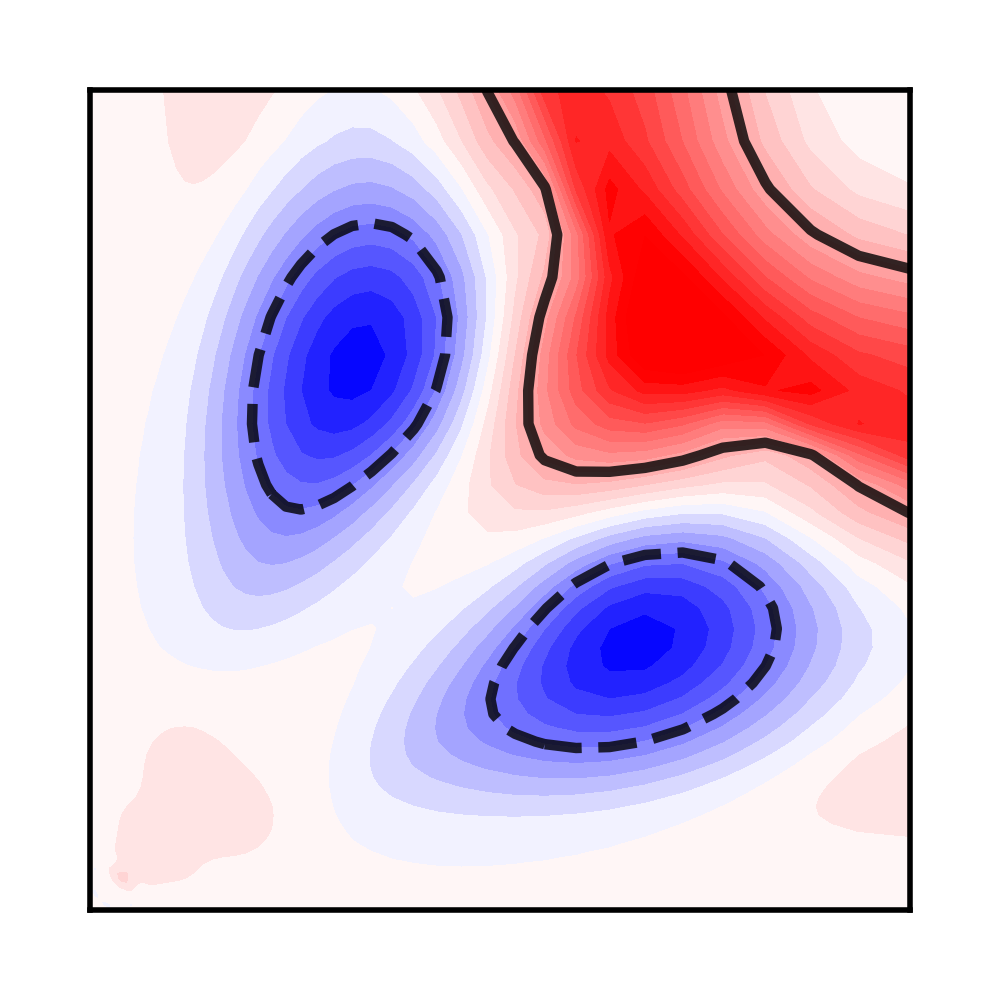}} &
        \raisebox{-.5\height}{\includegraphics[scale=0.45]{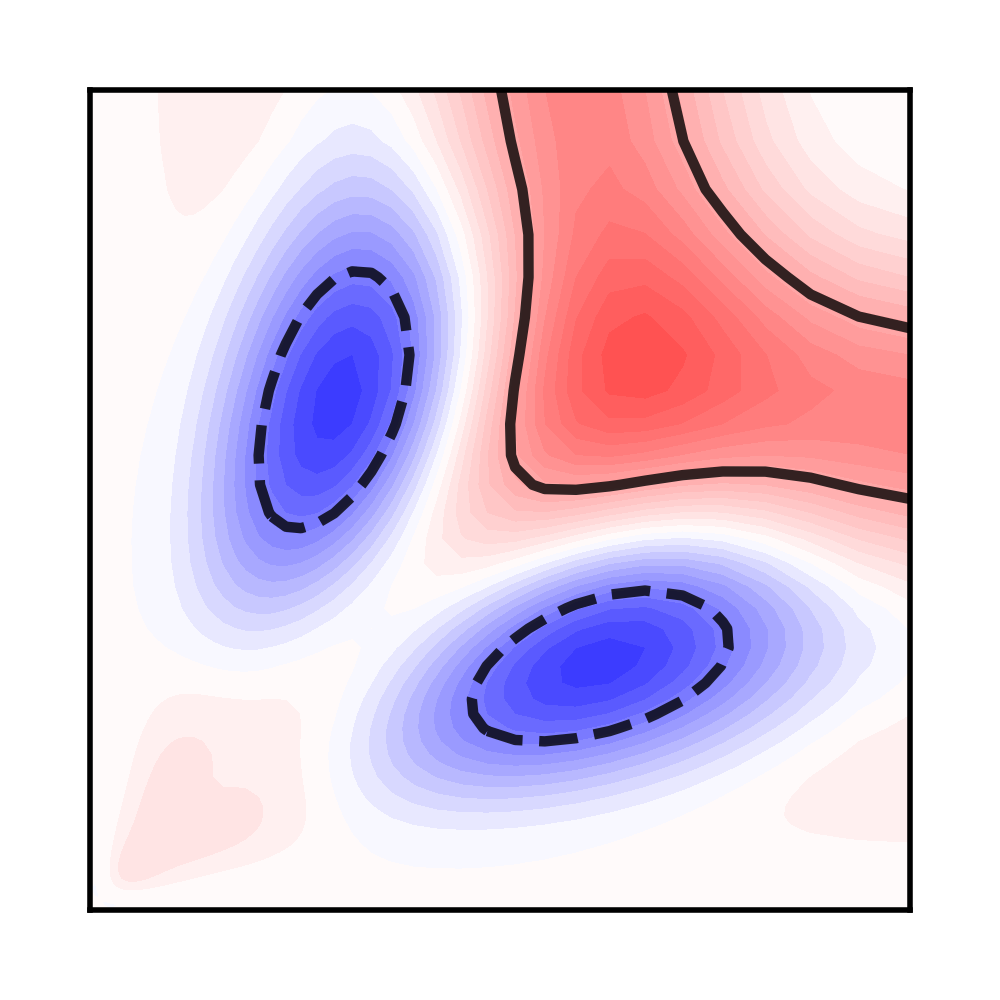}} &
        \raisebox{-.5\height}{\includegraphics[scale=0.45]{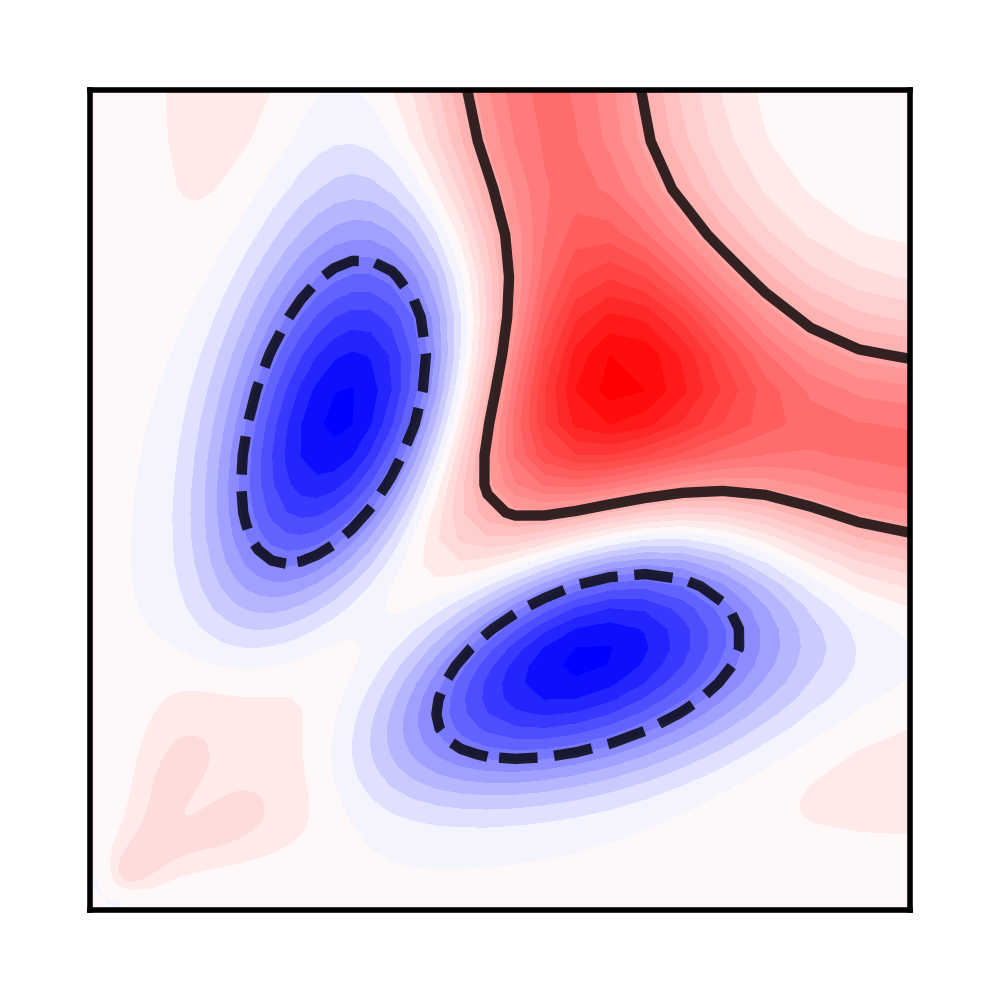}} & 
        \raisebox{-.5\height}{\includegraphics[scale=0.6]{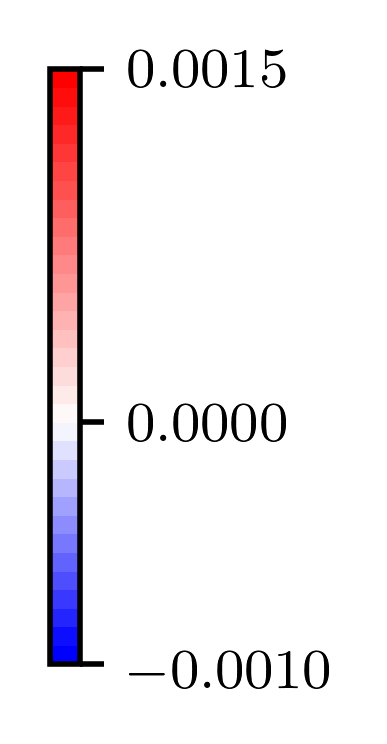}}
    \end{tabular}
    } 
    \subfloat[direct + indirect]{
    \includegraphics[width=0.35\textwidth]{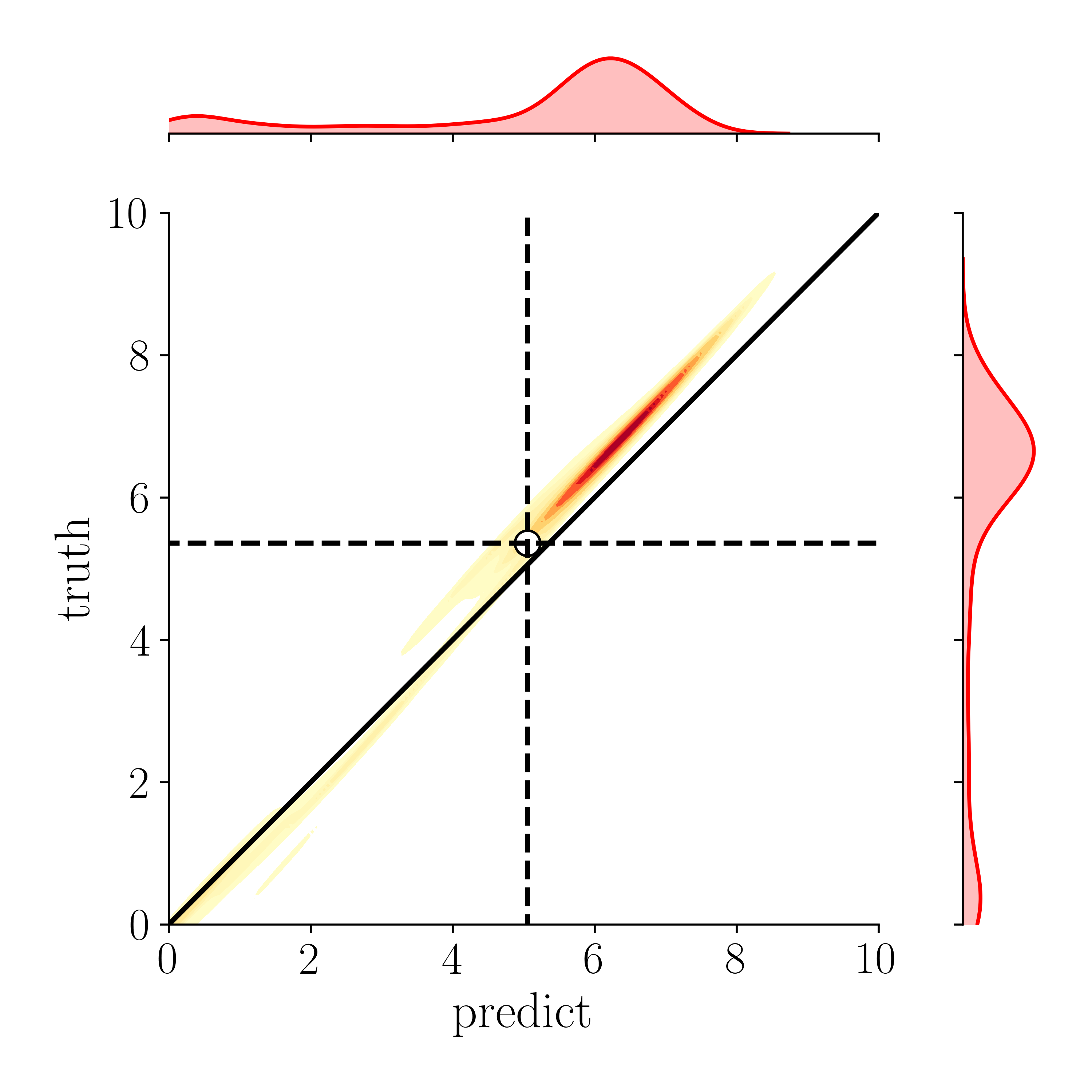}}
    \caption{
    Panel (a) shows contour plots of $\theta_1$ and $|\theta_1|-|\theta_2|$ with comparison among the model learned with indirect data, the model learned with indirect and direct data, and the truth; Panels (b) shows kernel density plots of $\theta_1$ from the truth and the model estimation for the square duct case, respectively.
    The contour lines indicate the levels of $2$ and $4$ for $\theta_1$ and $-0.0005$ and $0.0005$ for $ |\theta_1 | - | \theta_2 | $.
    The probability densities of the truth and the model estimation are plotted on the margins.
    The circle indicates the values of the $30 \%$ quantiles (i.e., $30\%$ of the cells have $\bm{\theta}$ smaller than this value in the magnitude) for the model learned from indirect and direct data. }
    \label{fig:theta_sd}
\end{figure}

Combining direct and indirect data is able to provide better estimation in the scalar invariants~$\boldsymbol{\theta}$, compared to using only indirect data.
Evidence is provided in Figure~\ref{fig:theta_sd} where the learned model with direct and indirect data better reconstructs the field of input features compared to the case with only indirect data.
Particularly at the region with small scalar invariants~$\boldsymbol{\theta}$, only using indirect data overestimates the scalar invariant~$\theta_1$.
It can be clearly seen from the plots of contour lines that learning from indirect and direct data provides more accurate reconstructions of scalar invariants than using only indirect data near the duct center.
The probability density function is plotted at the margin to indicate the distribution of the input features.
It shows that the $30 \%$ quantile of the cells is located around $5.5$, indicating that most cells have scalar invariants larger than this value.
The insufficient representation around the duct center poses additional difficulties to reconstruct the flow field with small scalar invariants.

\begin{figure}[!htb]
    \centering
    \includegraphics[width=0.6\textwidth]{duct_g_legend.png} \\
    \subfloat[$g^{(1)}$]{
    \includegraphics[width=0.3\textwidth]{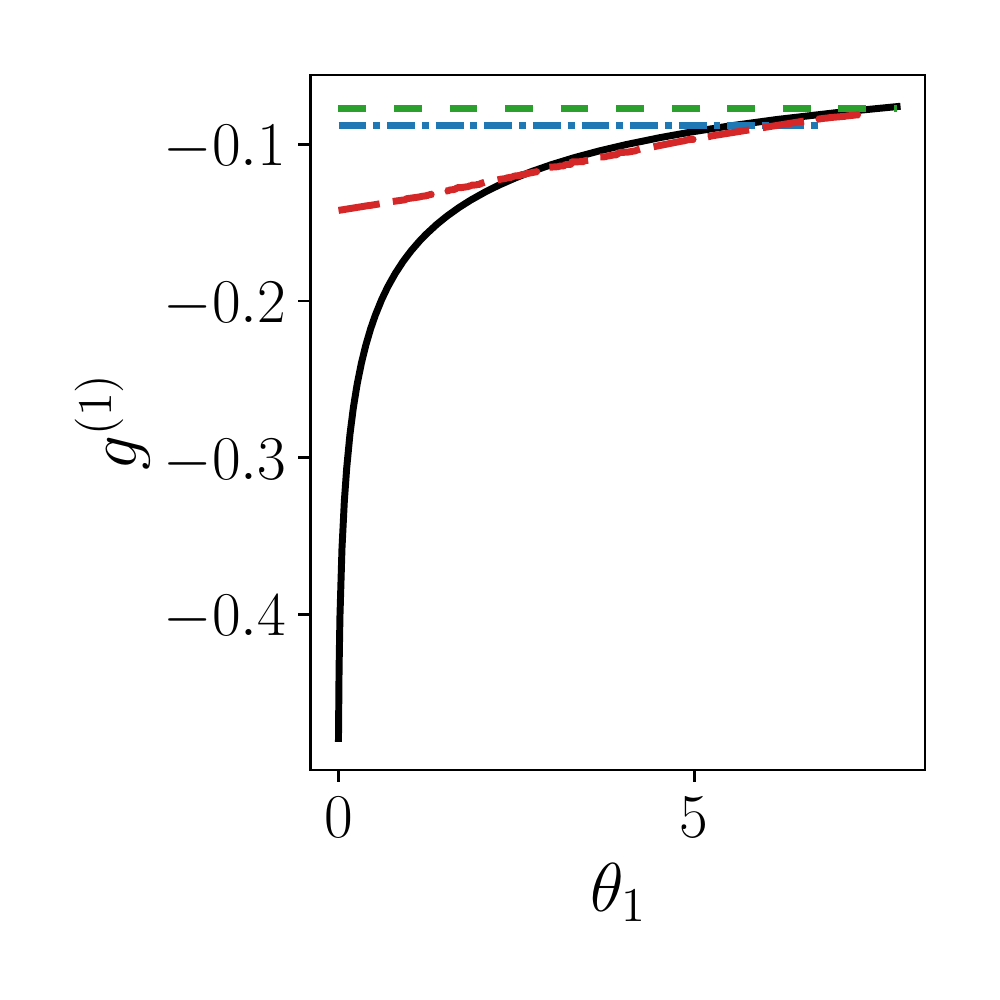}}
    \subfloat[$g^{(2)}-0.5g^{(3)}+0.5g^{(4)}$]{
    \includegraphics[width=0.3\textwidth]{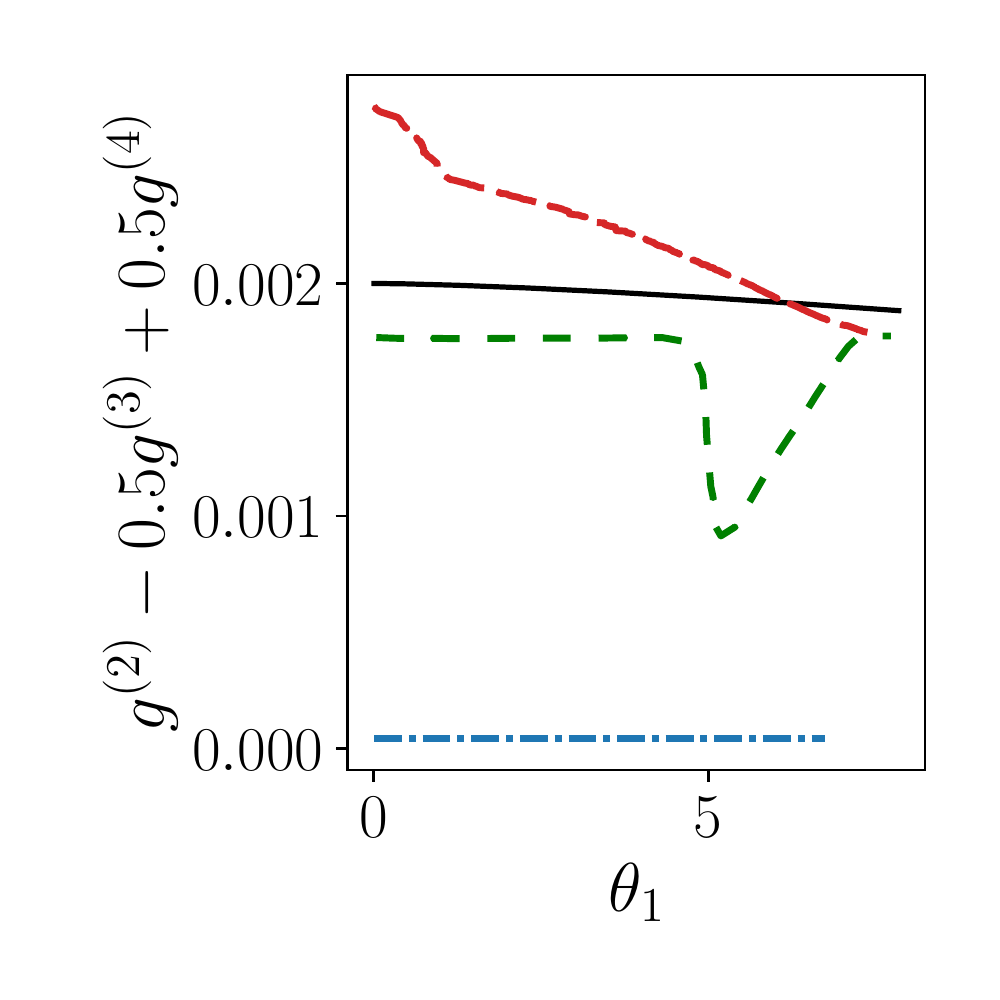}}
    \caption{Comparison of the $g$ functions among the baseline, the synthetic truth, the case with indirect data, and the case with direct and indirect data.}
    \label{fig:gfunc_sd}
\end{figure}

The combination of the direct and indirect data can enhance the discovery of underlying model functions.
It is supported by the plots of the $\boldsymbol{g}$ functions in Figure~\ref{fig:gfunc_sd}.
It can be seen that combining the direct data~$\mathbf{b}$ and the indirect data~$\boldsymbol{u}$ reconstructs better the model function than the case with only the indirect data~$\boldsymbol{u}$.
Specifically, using only indirect data learns almost constant $g^{(1)}$ function, which leads to large discrepancies, particularly in the range of $0$--$5$.
In contrast, learning from direct and indirect data provides relatively better agreement with the synthetic truth.
The combination $g^{(2)}-0.5g^{(3)}+0.5g^{(4)}$ affects the in-plane velocity, hence the plot is also provided for comparison.
For the combination of functions, the learned model with only indirect data leads to significant discrepancies in the range of $\theta_1 > 5$, while the model with both data sources provides relatively better estimations.
The combination of learned model functions~$g^{(2)}-0.5g^{(3)}+0.5g^{(4)}$ exhibits strong dependences on the scalar invariant~$\theta_1$. 
This is inconsistent with the synthetic truth that is almost constant as the scalar invariant~$\theta_1$ varies.
Such inaccurate dependencies are caused by the ill-conditioning of the training process.
Specifically, the learned $g$ functions are multiplied by the tensor bases associated with velocity gradients to construct the Reynolds stress based on Eq.~\eqref{eq:b}.
Hence, the learned model can produce arbitrary $g$ functions in regions with negligible velocity gradients.
The streamwise velocity typically dominates the in-plane velocity by several orders of magnitude, and its gradient approaches vanishing near the duct center, resulting in negligible tensor bases and further Reynolds stress.
Hence, the Reynolds stress and the velocity are not sensitive to the learned $g$ function near the duct center, which poses difficulties in learning accurate functional mapping for small scalar invariants.
To improve the accuracy of the learned model function, more data points need to be added near the duct center, and more weights can be assigned to observation data near the duct center compared to other regions.

Learning from direct and indirect data significantly increases the computational cost compared to learning from only indirect data for the square duct case.
That is because combining the two data sources requires performing additional back-propagation and regularization steps for each realization, which causes significant increases in CPU time.
Specifically, learning from the direct and indirect data requires $27.1$ core hours for the square duct case, while learning from indirect data only needs $5.8$ core hours.
Therefore, the proposed method provides more accurate models at relatively high computational costs than the ensemble Kalman method of using only indirect data.

\begin{figure}[!htbp]
    \centering
    \includegraphics[width=0.4\textwidth]{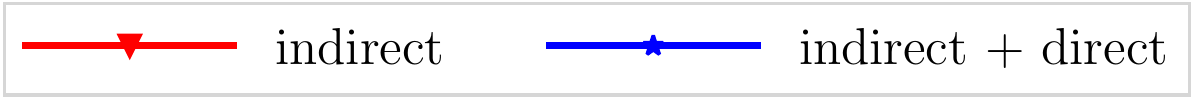}
    \subfloat[Reynolds stress]{
    \includegraphics[width=0.4\textwidth]{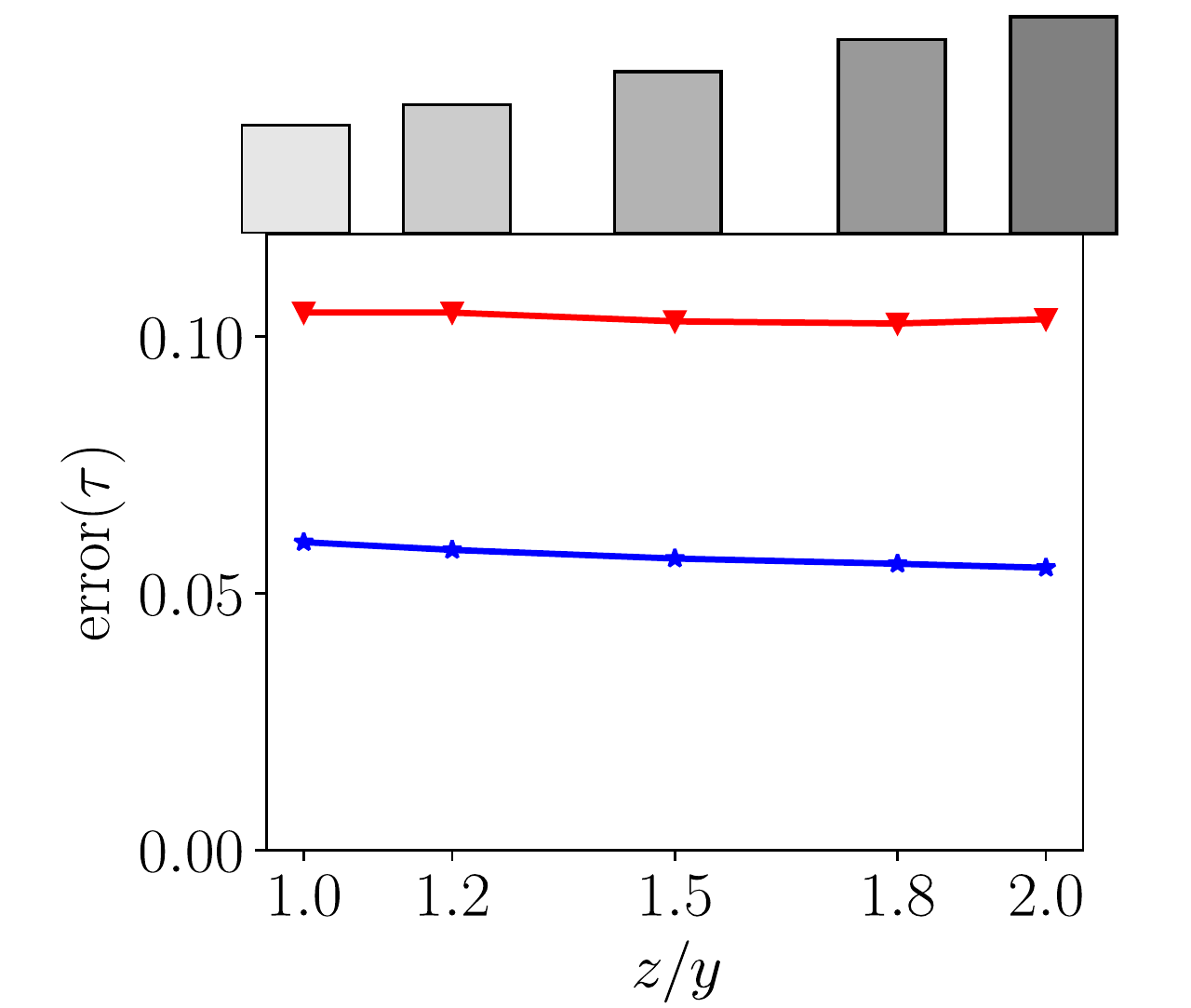}}
    \subfloat[In-plane velocity]{
    \includegraphics[width=0.4\textwidth]{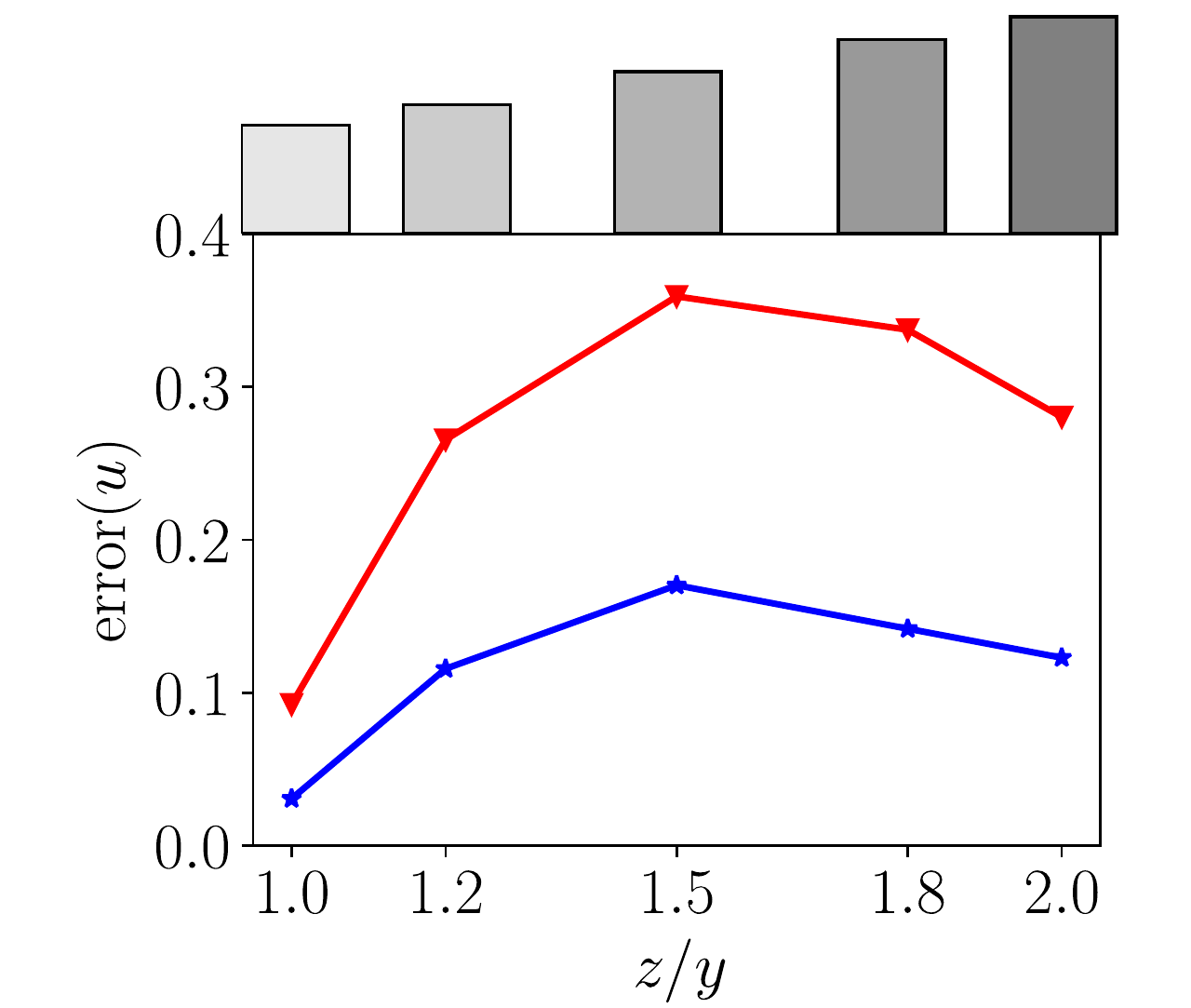}}
    \caption{Results of posterior tests on similar configurations with varying aspect ratios in the square duct case. The geometries of the configurations are indicated at the margin.}
    \label{fig:generalizability_sd}
\end{figure}

The posterior tests show that the combination of the direct and indirect data can empower the learned model with better generalizability in similar configurations. 
The prediction errors in Reynolds stress and in-plane velocity with the learned models are shown in Figure~\ref{fig:generalizability_sd}.
The error is computed over the entire computational domain as
\begin{equation}
    \text{error} = \frac{\| q_\text{truth} - q_\text{predict} \|}{\| q_\text{truth} \|} \text{.}
\end{equation}
The baseline $k$--$\varepsilon$ model is not able to capture the in-plane velocity, hence the plot is omitted for brevity.
The configurations with different aspect ratios~$z/y$ in the range of $[1.0, 2.0]$ are used to test the learned model, as indicated at the margin of Figure~\ref{fig:generalizability_sd}.
The training case of the aspect ratio $z/y=1$ is also provided.
It is noticeable that the learned model with the indirect data leads to larger discrepancies in terms of both Reynolds stress and in-plane velocity compared to the model learned from both the direct and indirect data.
Specifically, learning from only the indirect data leads to prediction errors in Reynolds stress at approximately $11\%$ for all the cases, while combining the direct and indirect data achieves relatively small errors around $6\%$.
As for the velocity prediction, the model learned from the direct and indirect data provides better prediction with errors around $0.1\%$.
The model learned from the indirect data provides larger discrepancies as the aspect ratio of the test case increases.
For both learned models, the predictive discrepancy increases as the test case deviates further from the training case of $z/y=1$, likely due to the limited extrapolation ability of data-driven models.
However, it is noteworthy that the maximum predictive error in the in-plane velocity occurs at $z/y=1.5$ for both learned models, and the prediction error decreases as the aspect ratio increases further.
This behavior may be caused by relatively large gradients of the Reynolds stress in the case of $z/y=1.5$, which renders the in-plane velocity more sensitive to local Reynolds stress discrepancies.
Specifically, the in-plane velocity is primarily influenced by the gradient of the Reynolds normal stress imbalance~$\frac{\partial}{\partial y \partial z}(\tau_{yy}-\tau_{zz})$~\cite{strofer2021end}.
The relatively larger gradients of the Reynolds stress imbalance in the test case of $z/y=1.5$ can cause larger discrepancies in the prediction of in-plane velocity, even though the Reynolds stresses exhibit similar global discrepancies over the entire field.

\begin{figure}[!htb]
    \centering
    \includegraphics[width=0.7\textwidth]{duct_g_legend.png} \\
    \subfloat[$u_y (z/y=1.5)$]{\includegraphics[width=0.3\textwidth]{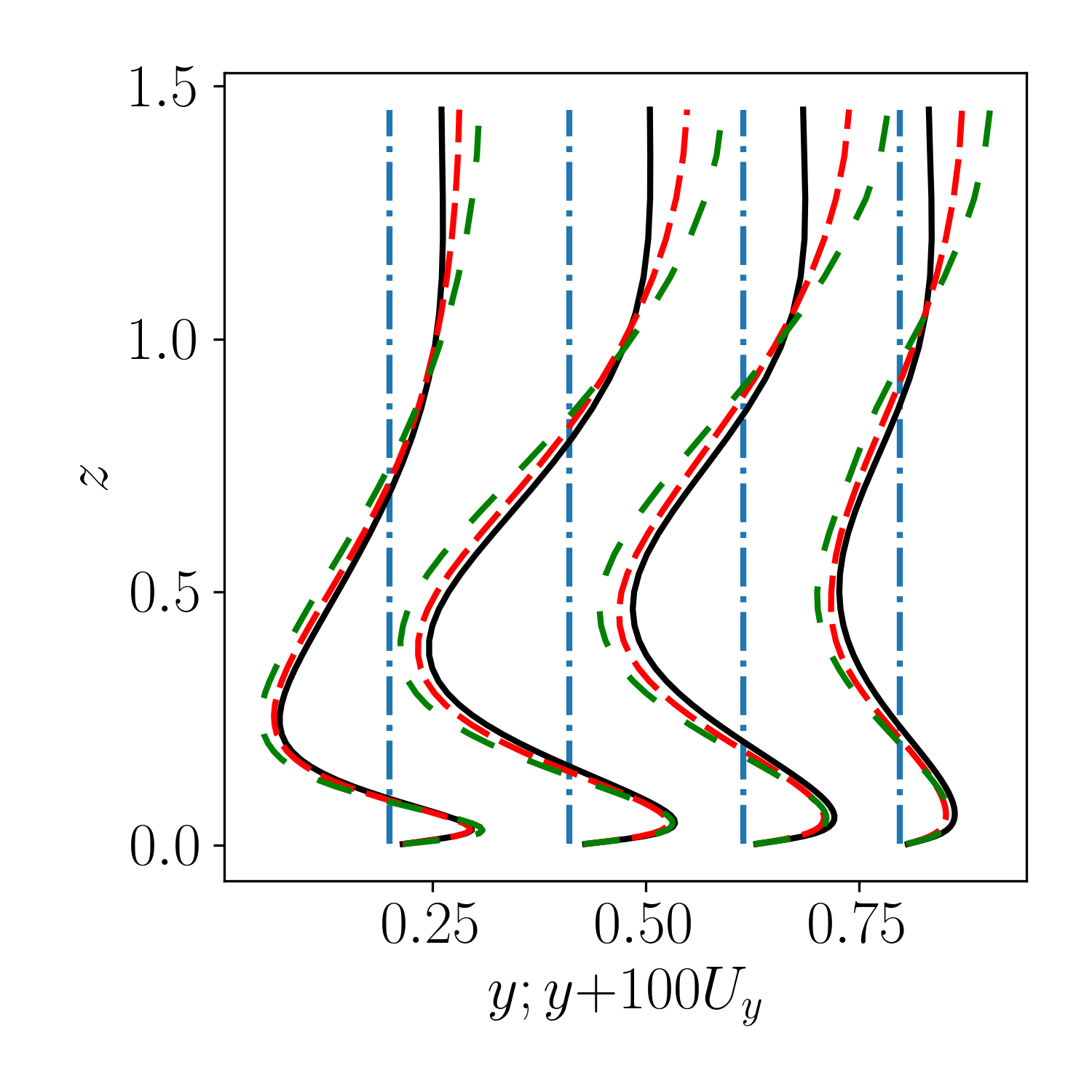}}
    \subfloat[$\tau_{yz} (z/y=1.5)$]{\includegraphics[width=0.3\textwidth]{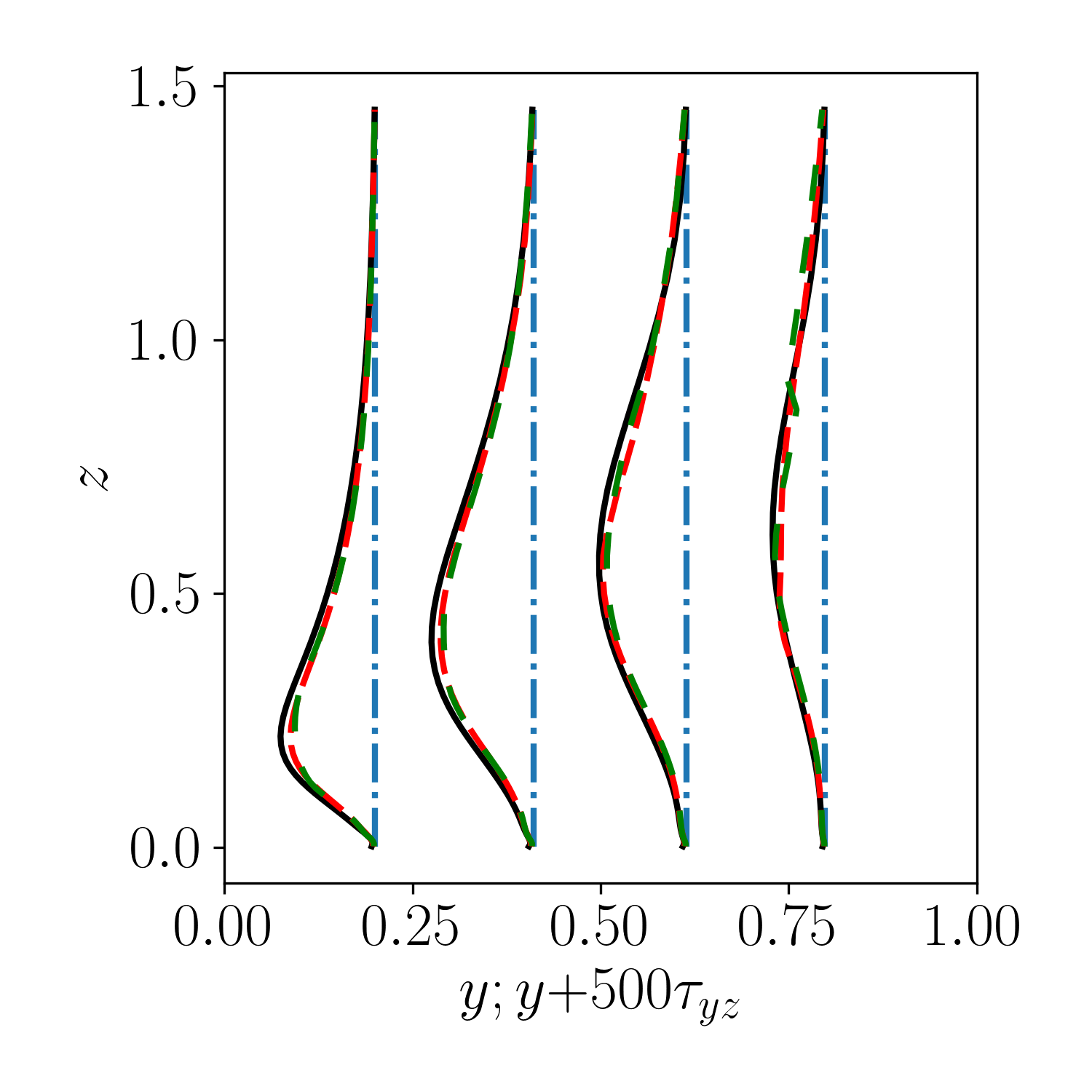}}
    \subfloat[$\tau_{yy}-\tau_{zz} (z/y=1.5)$]{\includegraphics[width=0.3\textwidth]{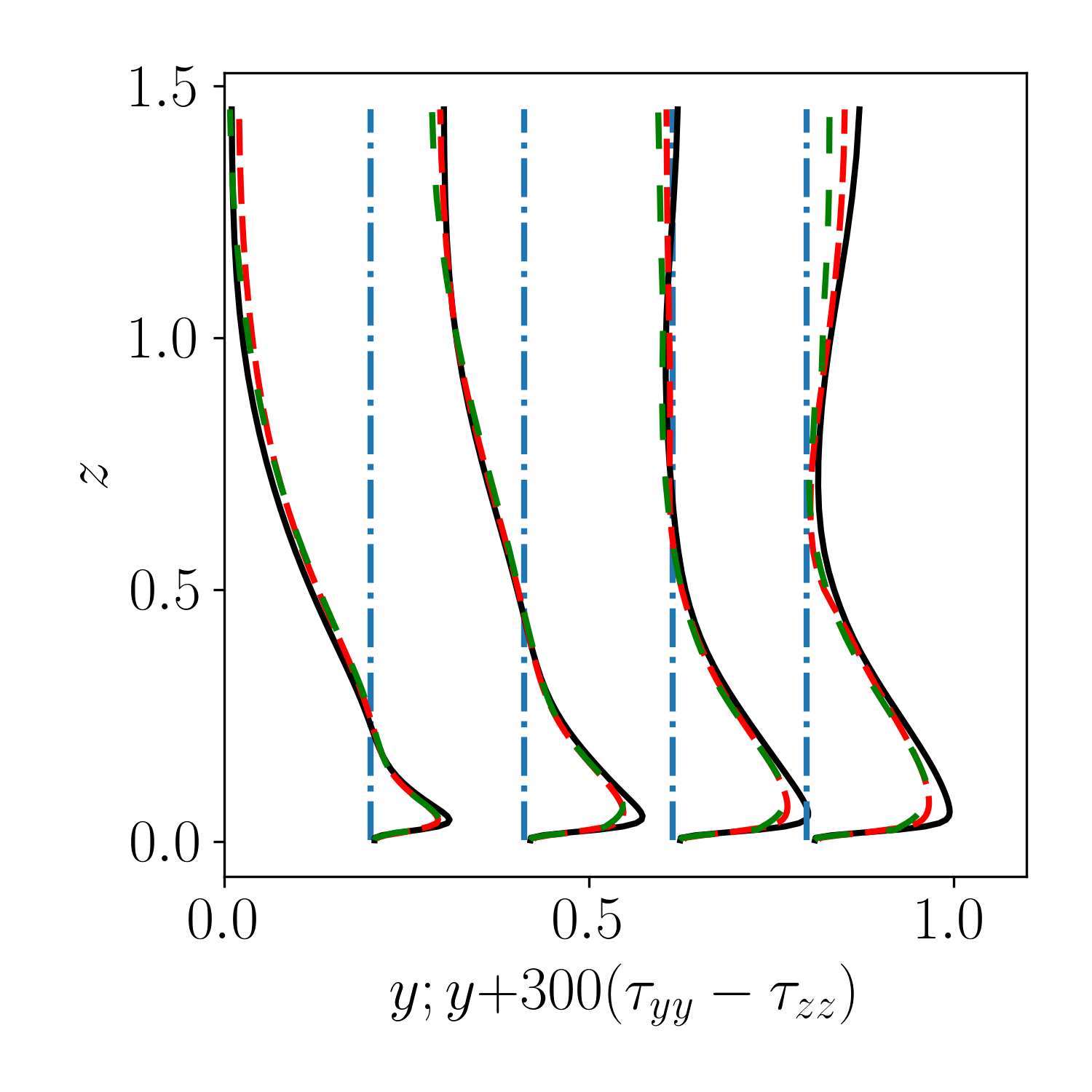}} \\
    \subfloat[$u_y (z/y=2)$]{\includegraphics[width=0.3\textwidth]{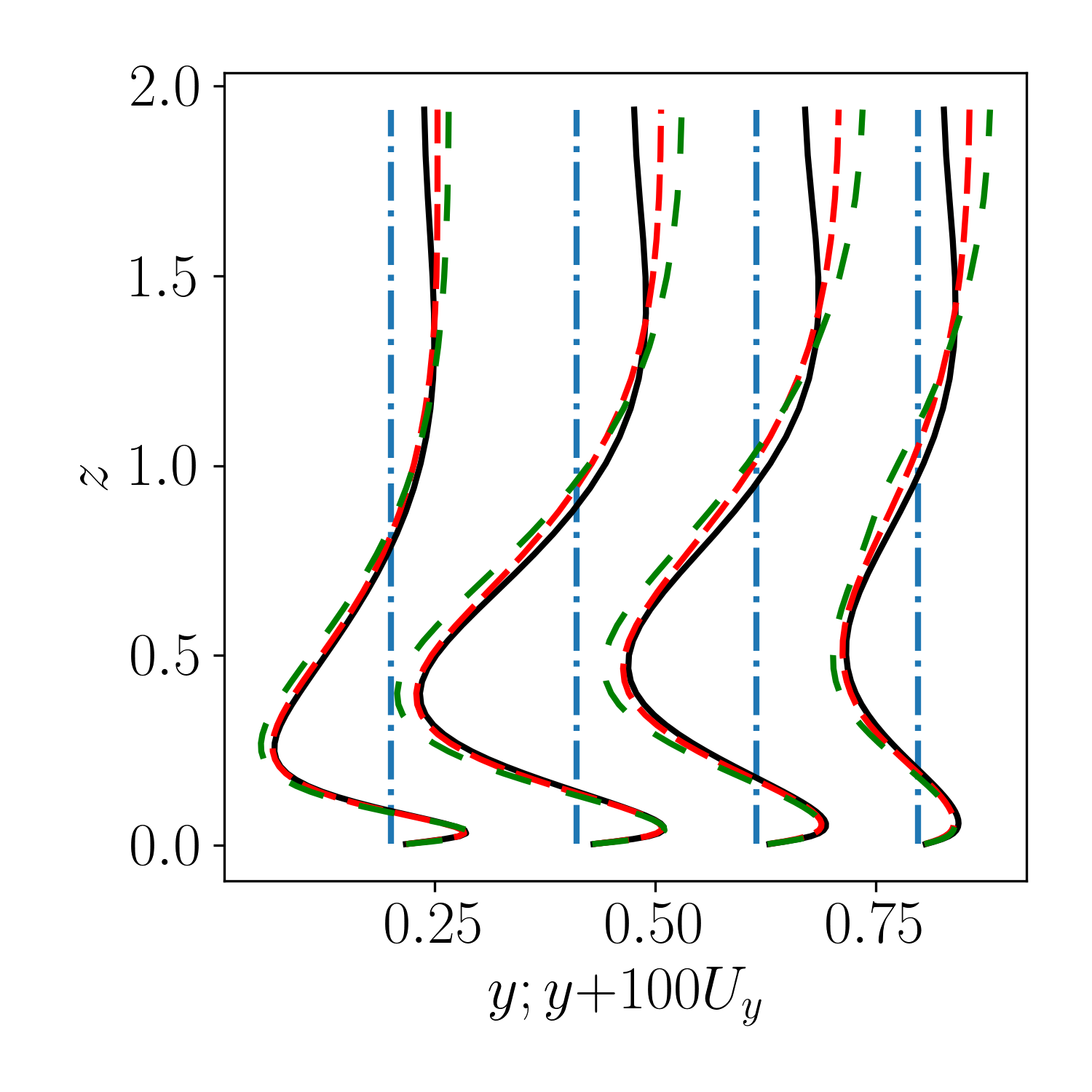}}
    \subfloat[$\tau_{yz} (z/y=2)$]{\includegraphics[width=0.3\textwidth]{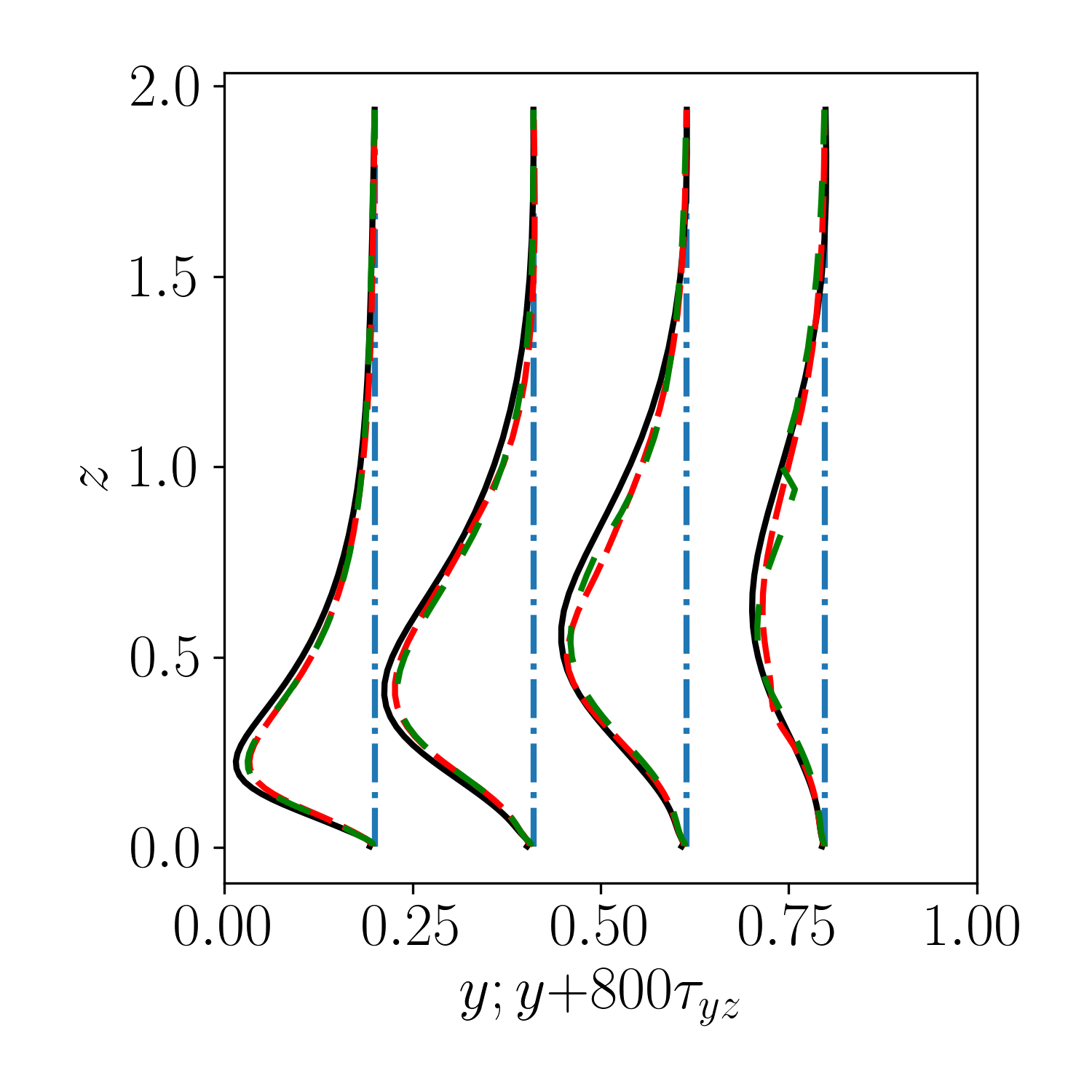}}
    \subfloat[$\tau_{yy}-\tau_{zz} (z/y=2)$]{\includegraphics[width=0.3\textwidth]{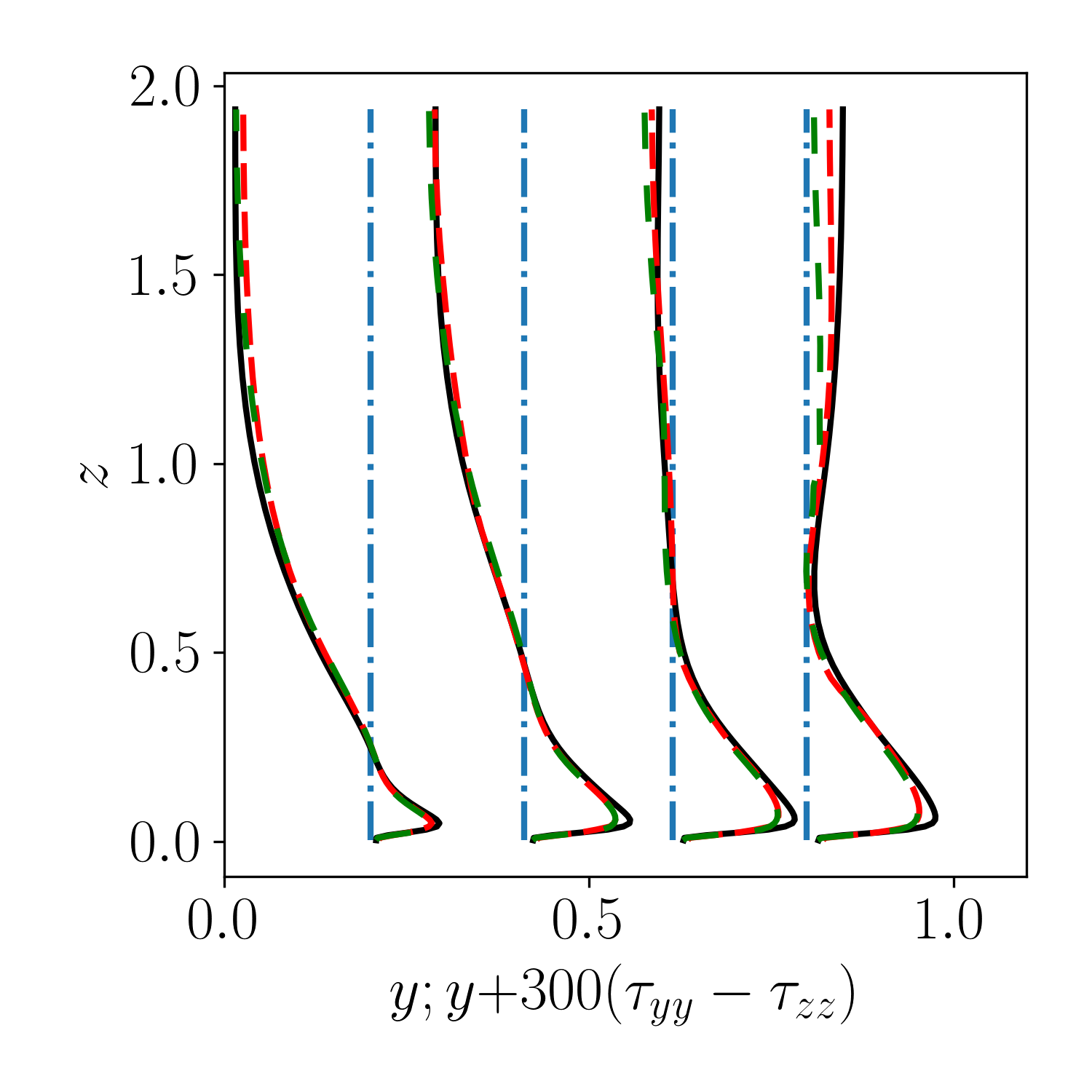}}
    \caption{Plots of in-plane velocity and Reynolds stress along profiles for test cases with different aspect ratios ($z/y=1.5,2$).
    }
    \label{fig:prof_sd_genalization}
\end{figure}

To clearly show the predictive ability of the learned models, we plot the predictions of the velocity and the Reynolds stress along different profiles ($y/H=0.2, 0.4, 0.6, 0.8$) with comparisons to the synthetic truth.
The results of the cases of $z/y = 1.5$ and $2.0$ are provided in Figure~\ref{fig:prof_sd_genalization} for brevity.
It can be seen that significant discrepancies in velocity mainly occur near the duct center for both cases of $z/y=1.5$ and $2$, which indicate the relatively high sensitivity of the velocity to the Reynolds stress in this region.
Especially, the case of $z/y=1.5$ has larger discrepancies in the velocity prediction than the case of $z/y=2$, while the discrepancies in the Reynolds stress are very similar between the two cases.
The relatively larger gradient of the Reynolds stress imbalance near the duct center for the case of $z/y=1.5$ is responsible for this observation.
Specifically, the magnitude of $ \frac{\partial}{\partial y \partial z} (\tau_{yy}-\tau_{zz}) $ in the range of $y/y_\text{max}>0.5$ and $z/z_\text{max}>0.5$ is 0.011 for the case of $z/y=1.5$, which is larger than other cases of, i.e., $0.007$, $0.010$, $0.009$, and $0.006$, respectively, as the aspect ratio increases.
This explains the similar discrepancy in the Reynolds stress of the two cases resulting in noticeable differences in the velocity prediction as shown in Figure~\ref{fig:generalizability_sd}.
One can further enhance the generalizability of the model by incorporating the data from different flow conditions, but the weighting strategies for each case need further investigation.

\subsection{Periodic hills}

\begin{figure}[!htb]
    \centering
    \begin{tabular}{ccccl}
        & baseline & learned model & LES & \\
        \rotatebox[origin=c]{90}{$u_x$} & 
        \raisebox{-.5\height}{\includegraphics[scale=0.64, trim=0 -5 0 -5, clip]{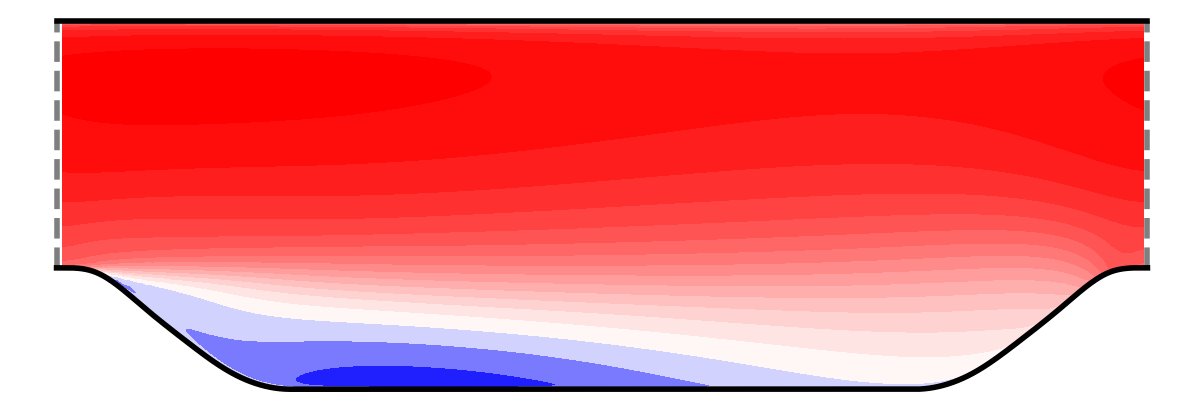}} &
        \raisebox{-.5\height}{\includegraphics[scale=0.64, trim=0 -5 0 -5, clip]{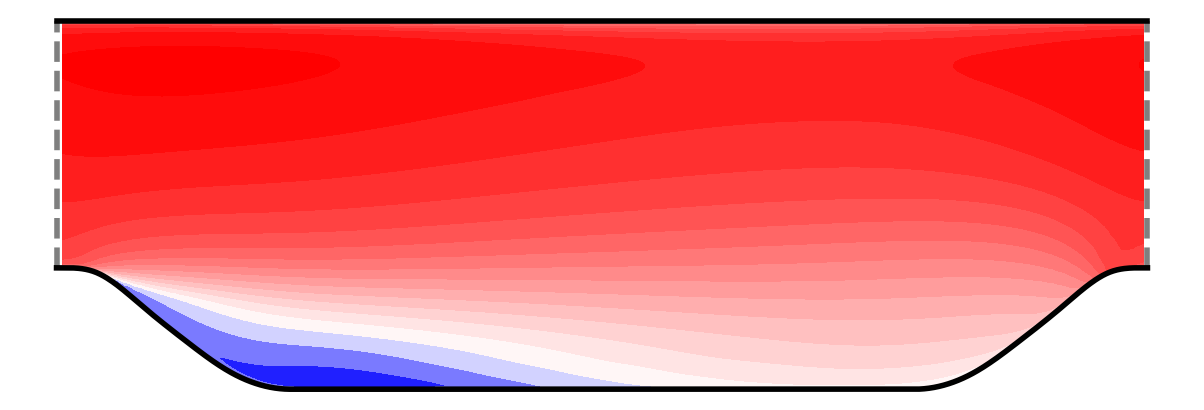}} &
        \raisebox{-.5\height}{\includegraphics[scale=0.64, trim=0 -5 0 -5, clip]{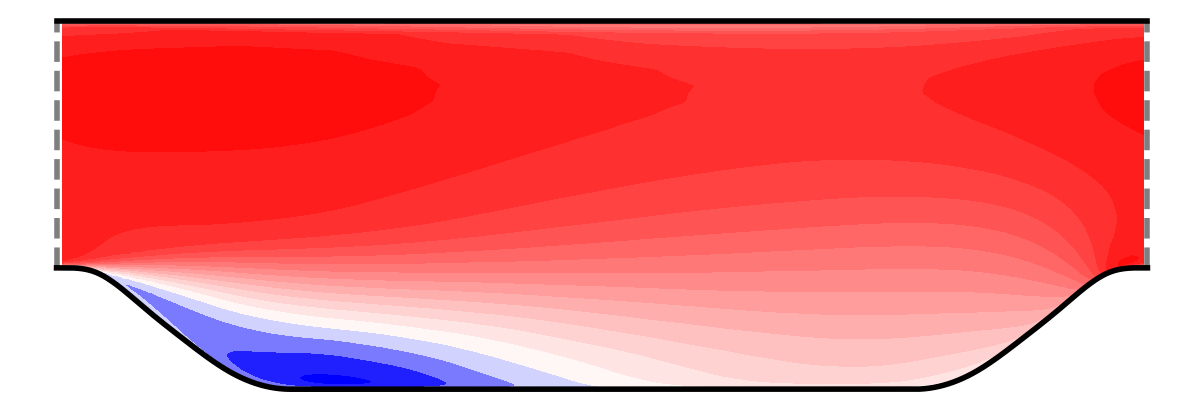}} &
        \raisebox{-.5\height}{\includegraphics[scale=0.4, clip]{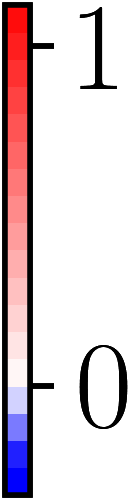}}
        \\
        \rotatebox[origin=c]{90}{$u_y$} & 
        \raisebox{-.5\height}{\includegraphics[scale=0.64, trim=0 -5 0 -5, clip]{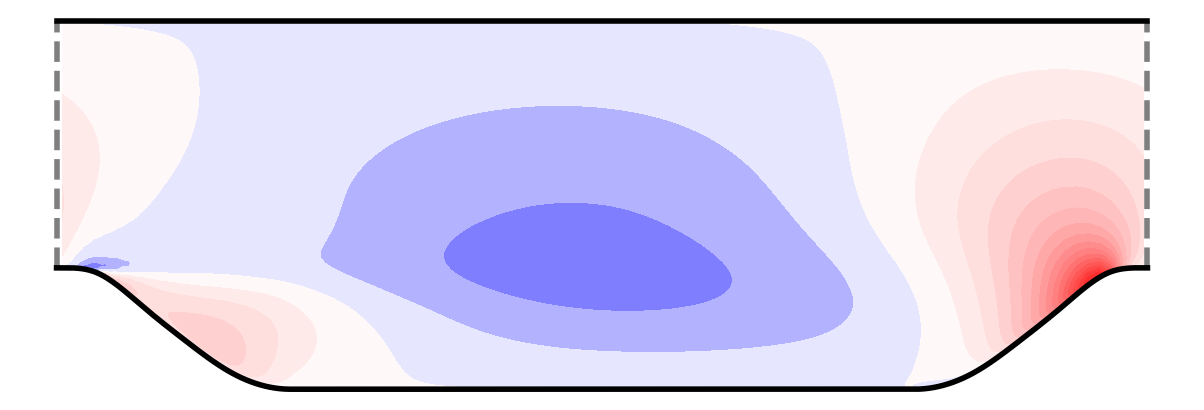}} &
        \raisebox{-.5\height}{\includegraphics[scale=0.64, trim=0 -5 0 -5, clip]{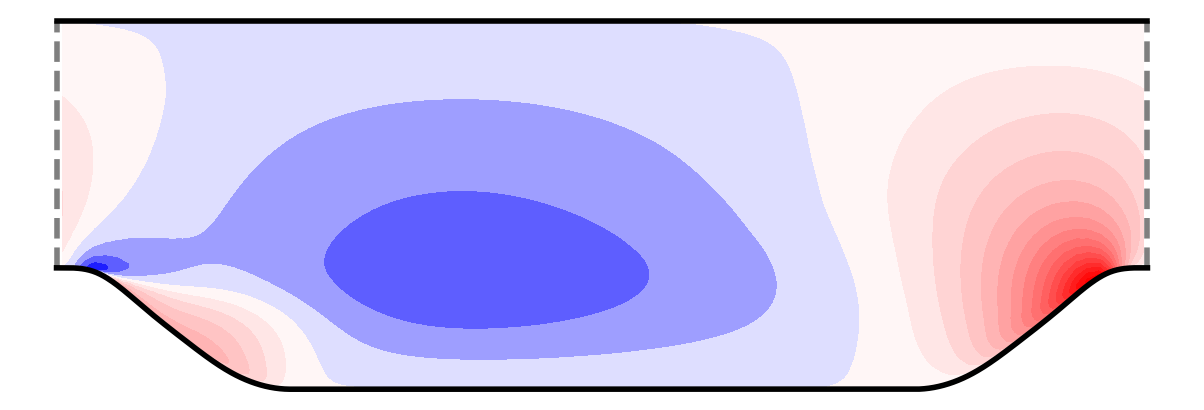}} &
        \raisebox{-.5\height}{\includegraphics[scale=0.64, trim=0 -5 0 -5, clip]{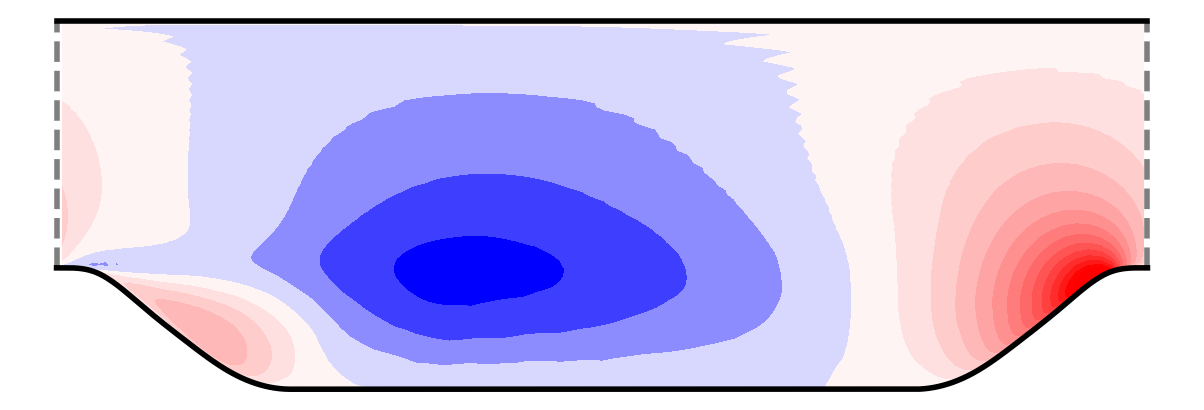}} &
        \raisebox{-.5\height}{\includegraphics[scale=0.4, clip]{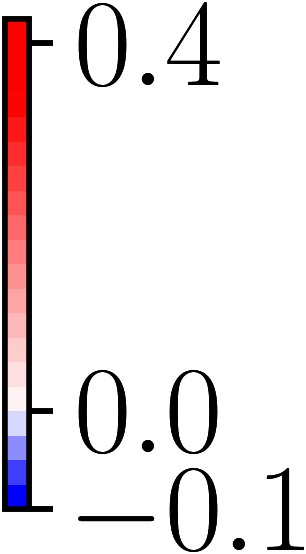}}
        \\
        \rotatebox[origin=c]{90}{$k$} & 
        \raisebox{-.5\height}{\includegraphics[scale=0.64, trim=0 -5 0 -5, clip]{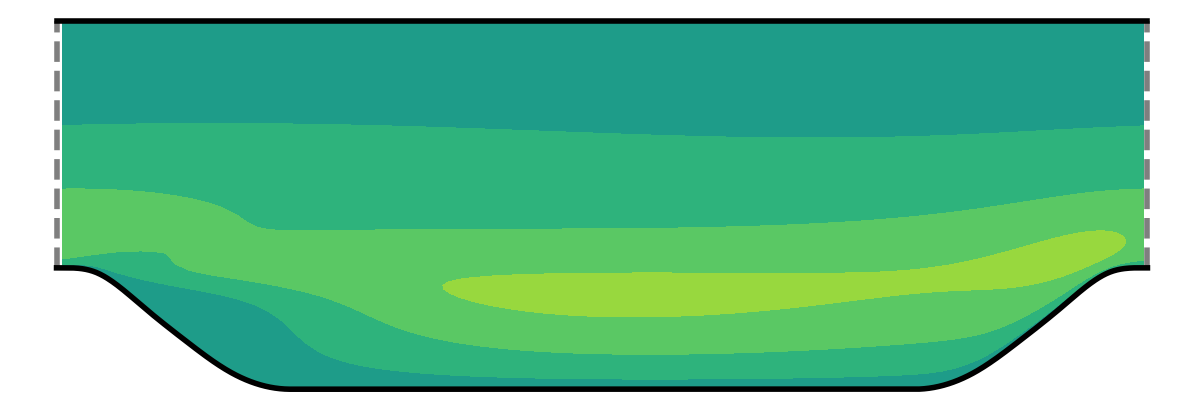}} &
        \raisebox{-.5\height}{\includegraphics[scale=0.64, trim=0 -5 0 -5, clip]{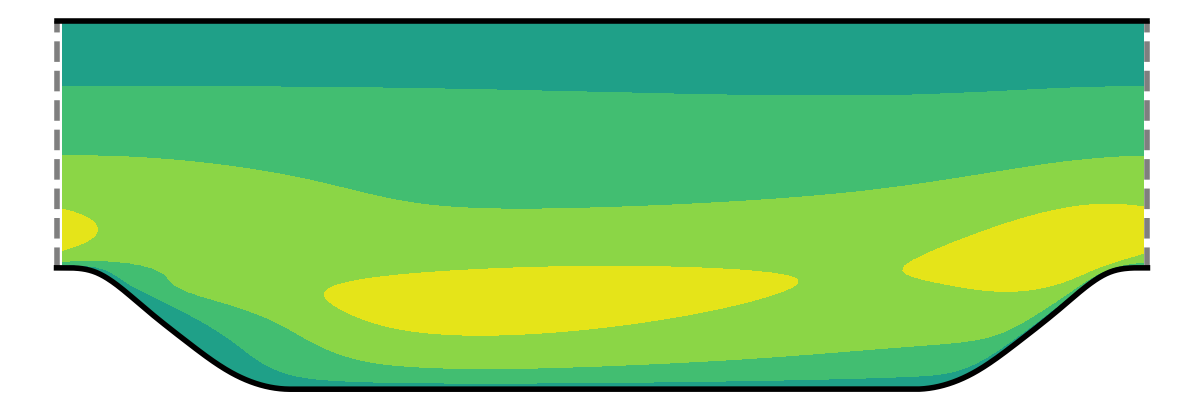}} &
        \raisebox{-.5\height}{\includegraphics[scale=0.64, trim=0 -5 0 -5, clip]{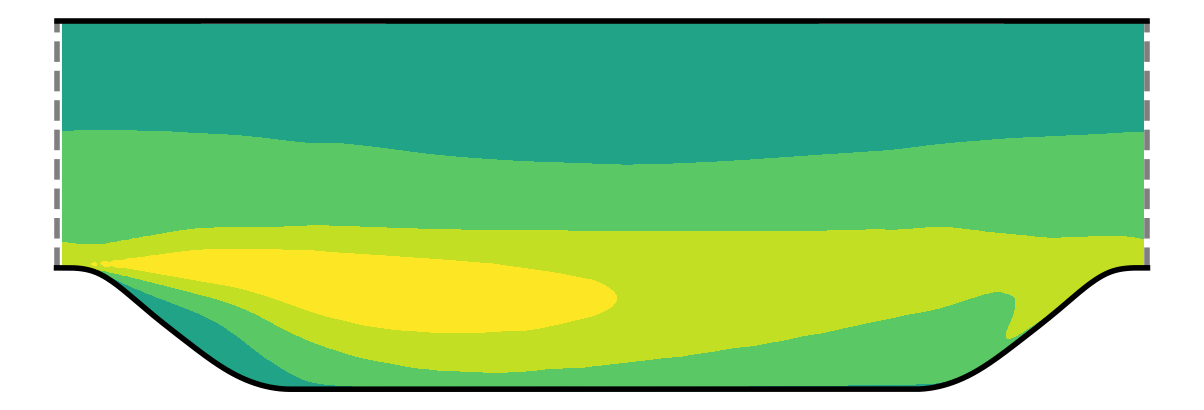}} &
        \raisebox{-.5\height}{\includegraphics[scale=0.4, clip]{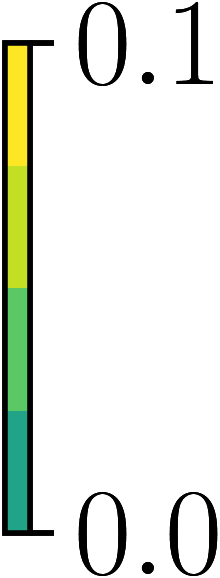}}
        \\
    \end{tabular}
    \caption{Contour plots of the velocity and turbulent kinetic energy with comparison among the baseline model, the learned model with the indirect and direct data, and the LES for the periodic hill case.}
    \label{fig:pehill_contour}
\end{figure}

The combination of the direct and indirect data can improve the prediction of both turbulent kinetic energy and velocity.
It is supported by the contour plots of velocity and TKE fields in Figure~\ref{fig:pehill_contour} with comparison among the baseline, the learned model, and the LES results.
The plots for the learned model show the results of combining the direct and indirect data.
It can be seen that the learned model improves the flow estimation in contrast to the baseline.
Specifically, the baseline model overestimates the flow separation, while the learned model provides a similar backflow area to the LES results.
Moreover, the learned model can also improve the estimation of vertical velocity~$u_y$ with similar patterns to the LES.
As for the TKE, the baseline model underestimates the TKE field, particularly in the recirculation zone.
In contrast, the learned model improves the TKE estimation compared to the baseline, although noticeable discrepancies still exist.

\begin{figure}[!htb]
    \centering
    \includegraphics[width=\textwidth]{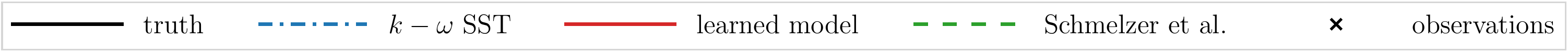} \\
    \subfloat[ $U_x$, indirect]{\includegraphics[width=0.49\textwidth, trim=0 0 0 0, clip]{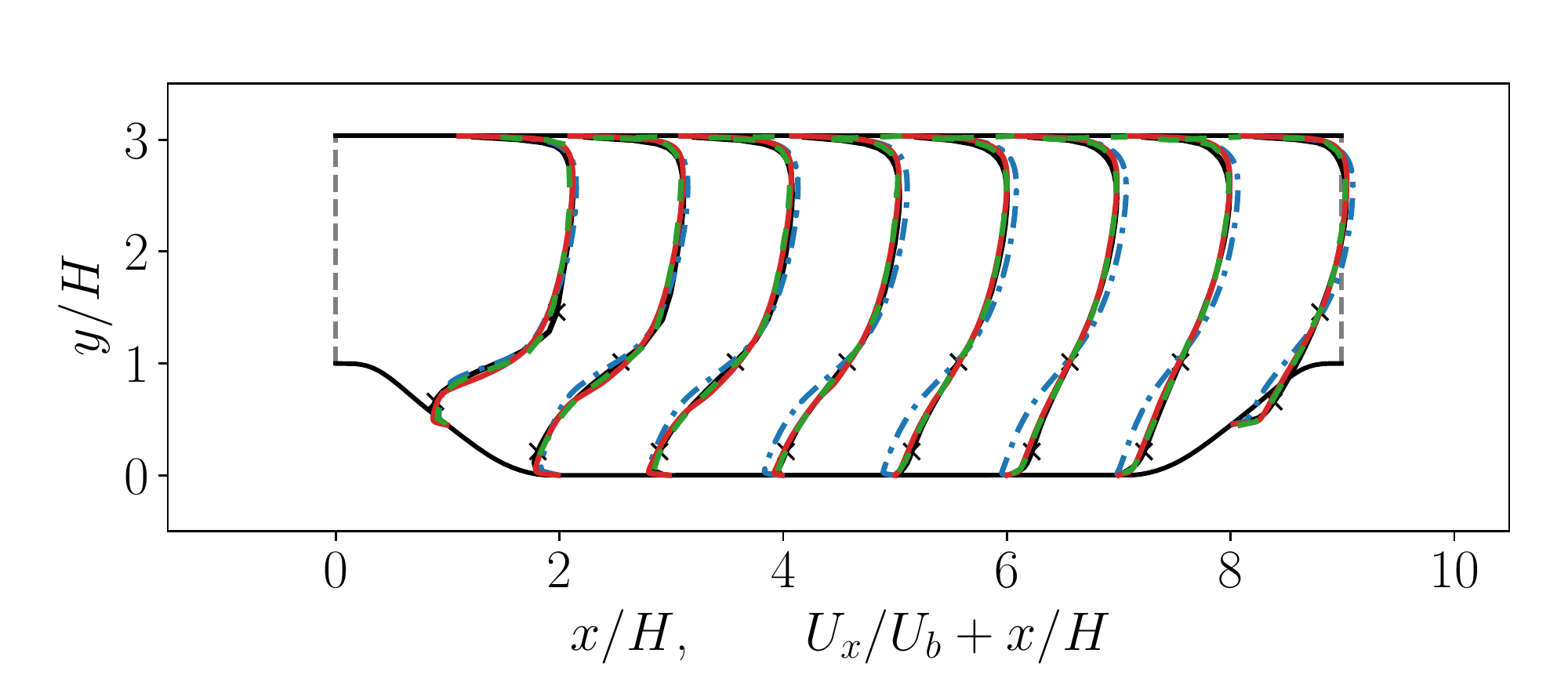}}
    \subfloat[ $U_x$, indirect + direct]{\includegraphics[width=0.49\textwidth, trim=0 0 0 0, clip]{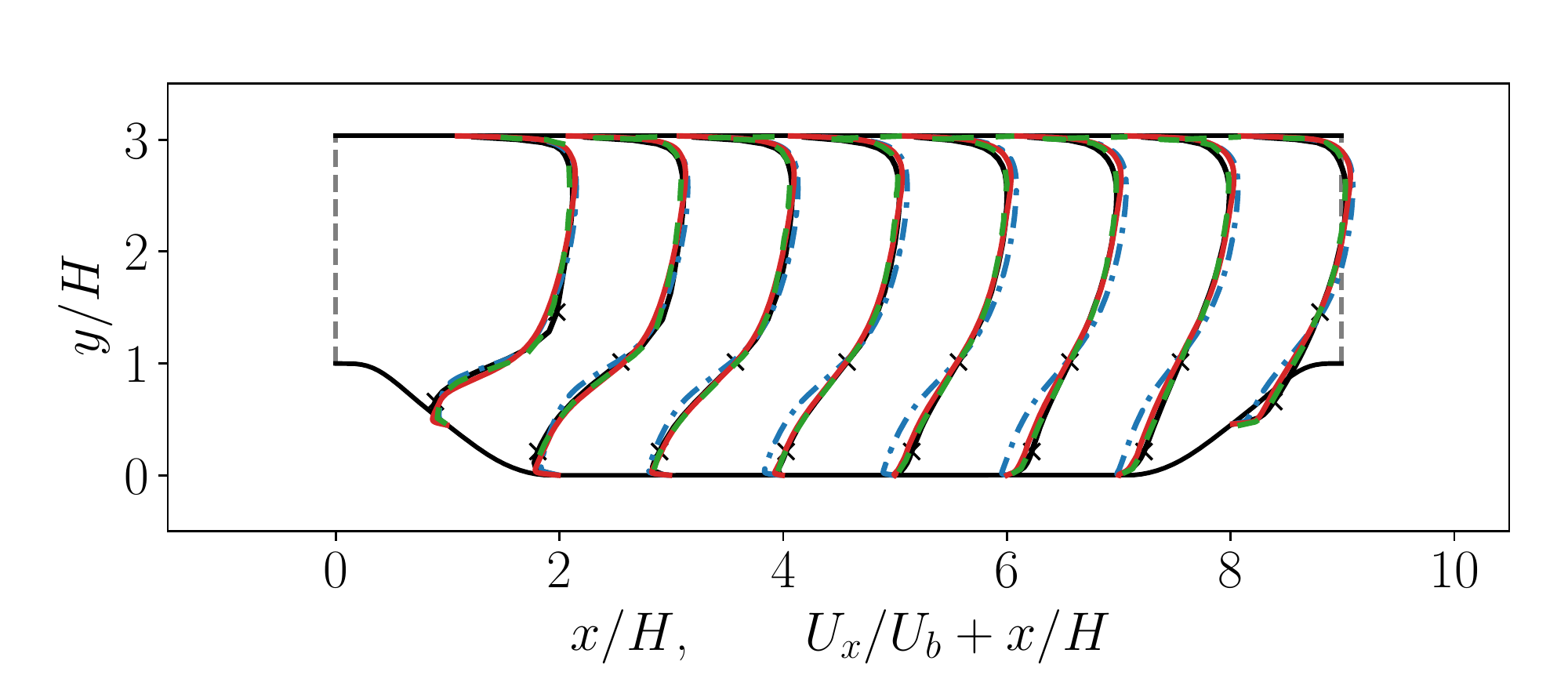}} \\
    \subfloat[$k$, indirect]{\includegraphics[width=0.49\textwidth, trim=0 0 0 0, clip]{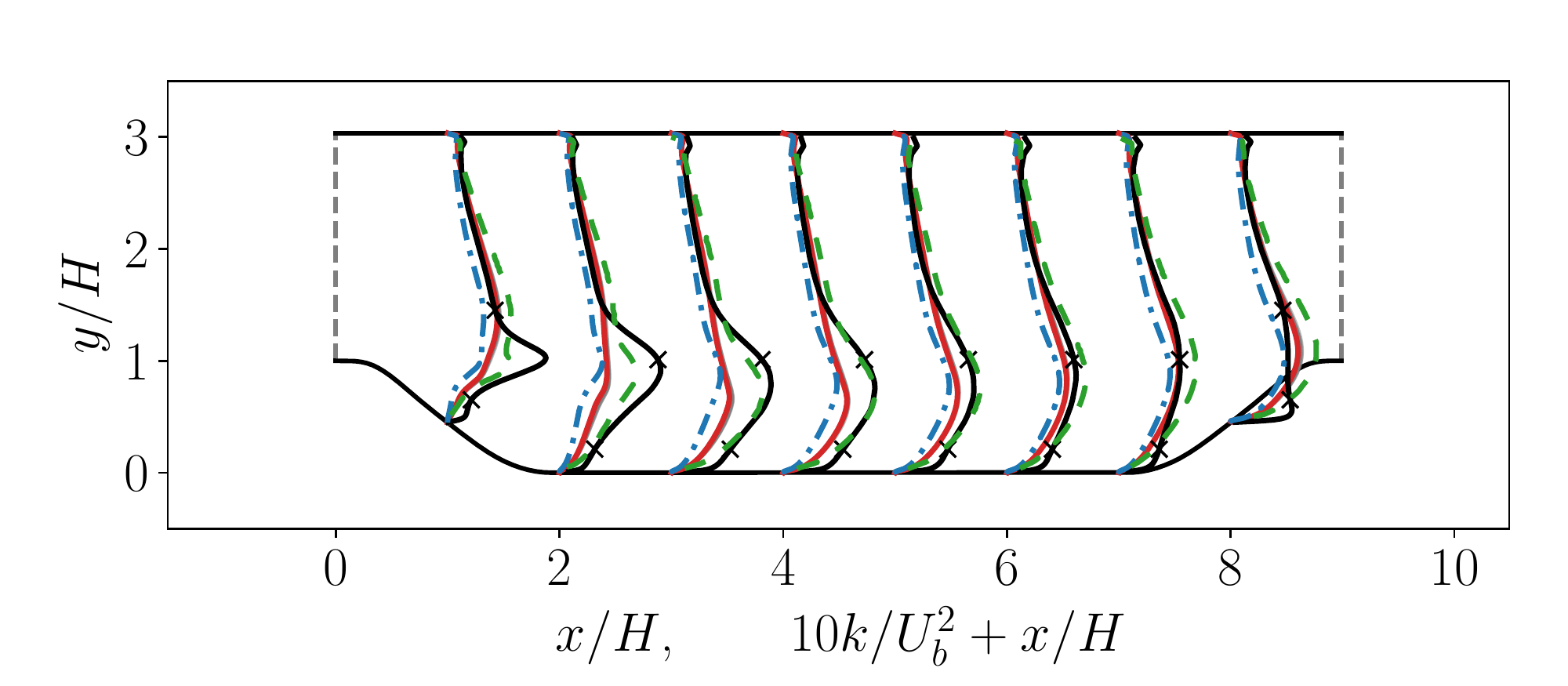}}
    \subfloat[$k$, indirect + direct]{\includegraphics[width=0.49\textwidth, trim=0 0 0 0, clip]{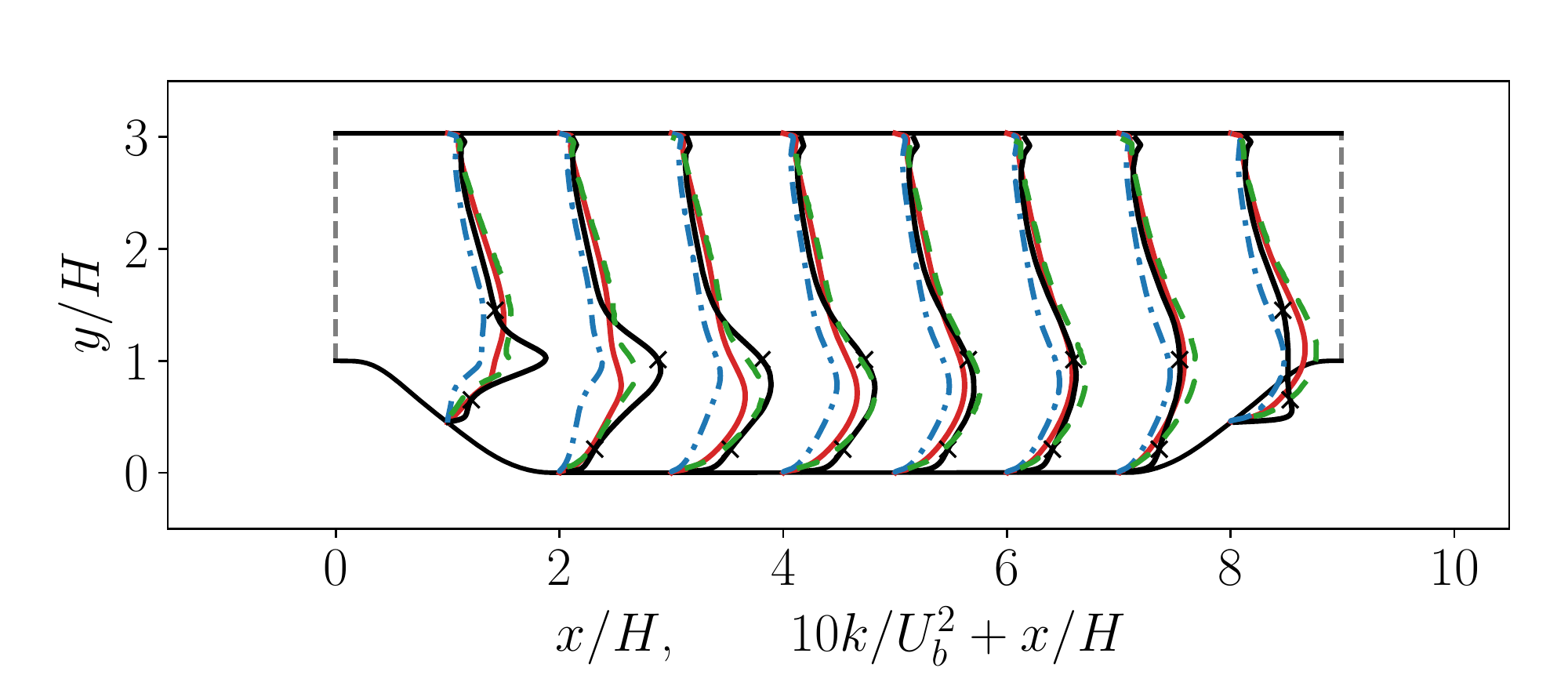}} \\
    \caption{Plots of the predicted velocity and turbulence kinetic energy with the models learned from indirect data and from both indirect and direct data. The truth, the baseline, and the results of Schmelzer et al.~\cite{schmelzer2020discovery} are also provided for comparison.}
    \label{fig:results_pehills}
\end{figure}

Learning with direct and indirect data can improve the reconstruction of flow fields, compared to learning with only indirect data.
The evidence is provided in Figure~\ref{fig:results_pehills} where the plots of velocity and TKE along profiles are provided.
Learning from only indirect data can also improve velocity estimation, while the reconstructed TKE is not varied much from the baseline and still exhibits large discrepancies compared to the LES results.
The discrepancies in TKE are likely due to the model inadequacy.
Specifically, the dissipation term in the TKE transport equation is in an \textit{ad hoc} form.
Moreover, the Reynolds stress anisotropy is expressed in the form of tensor bases under the weak equilibrium assumption.
Such data-model incompatibility would cause the issue of ill-posedness, i.e., different TKE fields can provide similar velocity predictions.
The results suggest that having additional data in TKE is necessary to improve the TKE estimation simultaneously.

The ensemble-based method is able to learn comparable models to the work of Schmelzer et al.~\cite{schmelzer2020discovery} in the predictions of velocity and TKE.
The plots of predicted velocity and TKE along profiles with the learned models in the present work and Schmelzer et al.~\cite{schmelzer2020discovery} are shown in Figure~\ref{fig:results_pehills}.
It can be seen that the ensemble-based method achieves very similar velocity predictions to the results of Schmelzer et al.~\cite{schmelzer2020discovery}.
As for the TKE, there are slight differences between the two models.
Specifically, at the profiles of~$x/H = 1, 2, 3, 4$, the ensemble method predicts the TKE less accurately than the reference~\cite{schmelzer2020discovery}, particularly at the near wall region. 
In contrast, at the profiles of $x/H = 6, 7, 8$, the ensemble method can achieve slightly better estimation than the reference results~\cite{schmelzer2020discovery}.
Here we note that the ensemble method is capable of providing a comparable model to the symbolic regression method, while the predictive accuracy may be inferior since very sparse data is used in the present framework.
Incorporating more data could further improve the predictive ability of the learned model, and the optimal strategy for placing additional data is of significant importance for further investigation.

\begin{figure}[!htb]
    \centering
    \subfloat[Contour of scalar invariants]{
    \begin{tabular}{ccccl} 
        & indirect & 
        indirect + direct & LES & \\
        \rotatebox[origin=c]{90}{$\theta_1$} & 
        \raisebox{-.5\height}{\includegraphics[scale=0.6, trim=0 0 0 0, clip]{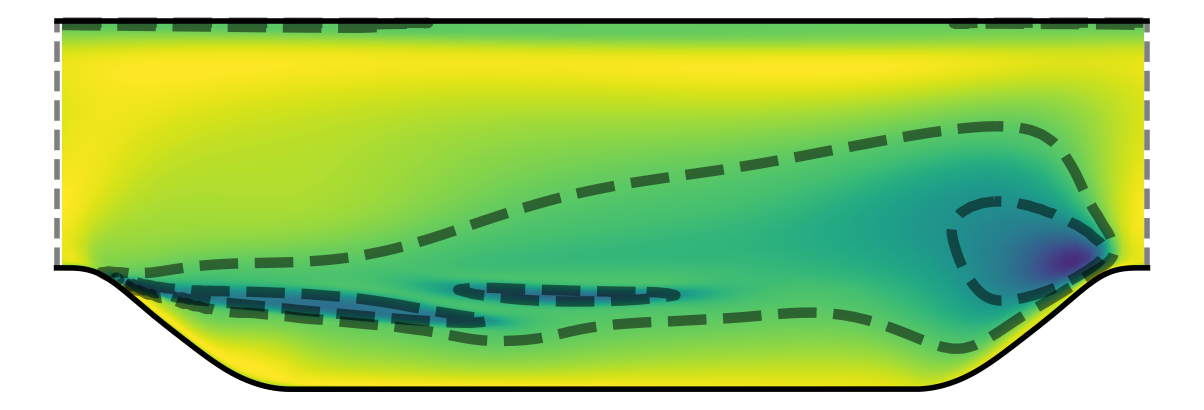}} &
        \raisebox{-.5\height}{\includegraphics[scale=0.6, trim=0 0 0 0, clip]{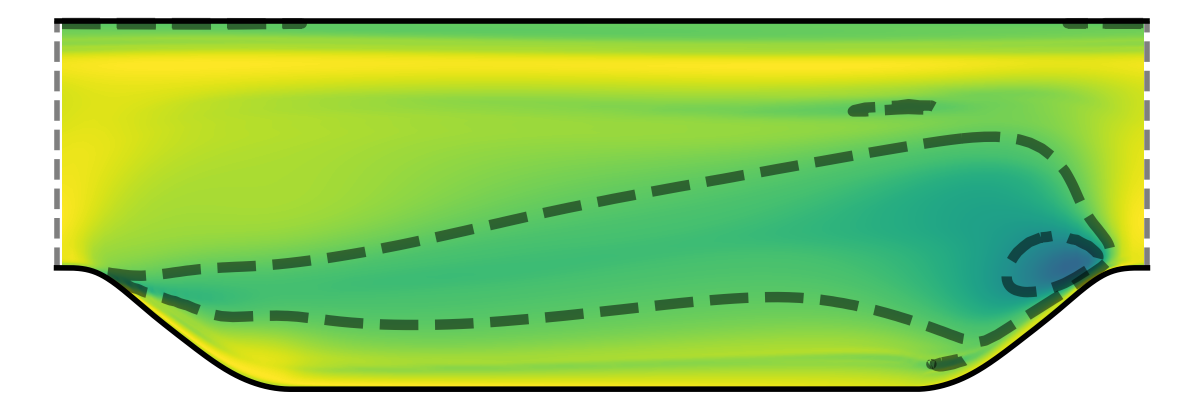}} &
        \raisebox{-.5\height}{\includegraphics[scale=0.6, trim=0 0 0 0, clip]{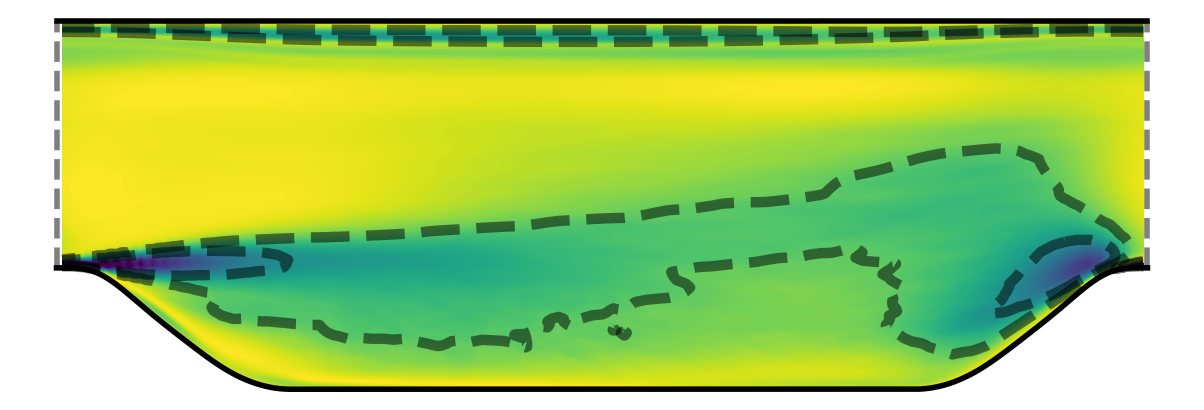}} &
        \raisebox{-.5\height}{\includegraphics[scale=0.5, clip]{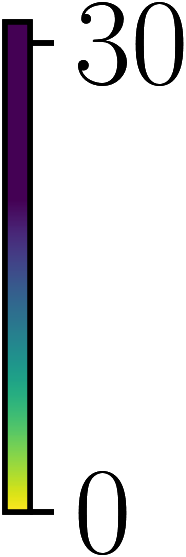}}
        \\
        \rotatebox[origin=c]{90}{$\theta_2$} & 
        \raisebox{-.5\height}{\includegraphics[scale=0.6, trim=0 0 0 0, clip]{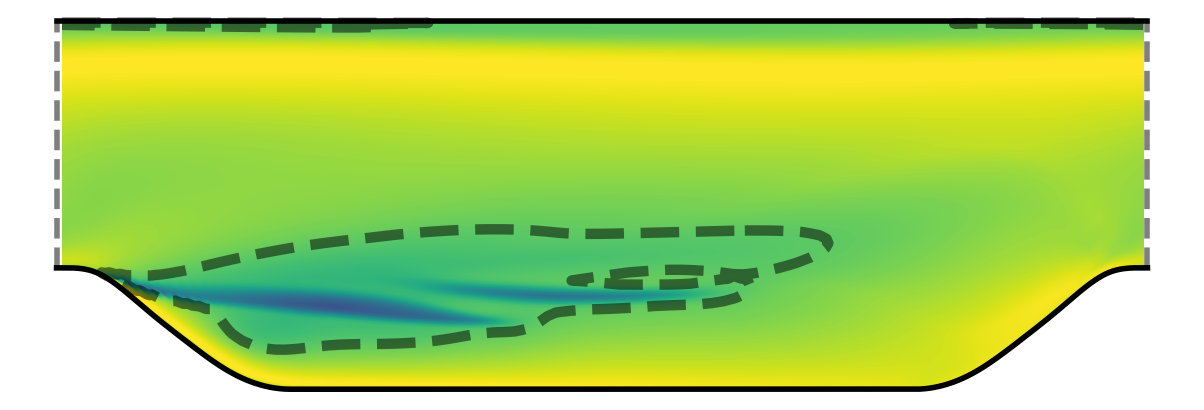}} &
        \raisebox{-.5\height}{\includegraphics[scale=0.6, trim=0 0 0 0, clip]{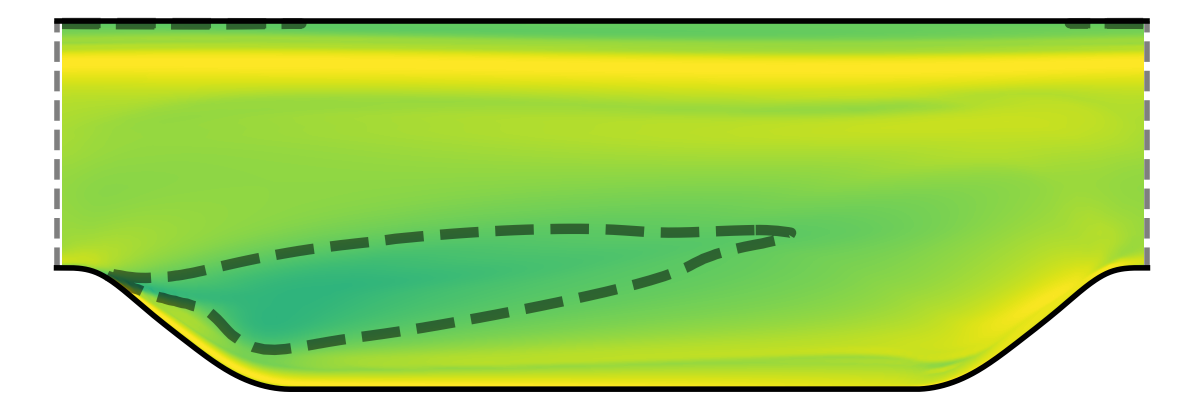}} &
        \raisebox{-.5\height}{\includegraphics[scale=0.6, trim=0 0 0 0, clip]{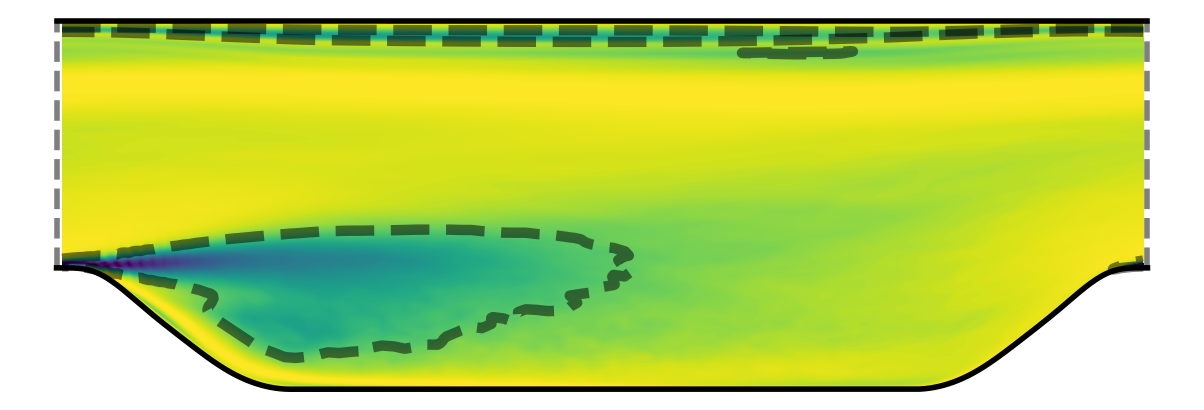}} & 
        \raisebox{-.5\height}{\includegraphics[scale=0.5, clip]{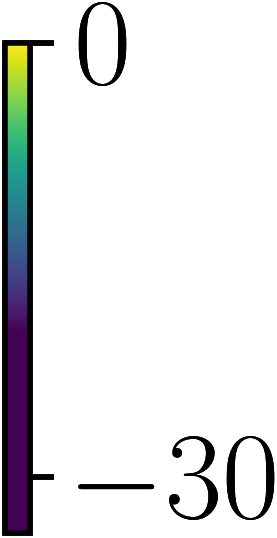}}
        \\
        \rotatebox[origin=c]{90}{$|\theta_1|-|\theta_2|$} &
        \raisebox{-.5\height}{\includegraphics[scale=0.6, trim=0 0 0 0, clip]{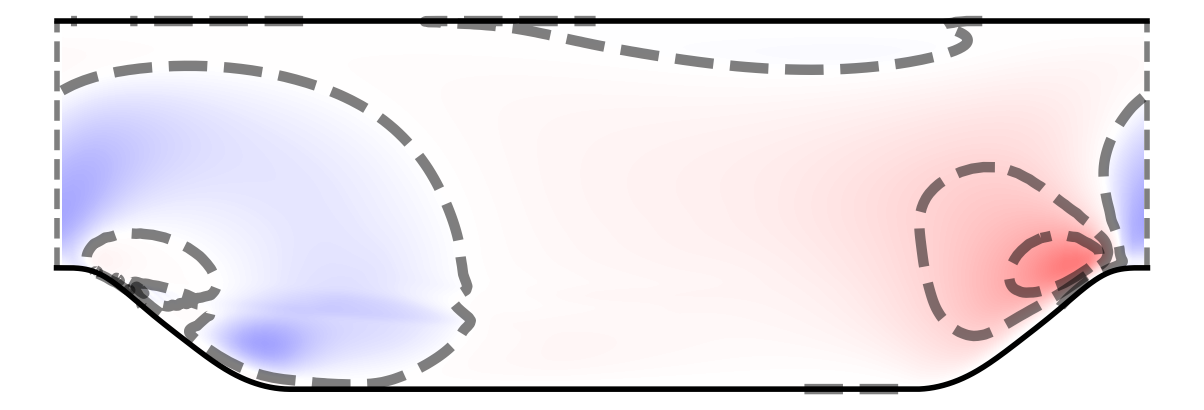}} &
        \raisebox{-.5\height}{\includegraphics[scale=0.6, trim=0 0 0 0, clip]{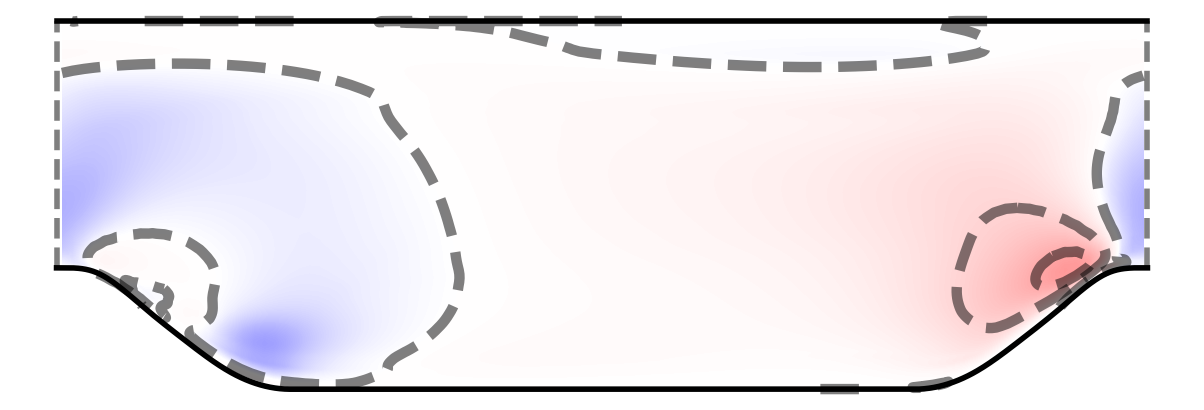}} &
        \raisebox{-.5\height}{\includegraphics[scale=0.6, trim=0 0 0 0, clip]{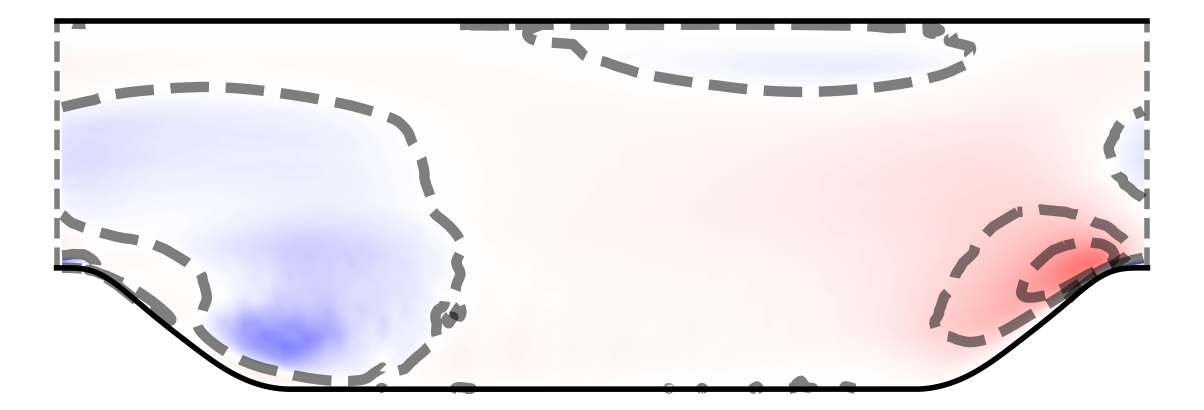}} & 
        \raisebox{-.5\height}{\includegraphics[scale=0.5, clip]{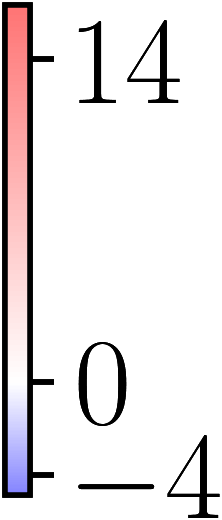}}
        \\
    \end{tabular}
    }\\
    \subfloat[Kernel density of $\theta_1$]{\includegraphics[width=0.4\textwidth]{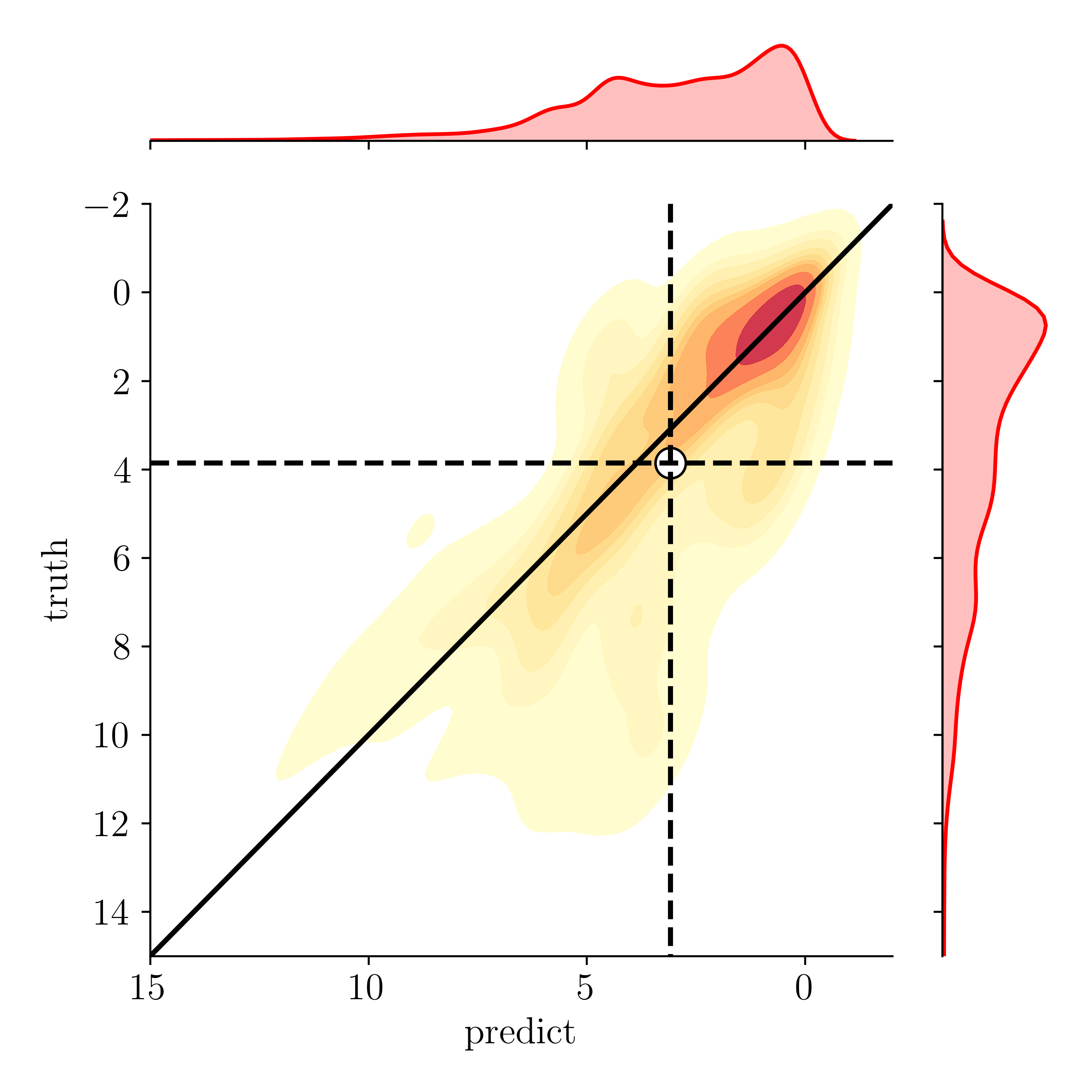}}
    \subfloat[Kernel density of $\theta_2$]{\includegraphics[width=0.4\textwidth]{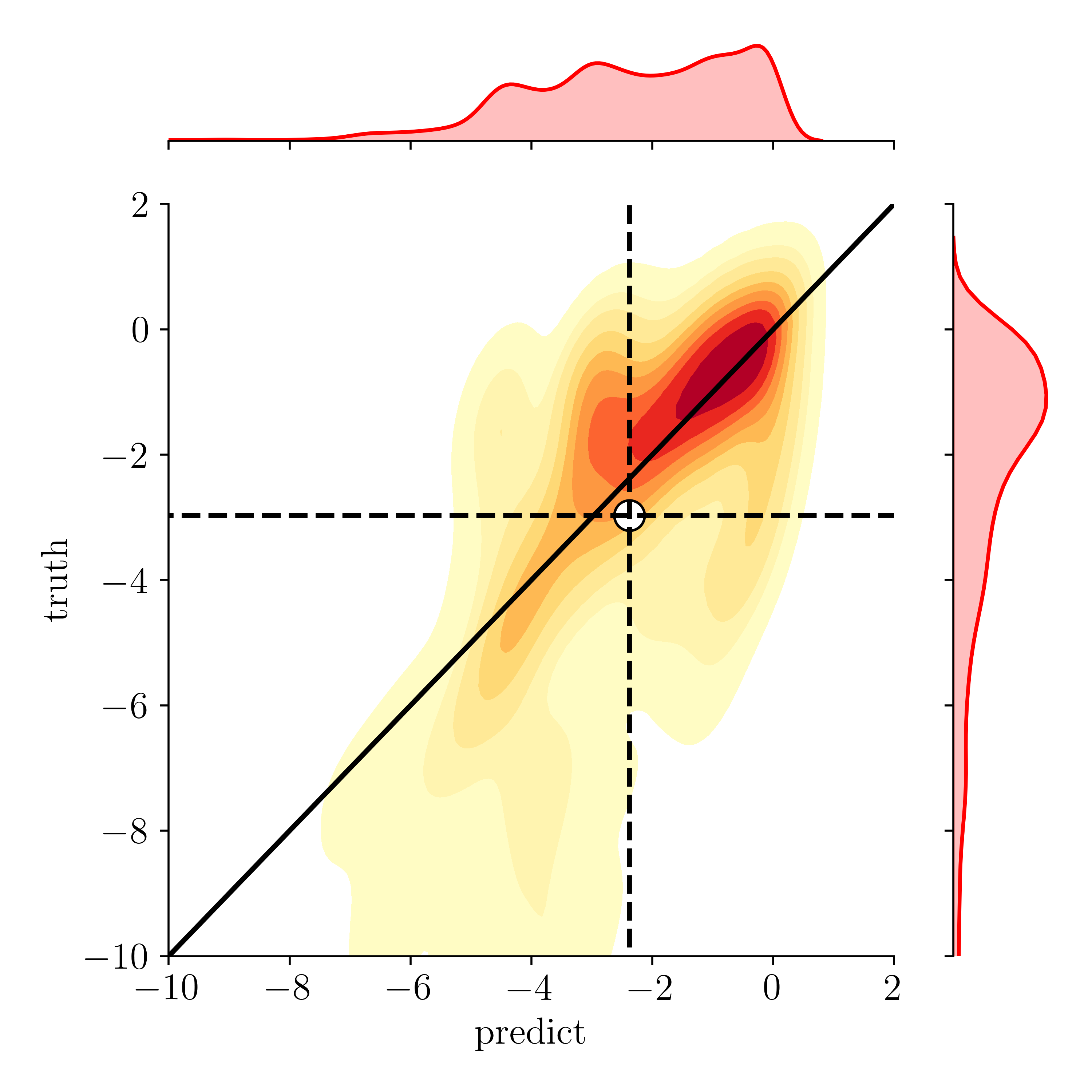}}
    \caption{Panel (a) shows contour plots of $\theta_1$, $\theta_2$, and $|\theta_1|-|\theta_2|$ with comparison among the model learned with indirect data, the model learned with direct and indirect data, and the LES; Panels (b) and (c) show kernel density plots of $\theta_1$ and $\theta_2$ from the truth and the model estimation for the periodic hill case, respectively.
    The probability densities of the truth and the model estimation are plotted on the margins.
    The circle indicates the values of the $30 \%$ quantiles (i.e., $30\%$ of the cells have $\bm{\theta}$ smaller than this value in the magnitude) for the model learned from indirect and direct data. 
    }
    \label{fig:hill_theta_pdf}
\end{figure}

The learned model with only indirect data and with the direct and indirect data can both estimate scalar invariants in good agreement with the LES results.
The plots of the scalar invariants are provided in Figure~\ref{fig:hill_theta_pdf} with a comparison among the learned model with indirect data, the learned model with direct and indirect data, and the LES results.
It can be seen that both learned models provide the scalar invariants $\theta_1$ and $\theta_2$ with similar patterns to the LES predictions.
That is likely because the two methods reconstruct similar velocity fields as shown in Figure~\ref{fig:results_pehills}.
Figures~\ref{fig:hill_theta_pdf}(b) and (c) present the kernel density of $\theta_1$ and $\theta_2$ based on the results of learning with the direct and indirect data.
The contours of the kernel density show the alignment with the diagonal line, which indicates good agreement between the truth and model predictions.
The probability density function (PDF) at the margin shows that the number of cells with large magnitudes is small, and these cells are mainly located at the shear layer from Figure~\ref{fig:hill_theta_pdf}(a).
The $30\%$ percentile is located at around $4$ and $-3$ for $\theta_1$ and $\theta_2$, respectively, which indicates that most cells are less than these values in magnitude.

\begin{figure}[!htbp]
    \centering
    \includegraphics[width=\textwidth]{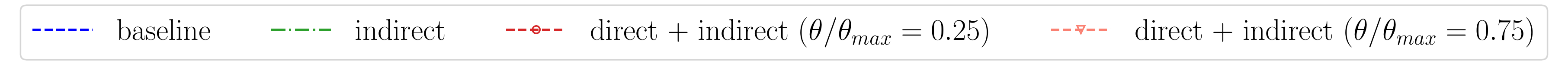}\\
    \subfloat[$g^{(1)}$ vs $\bm{\theta}$]{\includegraphics[width=0.44\textwidth]{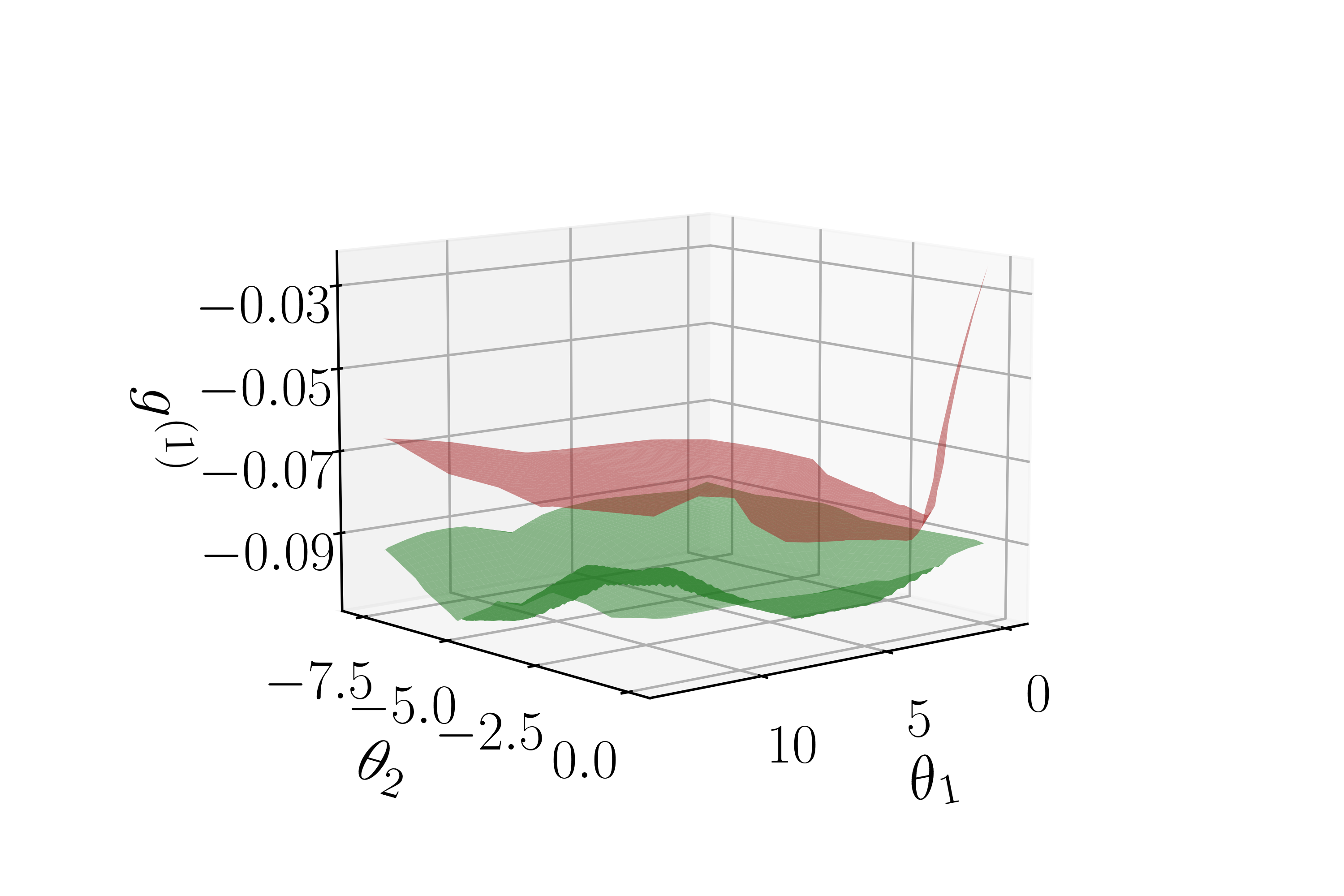}} 
    \subfloat[$g^{(1)}$ vs $\theta_1$]{\includegraphics[width=0.28\textwidth]{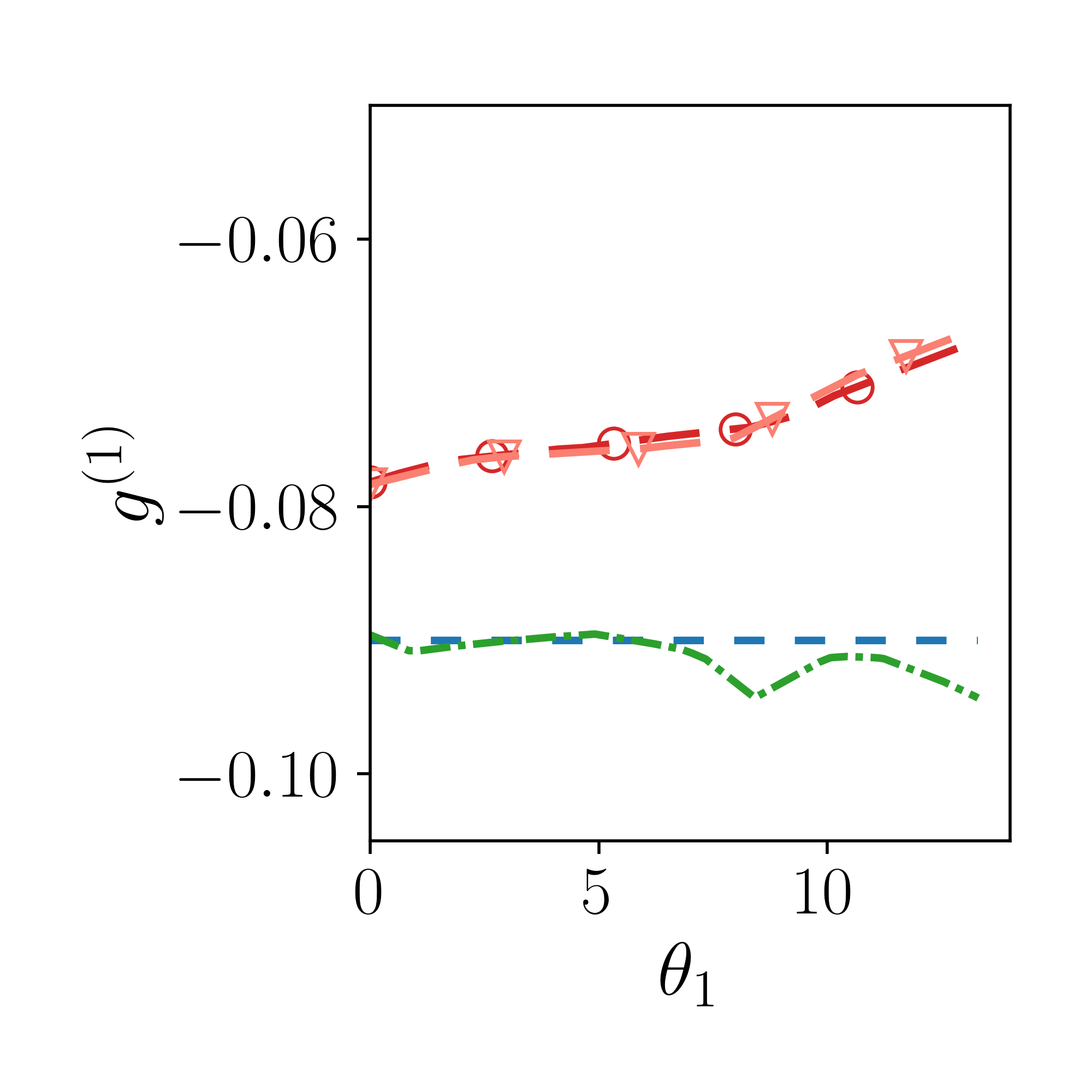}} 
    \subfloat[$g^{(1)}$ vs $\theta_2$]{\includegraphics[width=0.28\textwidth]{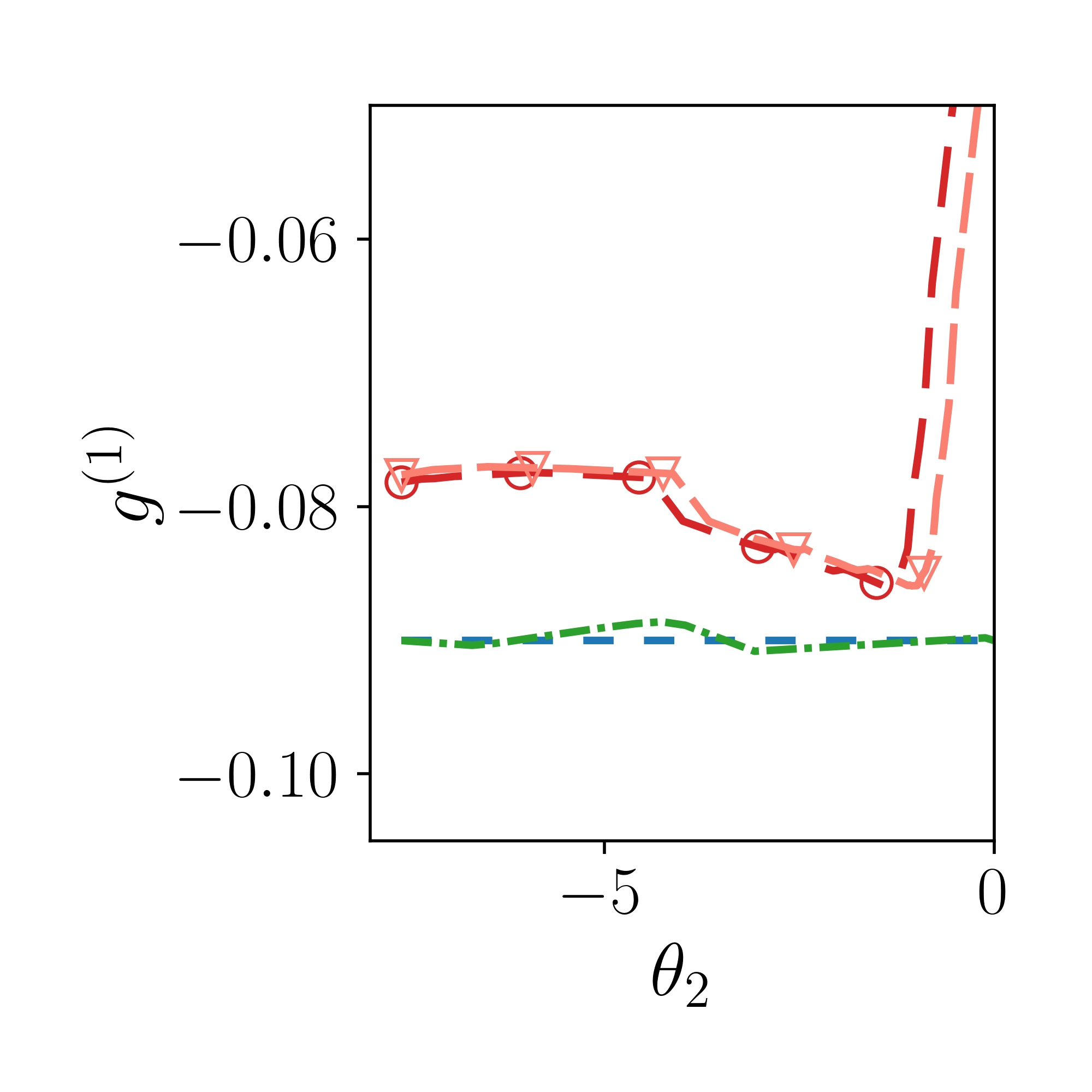}} \\
    \subfloat[$g^{(2)}$ vs $\bm{\theta}$]{\includegraphics[width=0.44\textwidth]{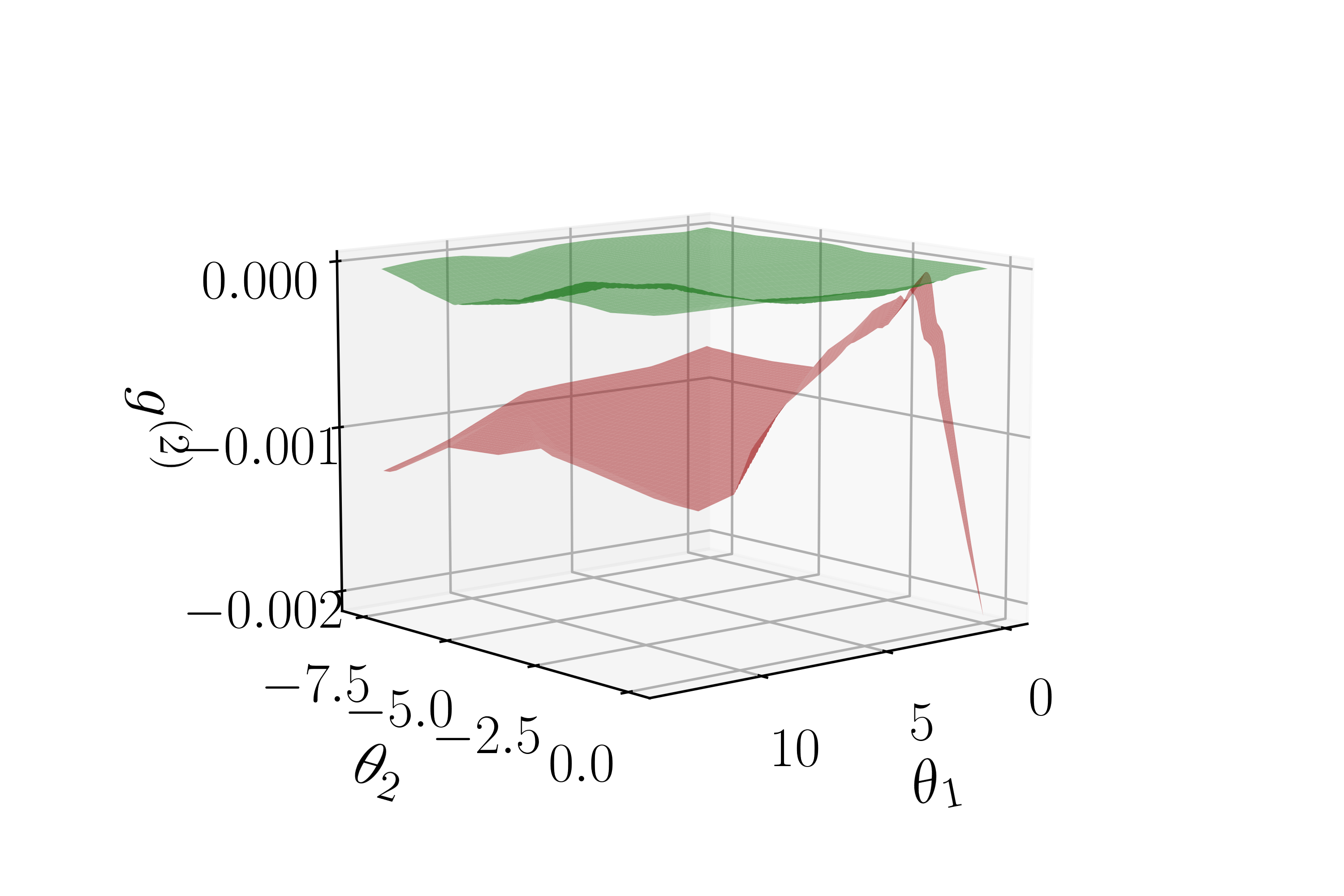}} 
    \subfloat[$g^{(2)}$ vs $\theta_1$]{\includegraphics[width=0.28\textwidth]{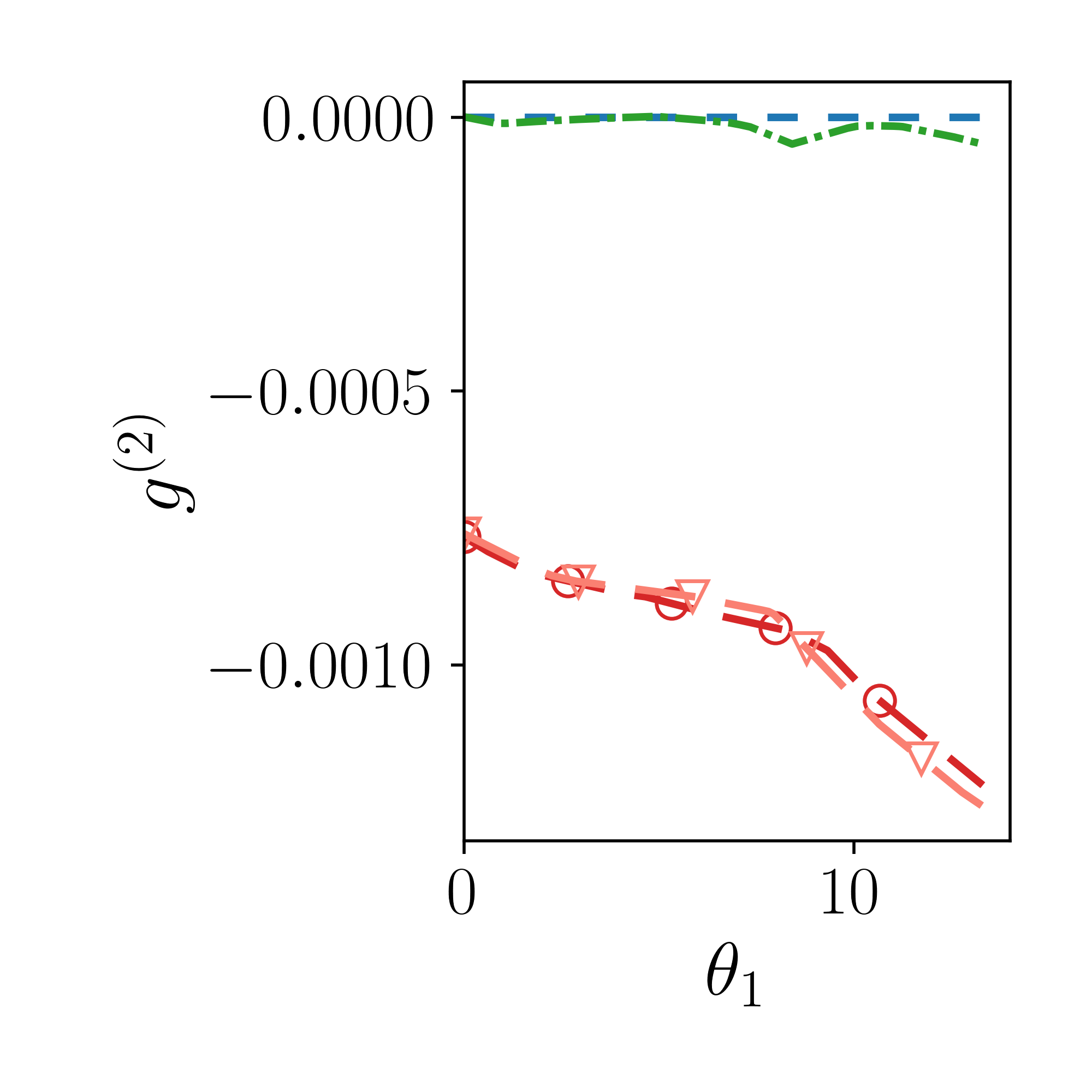}}
    \subfloat[$g^{(2)}$ vs $\theta_2$]{\includegraphics[width=0.28\textwidth]{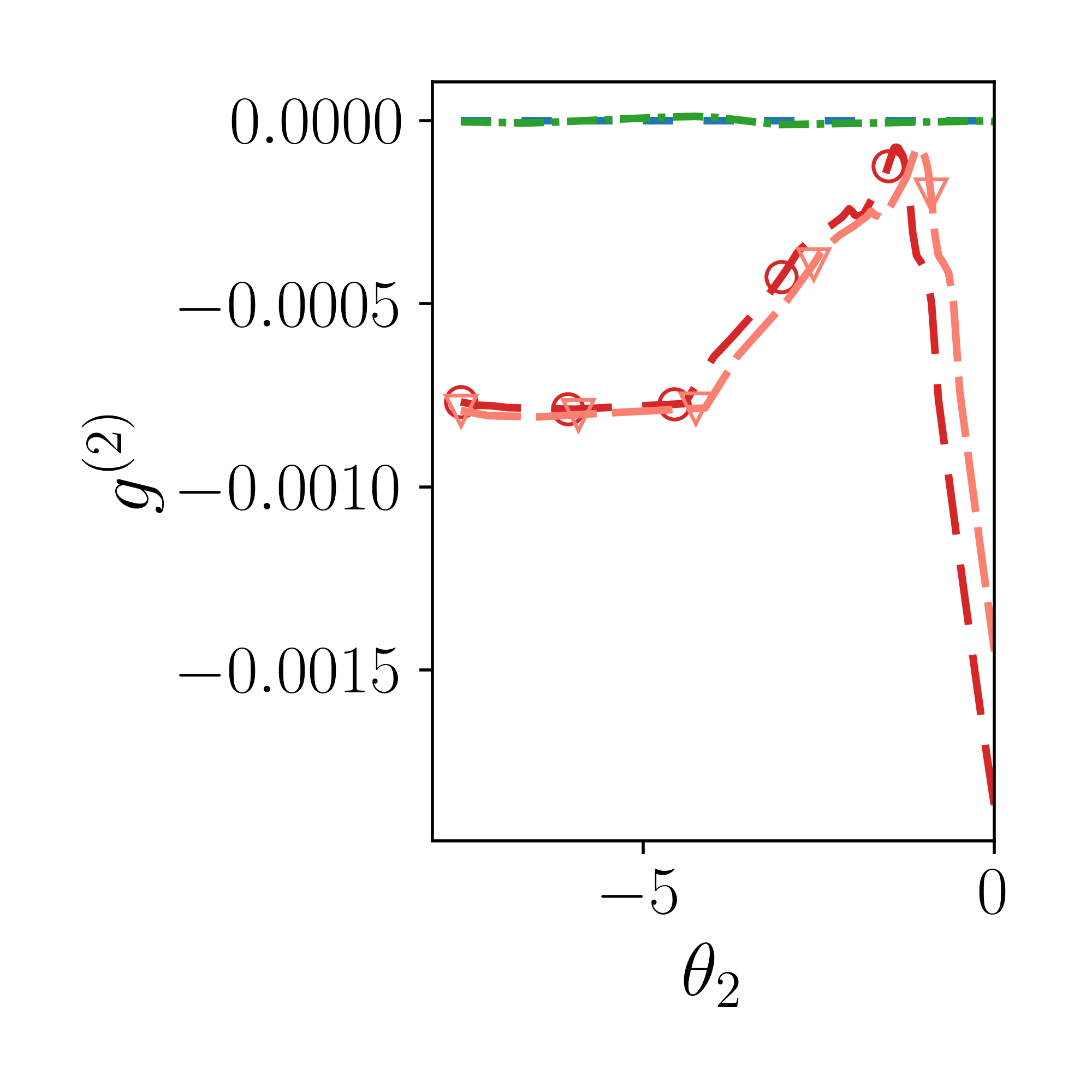}} \\
    \subfloat[$f^{(1)}$ vs $\bm{\theta}$]{\includegraphics[width=0.44\textwidth]{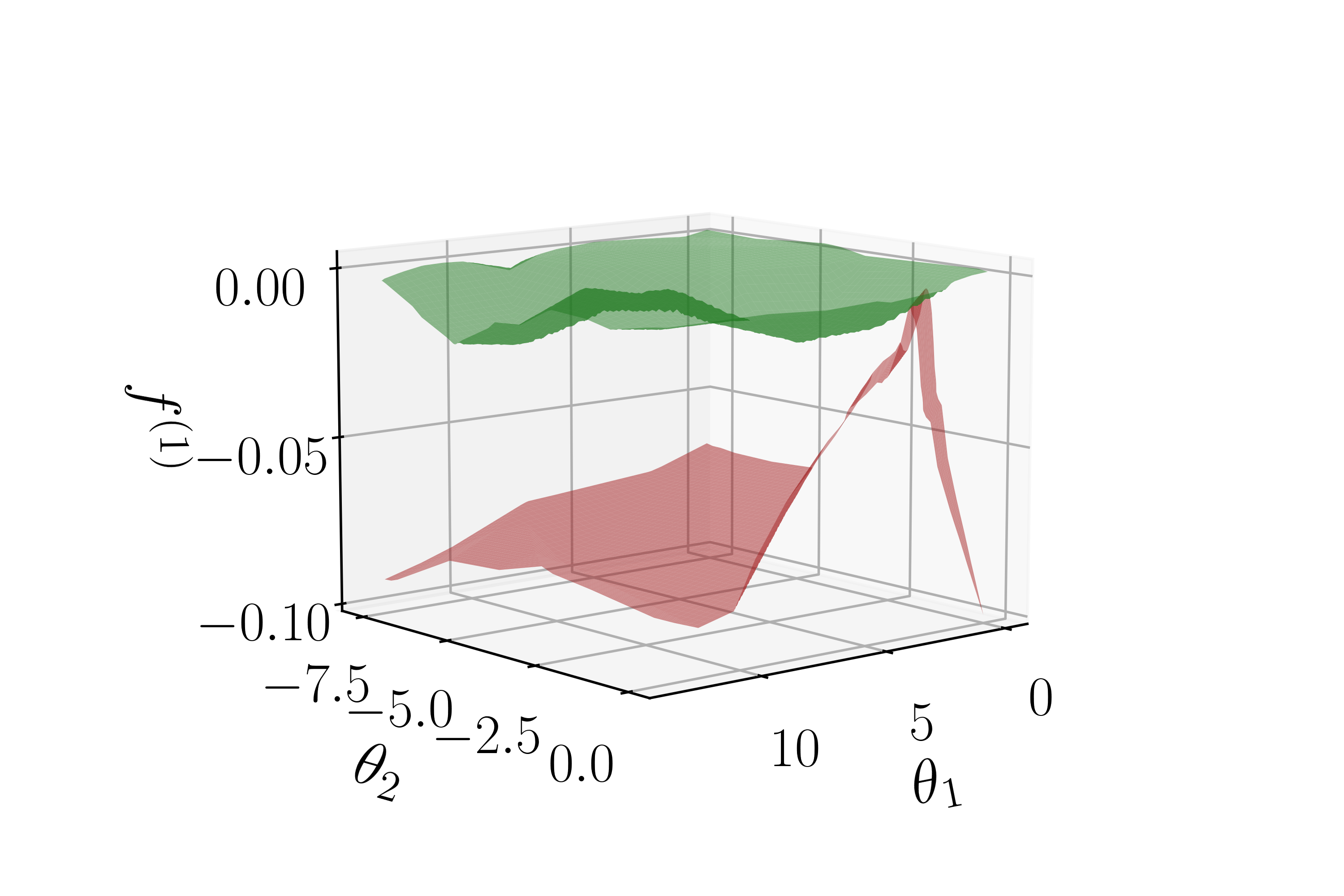}} 
    \subfloat[$f^{(1)}$ vs $\theta_1$]{\includegraphics[width=0.28\textwidth]{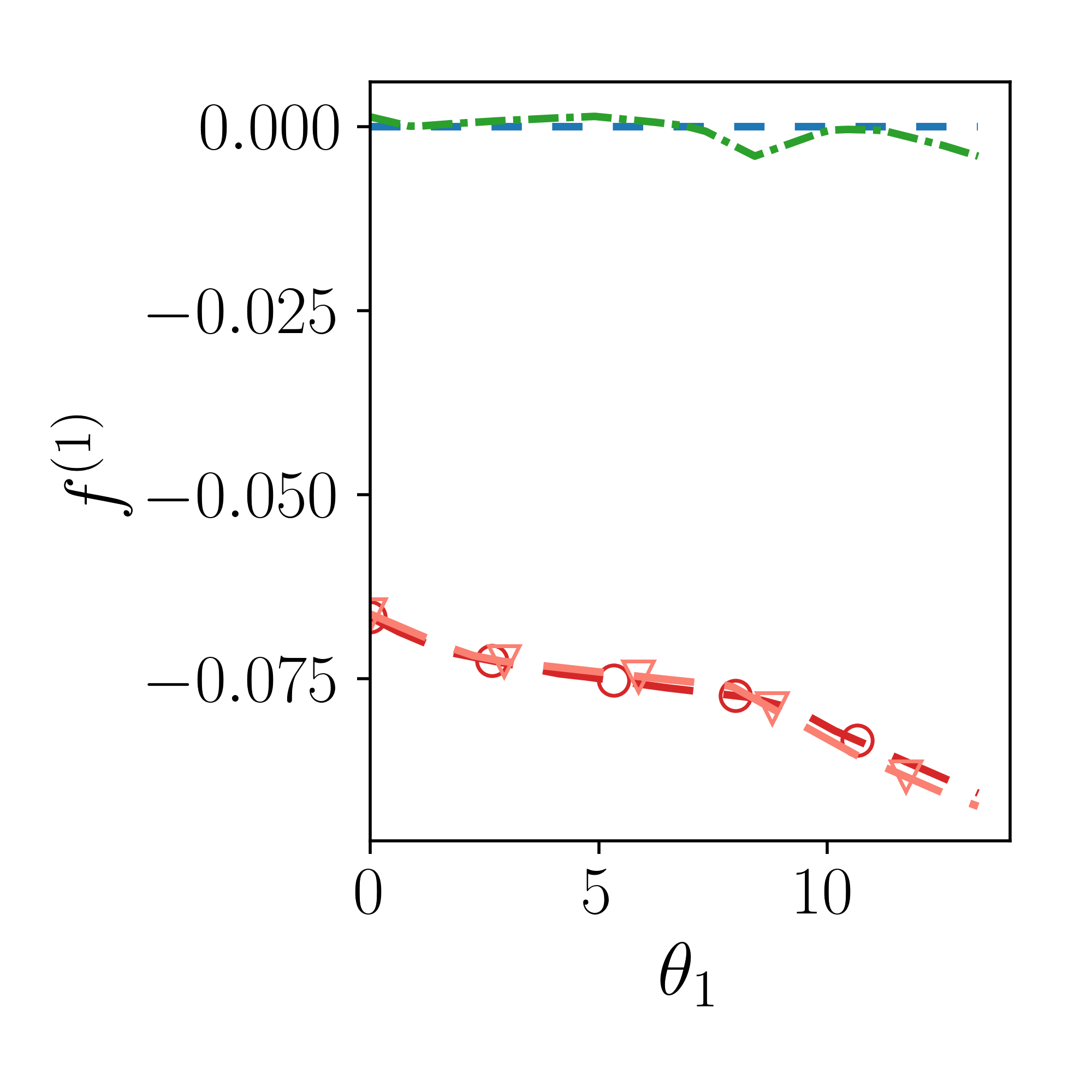}} 
    \subfloat[$f^{(1)}$ vs $\theta_2$]{\includegraphics[width=0.28\textwidth]{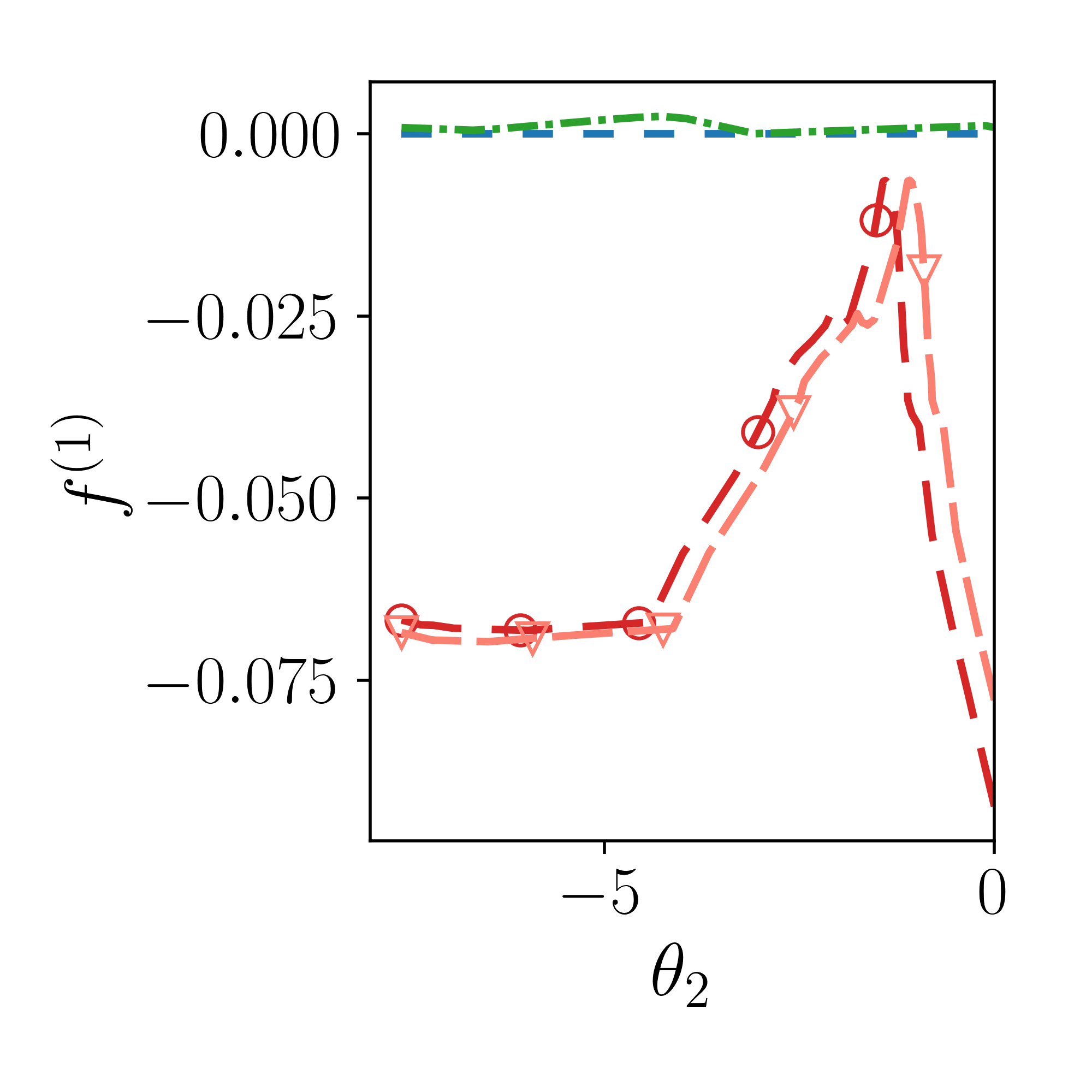}}
    \caption{Plots of the mapping between the scalar invariants~$\boldsymbol{\theta}$ and the tensor coefficient~$\boldsymbol{g}$ and $f$ with comparison among the baseline model, the model learned from indirect data, and the model learned from the direct and indirect data for the periodic hill case.
    Panels (a), (d), (g) show the model function of $g^{(1)}$, $g^{(2)}$, and $f^{(1)}$, respectively, where the red surface represents the learned model with direct and indirect data, and the green surface represents the learned model with indirect data.
    Panels (b), (c), (e), and (f) show the curve plots of the $\boldsymbol{g}$ function at specific planes, while panels (h) and (i) show the plots of the $f$ function.
    For learning with the direct and indirect data, the plots indicate the learned function at $\theta/\theta_\text{max}=0.25$ and $0.75$.
    For the baseline and the learned model from indirect data, the plots only show the learned function at $\theta/\theta_\text{max}=0.25$, since they are almost constant in the entire function space.
    }
    \label{fig:hill_g}
\end{figure}

Learning from the direct and indirect data provides the model functions with strong nonlinearity compared to learning from only indirect data.
The comparison between the learned functions with the two data sources is shown in Figure~\ref{fig:hill_g}.
It can be seen that the learned model with only the indirect data leads to almost constant values for the $\boldsymbol{g}$ and $\boldsymbol{f}$ functions.
On the contrary, the trained model with direct and indirect data varies over the entire function space.
The model function $\boldsymbol{g}$ learned from the direct and indirect data deviates from the baseline to a large extent, while the model with only indirect data varies slightly from the baseline value.
Specifically, the $g^{(1)}$ function learned from only indirect data is almost constant around $-0.09$.
Only for large scalar invariants, the magnitude of $g^{(1)}$ is slightly increased from Figure~\ref{fig:hill_g}(b).
Oppositely, the magnitude of the $g^{(1)}$ function with the direct and indirect data is decreased compared to the baseline.
Moreover, with only indirect data, the learned model is almost a linear eddy viscosity model since the nonlinear term $g^{(2)}$ and the correction term~$f^{(1)}$ is almost zero.
By contrast, with the indirect data and direct data, the learned model provides a nonlinear eddy viscosity model where $g^{(2)}$ function is varied in the range of $[-0.002, 0]$.
Besides, the correction term~$f^{(1)}$ is also varied between $-0.1$ and $0$ to achieve good predictions of the velocity and TKE simultaneously.
Learning from only indirect data leads to the correction term~$f^{(1)}$ around $0$, which means that the velocity field can be reconstructed well  without activating the correction in the TKE equation.
In contrast, learning from the direct and indirect data leads to a correction~$f^{(1)}$ with a large magnitude.
The learned model with direct and indirect data exhibits abrupt variations around $\bm{\theta}=0$, likely due to the specific parameter sets used in the training.
Our numerical test indicates that the standard deviation of indirect data of $0.02$ can provide similar training performance without such abrupt changes, and the plots are omitted for brevity.
Moreover, these abrupt changes mainly occur around $\theta_1=\theta_2=0$ where the learned $g$ functions have negligible effects on the Reynolds stress and further velocity.
Hence, the training method can provide different $\boldsymbol{g}$ functions in the range of small scalar invariants that make accurate velocity predictions.

The ensemble Kalman method can train the model with direct and indirect data efficiently.
Specifically, the ensemble method trains the model in around $2500$ core hours, while the genetic programming method needs $4600$ core hours in the work of Waschkowski et al.~\cite{waschkowski2022multi}.
Both methods achieve comparable predictions to the benchmark results~\cite{schmelzer2020discovery}.
It is noted that the model learning frameworks of these two works are different.
That is, the work of~\cite{waschkowski2022multi} uses a genetic programming method to learn the model in the symbolic form, while the present work uses the ensemble method to train the tensor basis neural networks.
Hence, the comparison only presents the efficiency of the two learning approaches to find the optimal turbulence closures instead of a comprehensive comparison between the ensemble method and genetic programming. 
The relatively high training efficiency of the ensemble method is likely due to the fact that the present method amounts to finding a local minimum based on the ensemble-based approximated gradient.
In contrast, genetic programming is a global optimization method that uses the evolutionary algorithm to find a global optimum.
Moreover, the ensemble method uses the low-rank approximated Hessian information to achieve a second-order optimization, thereby accelerating the training convergence.
The computational costs of combining the two data sources are not varied much from that of using only indirect data for the periodic hill case. 
This is because observation augmentation is used in this case without involving additional regularization steps, which does not significantly increase the computational complexity of the Kalman updating scheme.

\begin{figure}[!htb]
    \centering
    \includegraphics[width=0.8\textwidth]{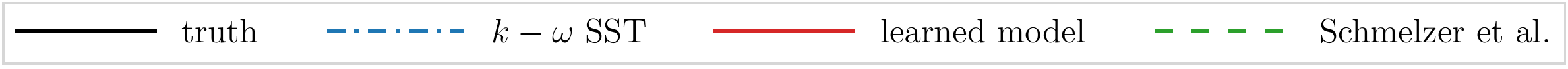} \\
    \subfloat[indirect ($U_x$)]{\includegraphics[width=0.5\textwidth]{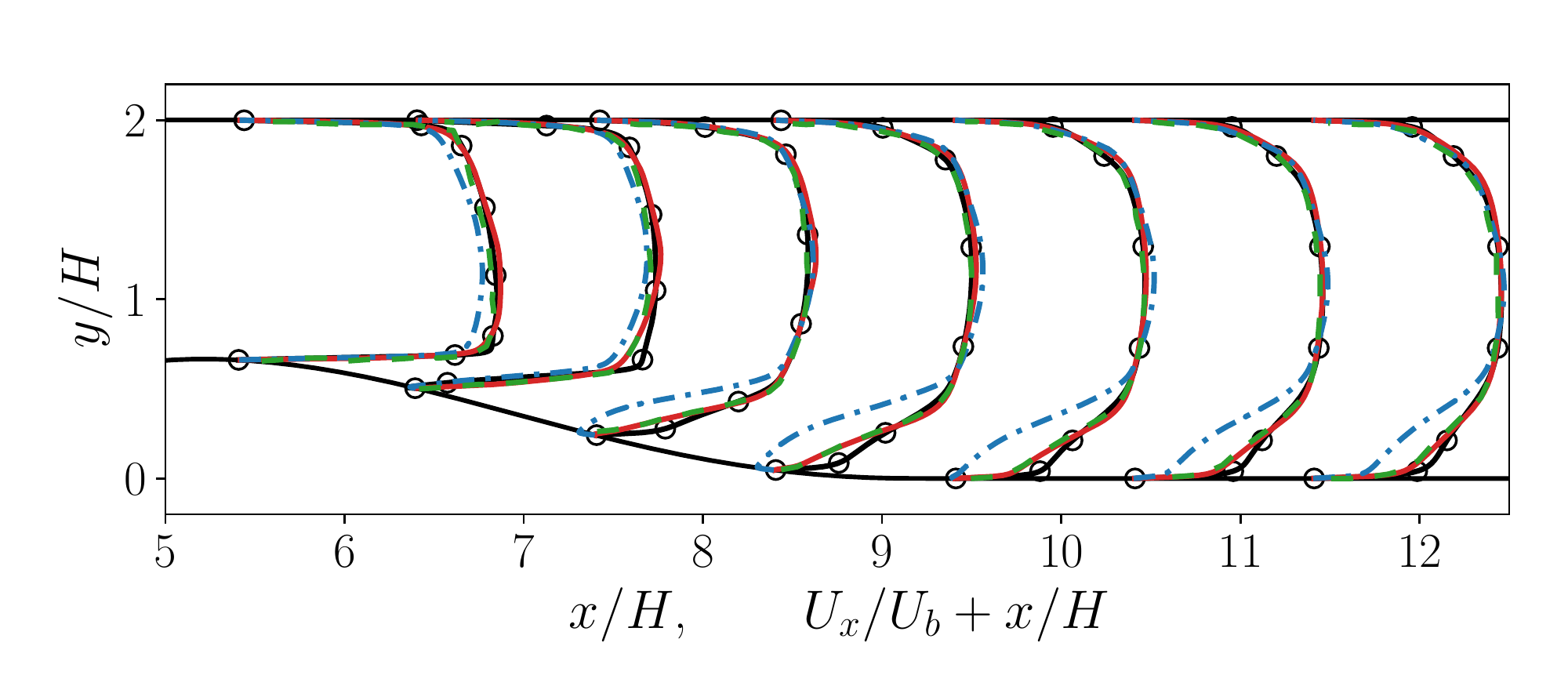}}
    \subfloat[indirect and direct ($U_x$)]{\includegraphics[width=0.5\textwidth]{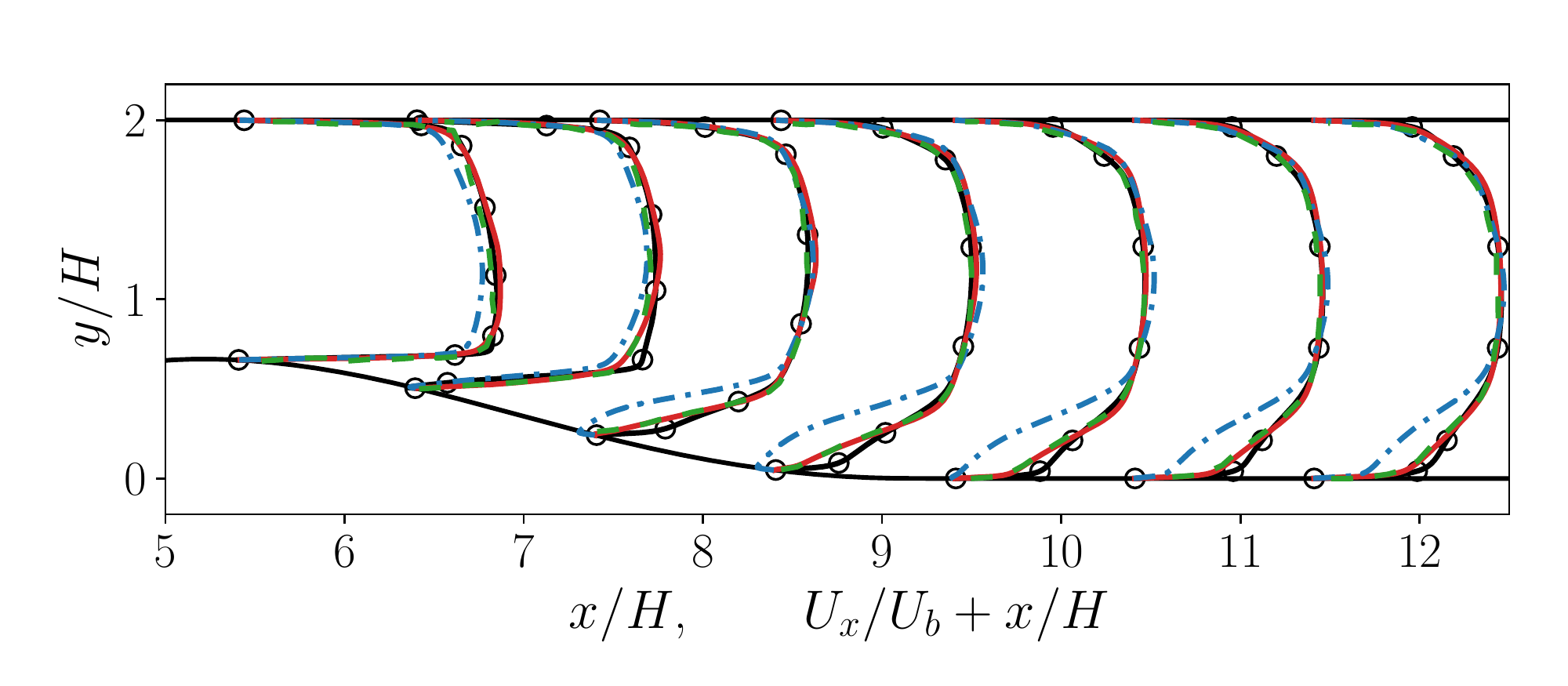}} \\
    \subfloat[indirect (k)]{\includegraphics[width=0.5\textwidth]{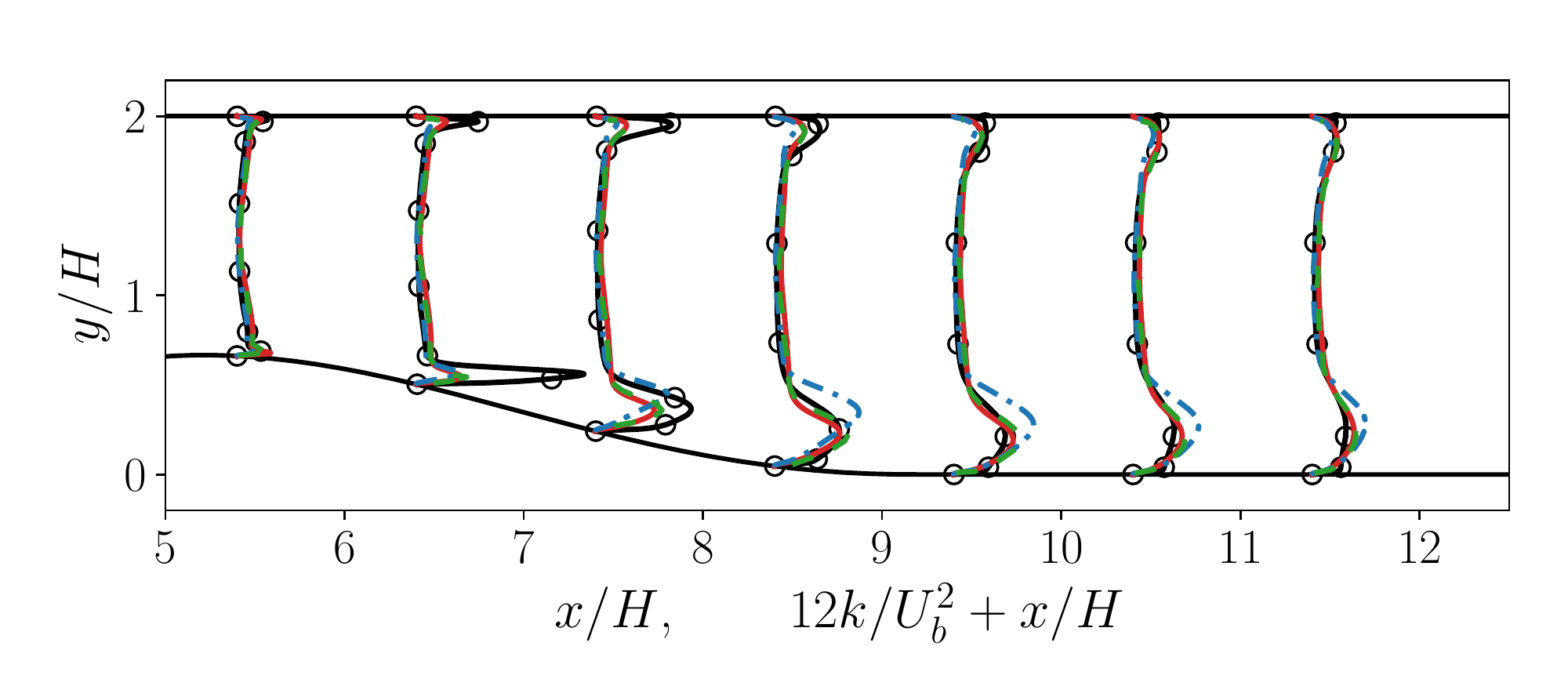}}
    \subfloat[indirect and direct (k)]{\includegraphics[width=0.5\textwidth]{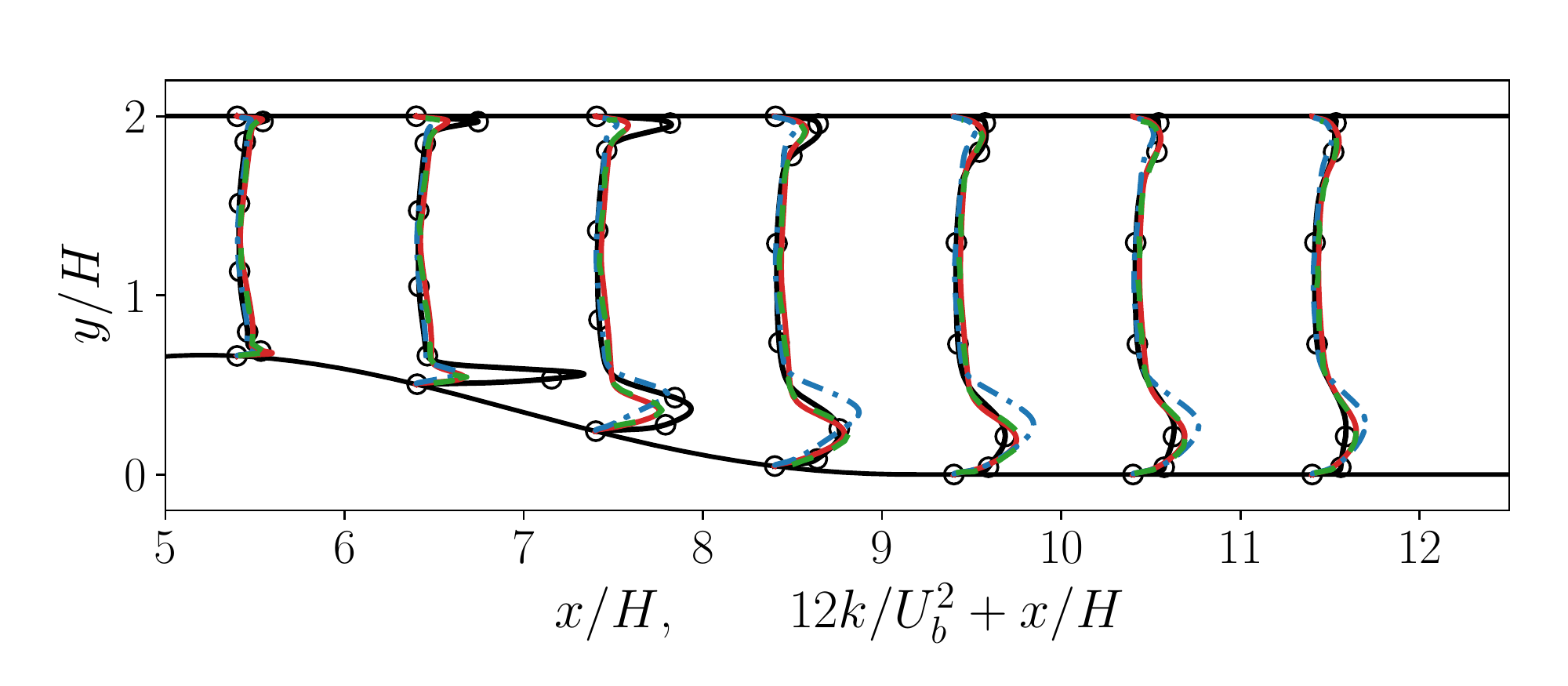}} \\
    \caption{Results of posterior test on flows in a convergent-divergent channel. The left column shows the velocity and turbulent kinetic energy predictions with the model learned from indirect data, while the right column shows the prediction with the model learned from the direct and indirect data.}
    \label{fig:generalizability_convdiv}
\end{figure}

Our posterior tests show that the model learned with direct and indirect data can be better generalized to similar configurations than the model learned from only indirect data.
The results of velocity and turbulent kinetic energy for the ConvDiv case are shown in Figure~\ref{fig:generalizability_convdiv}.
It can be seen that both learned models provide good agreements with the LES data in the velocity and turbulent kinetic energy.
The ConvDiv case has very small separation bubbles near the throat, and such features are slightly different from the periodic hill case.
In the region with adverse pressure gradients, both learned models are not able to provide satisfactory predictions in the TKE, while in the region with zero pressure gradients, the two models can predict velocity and TKE accurately compared to the DNS results.
The prediction errors of the two learned models are provided in Table~\ref{tab:error}.
It shows that velocity predictions of the two models are similar, both of which lead to the prediction error at $8.6\%$.
In contrast, combining the direct and indirect data can achieve slightly better predictions of TKE than learning with only indirect data.
The prediction error of TKE with combined direct and indirect data is around~$53.0 \%$, while that for learning with only indirect data is around~$54.5 \%$.
In addition, the model learned with direct and indirect data provides similar predictions to that of Schmelzer et al.~\cite{schmelzer2020discovery}, which indicates that the ensemble method is able to train the comparable model in the predictive ability for the ConvDiv case.

\begin{figure}[!htb]
    \centering
    \includegraphics[width=0.8\textwidth]{legend.pdf} \\
    \subfloat[indirect ($U_x$)]{\includegraphics[width=0.49\textwidth, trim=0 0 0 0, clip]{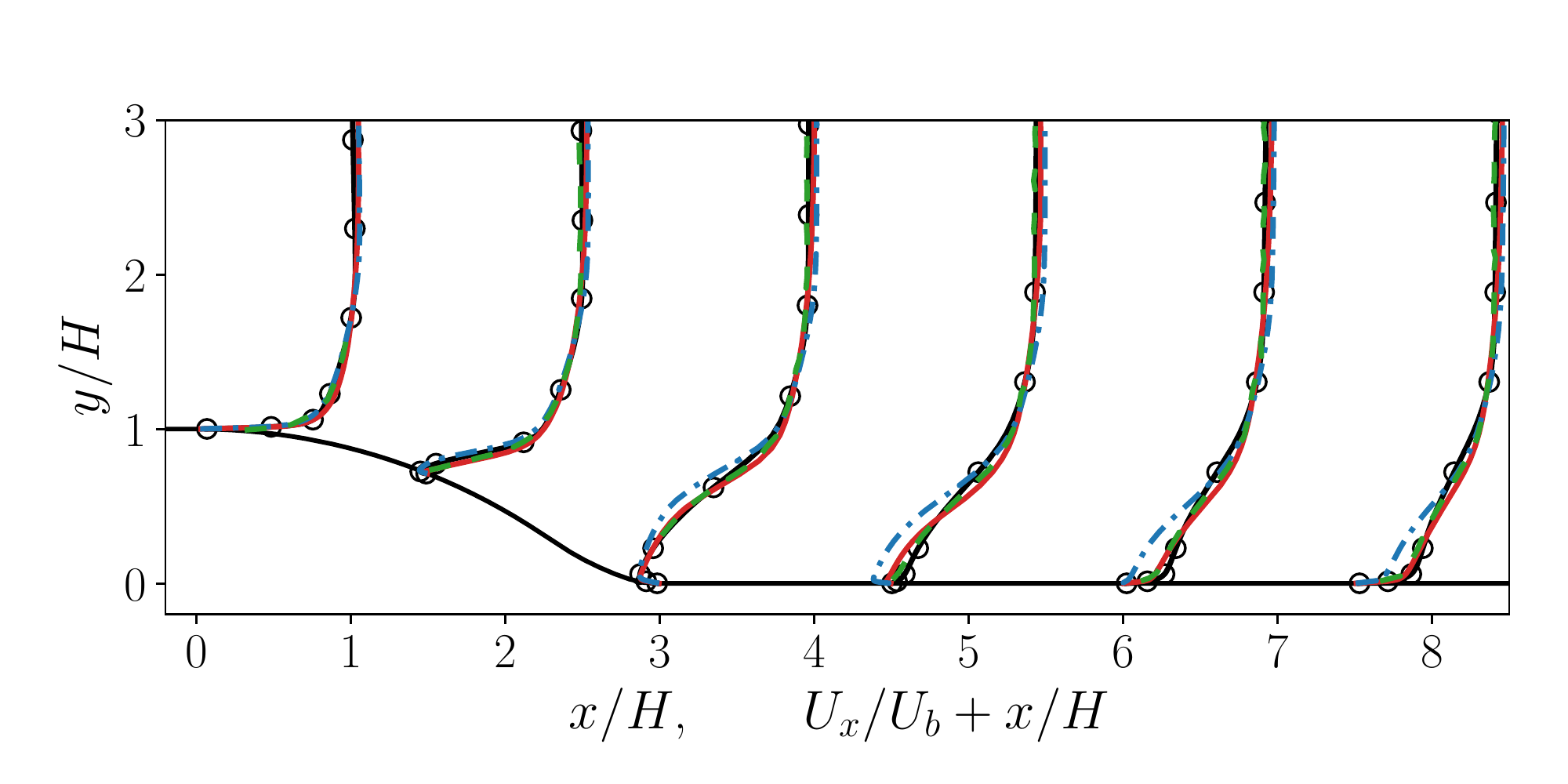}}
    \subfloat[indirect + direct ($U_x$)]{\includegraphics[width=0.49\textwidth, trim=0 0 0 0, clip]{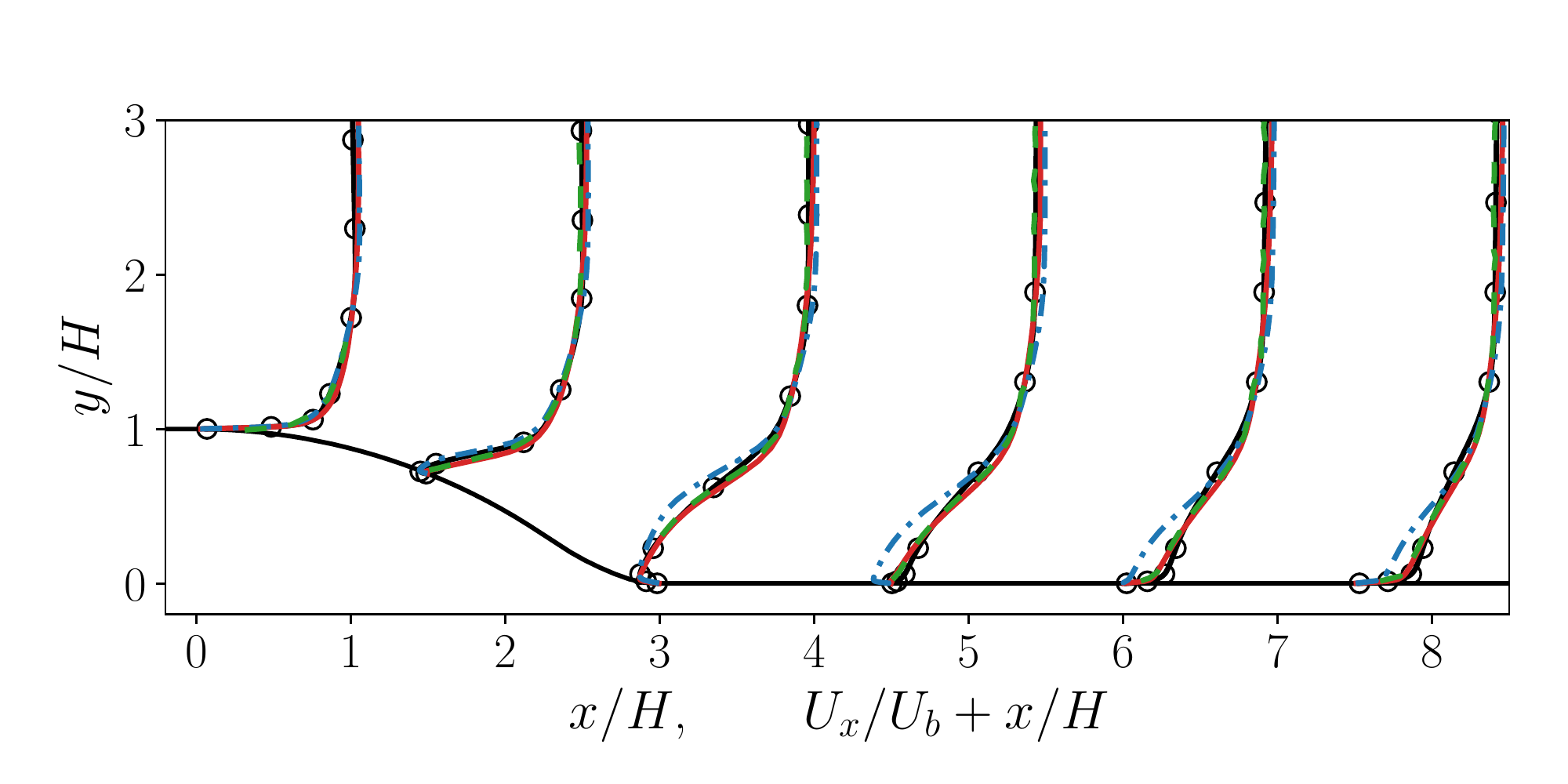}} \\
    \subfloat[indirect ($k$)]{\includegraphics[width=0.49\textwidth, trim=0 0 0 0, clip]{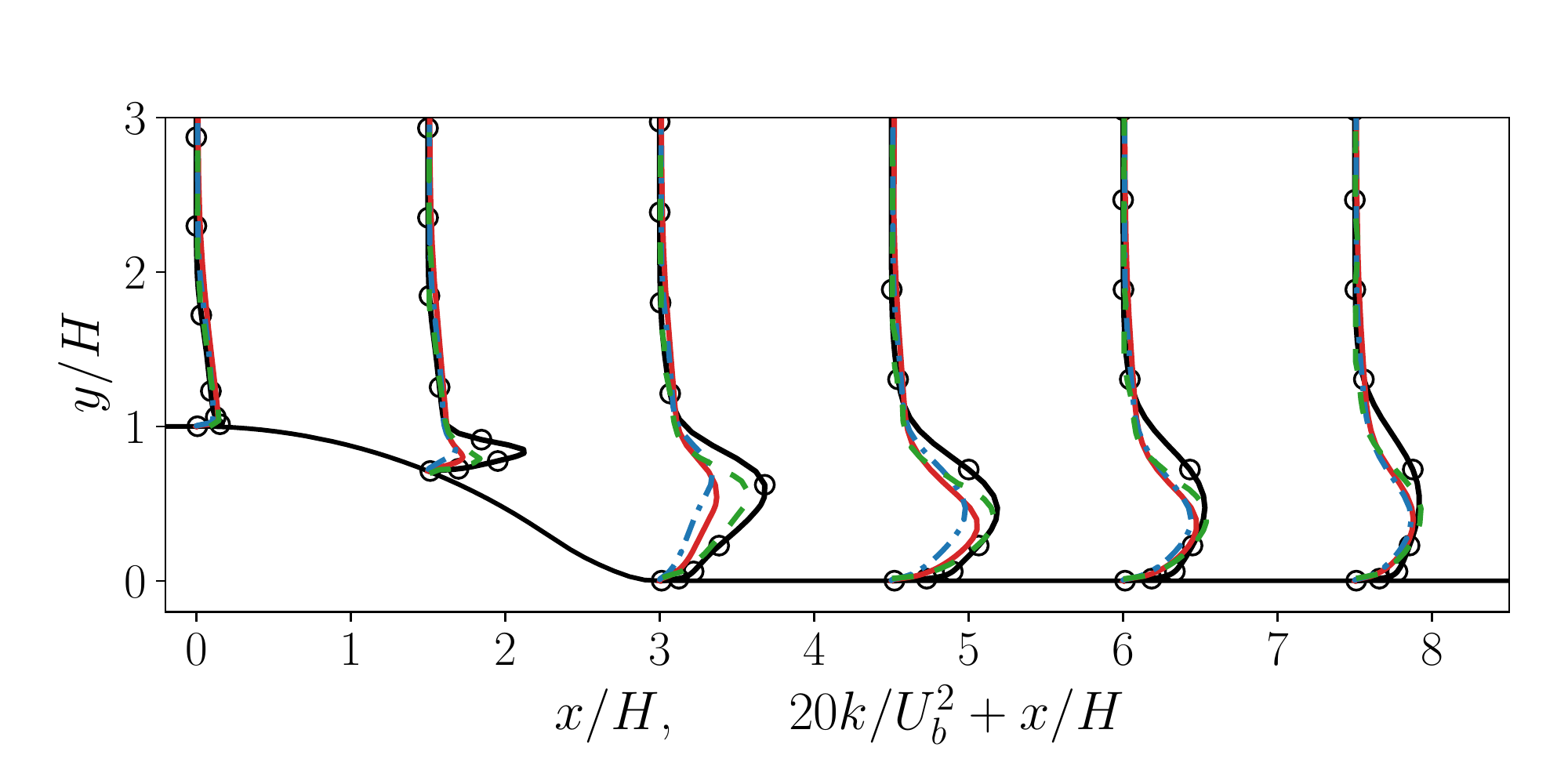}}
    \subfloat[indirect + direct ($k$)]{\includegraphics[width=0.49\textwidth, trim=0 0 0 0, clip]{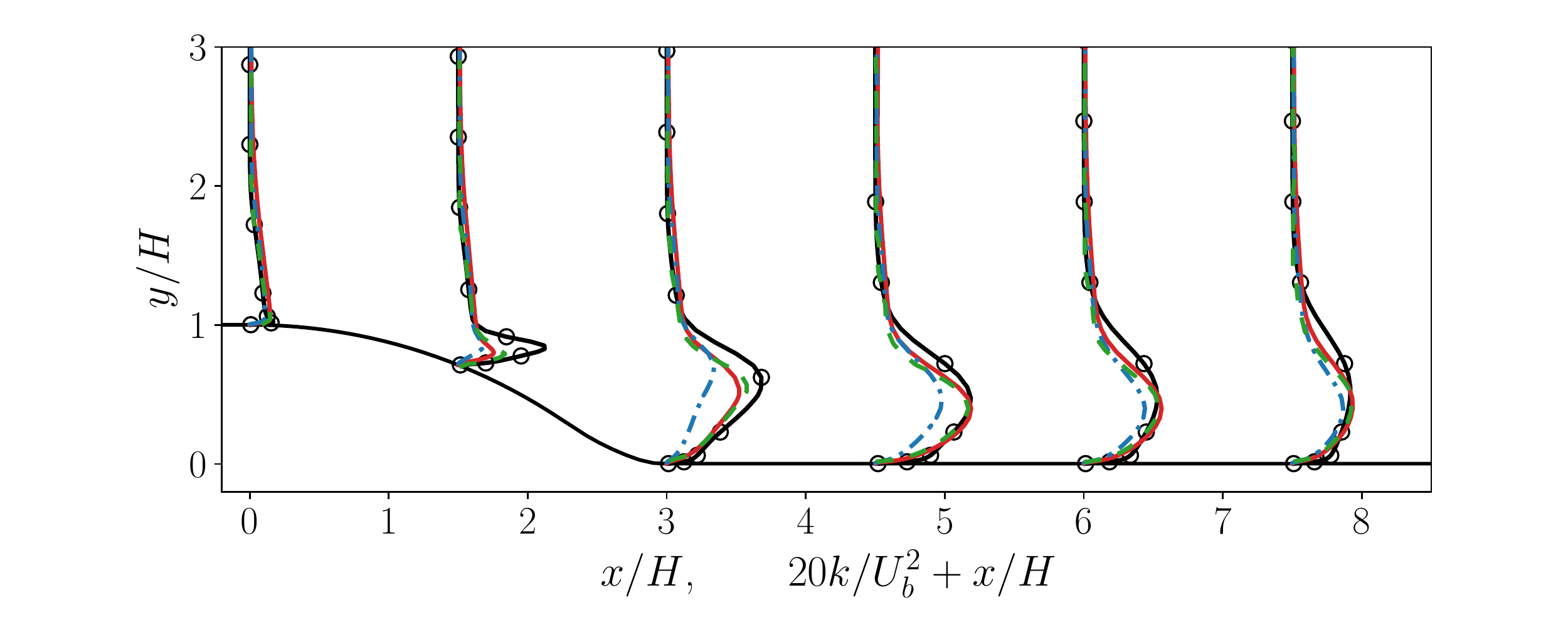}} \\
    \caption{Results of posterior test on flows over a curved back-forward step with a comparison to Schmelzer et al.~\cite{schmelzer2020discovery}. The left column shows the predictions of velocity ($u_x$, top row) and turbulent kinetic energy ($k$, bottom row) with the model learned from indirect data only, while the right column shows the prediction with the model learned from a combination of direct and indirect data.}
    \label{fig:generalizability_cbfs}
\end{figure}

The posterior results in the CBFS case further suggest that the model learned with the direct and indirect data can improve the predictive ability of the learned models.
The results of velocity and turbulent kinetic energy for the CBFS case are shown in Figure~\ref{fig:generalizability_cbfs}.
It can be seen that both models provide improved velocity predictions compared to the LES data.
The model learned from the indirect and direct data is able to achieve better reconstructions of the velocity compared to the case with only indirect data.
As for the turbulent kinetic energy, the prediction with the learned model from the indirect data varies slightly from the baseline $k$--$\omega$ SST model.
In contrast, the model learned from the indirect and direct data can provide the predictions of TKE in relatively better agreement with the LES data.
The prediction errors with different models are provided in Table~\ref{tab:error}.
It can be clearly seen that the velocity predictions are similar for the two learned models, while the model learned from the direct and indirect data achieve better prediction of the TKE compared to the model learned with only indirect data.
Specifically, the prediction error in TKE for the model learned from the direct and indirect data is $39.7 \%$, while that for the model with only indirect data is $45.3 \%$.
It further demonstrates that the ensemble method can effectively combine the direct and indirect data to enhance the predictive ability of data-driven turbulence models.
Our results demonstrate the effectiveness of the proposed method in combining direct and indirect sparse data.
However, we observe only a marginal improvement in global TKE prediction, possibly due to the dominant discrepancy near the throat.
While combining direct and indirect data improves the TKE prediction downstream, it provides similar results to using indirect data near the throat, which exhibits significant discrepancies from the DNS data.
Note that the prediction error is computed over the entire domain.
The large discrepancies near the throat may lead to unnoticeable global improvements, while local improvements can be observed from the profile plots, particularly in Figure~\ref{fig:generalizability_cbfs}.

\begin{table}[!htb]
	\centering
	\begin{tabular}{cccc}
		\hline
		cases       &  method           & err($\boldsymbol{u}$) & err($k$) \\
        \hline
		& baseline  & 17.9 \%   & 48.7 \%  \\
		Periodic hills 
		            & indirect          & 13.0 \%     & 33.7 \% \\
		            & indirect + direct & 13.2 \%     & 28.5 \%  \\
		            \hline
		            & baseline          & 19.2 \%     & 64.6 \%  \\
		Convergent-divergent channel     & indirect          & 8.6 \%      & 54.5 \% \\
					& indirect + direct & 8.6 \%      & 53.0 \%  \\
					\hline
		            & baseline          & 10.8 \%     & 52.1 \%  \\
		Curved back-forward step        & indirect          & 9.1 \%      & 45.3 \% \\
					& indirect + direct & 9.1 \%      & 39.7 \%  \\		            
		\hline
	\end{tabular}
	\caption{Summary of prediction errors over the entire domain}
	\label{tab:error}
\end{table}

\section{Conclusion}
\label{sec:conclusion}

This work presents an ensemble-based framework to combine the direct and indirect observations for learning turbulence models.
The regularized ensemble Kalman method and the observation augmentation are introduced to combine the indirect data of velocity with different direct data.
Specifically, the regularized ensemble Kalman method is used to combine the direct data of the Reynolds stress anisotropy and the indirect data of velocity by integrating the back-propagation technique and the ensemble-based gradient approximation.
The ensemble method with observation augmentation is used to combine the direct data of the turbulent kinetic energy and the indirect velocity data.
Combining the direct and indirect data can ensure the consistency between the training and prediction environments and thus improve the predictive ability of the learned model.
Moreover, the two data sources can take synergy effects to enable model learning from very sparse data.
The present method is non-intrusive and efficient in combining direct and indirect data by leveraging the available analytical and ensemble-based approximated gradients.

Two test cases, i.e., flows in a square duct and flows over periodic hills, are used to demonstrate the capability of the ensemble method.
The square duct case shows the capability of the present method to learn underlying model functions from the direct and indirect sparse data.
The periodic hill case demonstrates the performance of the method to learn the neural network-based models with improved predictions in both velocity and turbulent kinetic energy.
Both cases show that the ensemble method can combine the direct and indirect sparse data to improve the model's generalizability in similar configurations, compared to using only indirect data.
The ensemble method is very promising for learning generalizable turbulence models from multiple data sources due to its non-derivative nature.

\appendix
\counterwithin{figure}{section}
\counterwithin{table}{section} %
\counterwithin{equation}{section} %
\renewcommand\thefigure{\Alph{section}.\arabic{figure}}
\renewcommand\thetable{\Alph{section}.\arabic{table}}
\renewcommand\theequation{\Alph{section}.\arabic{equation}}

\section{Blending functions of $k$--$\omega$ SST model}
\label{sec:sst}

In this section, the formulation of the blending functions used in the $k$--$\omega$ SST model is introduced.
The formulation of the term~$Q_\text{SST}$ in Eq.~\eqref{eq:omega} of $k$--$\omega$ SST model is to combine the $k$--$\varepsilon$ model and the $k$--$\omega$ model, which can be written as
\begin{equation}
Q_\text{SST} = 2(1-F_{1}) \sigma_{\omega 2} \frac{1}{\omega} \frac{\partial k}{\partial x_j} \frac{\partial \omega}{\partial x_j} \text{.}
\end{equation}
In the formula above, $\sigma_{\omega2}$ is a model constant, and the blending function $F_1$ is formulated as
\begin{equation}
\begin{aligned}
&F_{1}=\tanh \left[\left(\min \left[\max \left(\frac{\sqrt{k}}{\beta^{*} \omega y}, \frac{500 \nu}{y^{2} \omega}\right), \frac{4 \sigma_{\omega 2} k}{\text{CD}_{k \omega} y^{2}} \right] \right)^{4} \right] \\
& \text{CD}_{k \omega}=\max \left(2 \sigma_{\omega 2} \frac{1}{\omega} \frac{\partial k}{\partial x_{j}} \frac{\partial \omega}{\partial x_{j}}, 10^{-10} \right) \text{,}
\end{aligned}
\end{equation}
where $\beta^*$ is a model constant, $y$ is the distance to the nearest wall, and $\nu$ is molecular viscosity.
The eddy viscosity of $k$--$\omega$ SST model is written as
\begin{equation}
    \nu_{t}=\frac{a_{1} k}{\max \left(a_{1} \omega, S F_{2}\right)}
\end{equation}
where $a_1$ is a model constant, $S$ represents the magnitude of the mean strain rate, and $F_2$ is a second blending function that confines the Reynolds shear stress based on Bradshaw's assumption within the boundary layer.
The blending function can be formulated as
\begin{equation}
    F_{2}=\tanh \left[\left[\max \left(\frac{2 \sqrt{k}}{\beta^{*} \omega y}, \frac{500 \nu}{y^{2} \omega}\right)\right]^{2}\right] \text{.}
\end{equation}
The model coefficients are obtained by blending the coefficients of the $k$--$\varepsilon$ and $k$--$\omega$ models as $\phi = \phi_2 + F_1 (\phi_1 - \phi_2)$, where $\phi$ represents the coefficient.
The whole set of the coefficients is
\begin{equation}
\begin{array}{ll}
\alpha_{1}=0.553, & \alpha_{2}=0.44, \quad \beta_{1}=0.075, \quad \beta_{2}=0.0828 \\
\sigma_{k 1}=0.85, & \sigma_{k 2}=1.0, \quad \sigma_{\omega 1}=0.5, \quad \sigma_{\omega 2}=0.856, \\
\beta^{*}=0.09, & \quad a_{1}=0.31 .
\end{array}
\end{equation}

\section{Hybrid gradient}
\label{sec:hybrid}
We present the hybrid gradient method~\cite{oliver2022hybrid} to learn the neural network-based turbulence model from indirect data in this section.
The ensemble Kalman method uses the ensemble-based gradient, i.e.,~$\mathsf{H} = {\partial \boldsymbol{u}}/{\partial \boldsymbol{w}} \approx \Delta \mathsf{U} (\Delta \mathsf{W})^{-1}$, to update the neural network weights, where the readily available gradient of the neural network is not used.
The hybrid gradient approach can combine the analytical gradient and the ensemble-based gradient to train neural network models with indirect data.
Specifically, the local gradient is decomposed into two parts based on the chain rule, i.e., 
\begin{equation}
    \mathsf{H} = \frac{\partial \boldsymbol{u}}{\partial \boldsymbol{w}} = \frac{\partial \boldsymbol{u}}{\partial \boldsymbol{\tau}} \frac{\partial \boldsymbol{\tau}}{\partial \boldsymbol{w}} \text{.}
\end{equation}
The gradient ${\partial \boldsymbol{u}}/{\partial \boldsymbol{\tau}}$ is obtained via the ensemble-based approximation as ${\partial \boldsymbol{u}}/{\partial \boldsymbol{\tau}} \approx \Delta \mathsf{U} (\Delta \boldsymbol{\tau})^{-1}$, while the gradient ${\partial \boldsymbol{\tau}}/{\partial \boldsymbol{w}}$ is obtained based on the back-propagation technique.
Thus the full gradient can be formulated as
\begin{equation}
    \frac{\partial \boldsymbol{u}}{\partial \boldsymbol{w}} \approx \Delta \mathsf{U} (\Delta \boldsymbol{\tau})^{-1} \frac{\partial \boldsymbol{\tau}}{\partial \boldsymbol{w}} \text{.}
\end{equation}
Further, the hybrid gradient is used to substitute the ensemble-based approximated gradient in the ensemble Kalman method.
It is noted that the local gradient $\partial \boldsymbol{u} / \partial \boldsymbol{w}$ is pre-multiplied by the model error covariance~$\mathsf{P}$ in the ensemble-based update scheme.
As such, the hybrid method requires computing the pseudo inversion~$(\Delta \boldsymbol{\tau})^{-1}$, while the ensemble method can avoid computing the pseudo inversion by reformulating the scheme by $$\mathsf{P}\mathsf{H}^\top \propto \Delta \mathsf{W} (\Delta \mathsf{W})^\top \left(\Delta \mathsf{U} (\Delta \mathsf{W})^{-1}\right)^\top = \Delta \mathsf{W} \Delta \mathsf{U}^\top \text{.}$$

\begin{figure}[!htb]
    \centering
    \begin{tabular}{ccccc}
        & $u_y$ & $\tau_{xy}$ & $\tau_{yz}$ & $\tau_{yy}-\tau_{zz}$ \\
        \rotatebox[origin=c]{90}{ground truth} & 
        \raisebox{-.5\height}{\includegraphics[scale=0.5]{duct_Uy_t.png}} &
        \raisebox{-.5\height}{\includegraphics[scale=0.5]{duct_tau12_t.png}} &
        \raisebox{-.5\height}{\includegraphics[scale=0.5]{duct_tau23_t.png}} &
        \raisebox{-.5\height}{\includegraphics[scale=0.5]{duct_tau22-tau33_t.png}}
        \\
        \rotatebox[origin=c]{90}{ensemble} & 
        \raisebox{-.5\height}{\includegraphics[scale=0.5]{duct_Uy_indirect.png}} &
        \raisebox{-.5\height}{\includegraphics[scale=0.5]{duct_tau12_indirect.png}} &
        \raisebox{-.5\height}{\includegraphics[scale=0.5]{duct_tau23_indirect.png}} &
        \raisebox{-.5\height}{\includegraphics[scale=0.5]{duct_tau22-tau33_indirect.png}}
        \\
        \rotatebox[origin=c]{90}{hybrid} &
        \raisebox{-.5\height}{\includegraphics[scale=0.5]{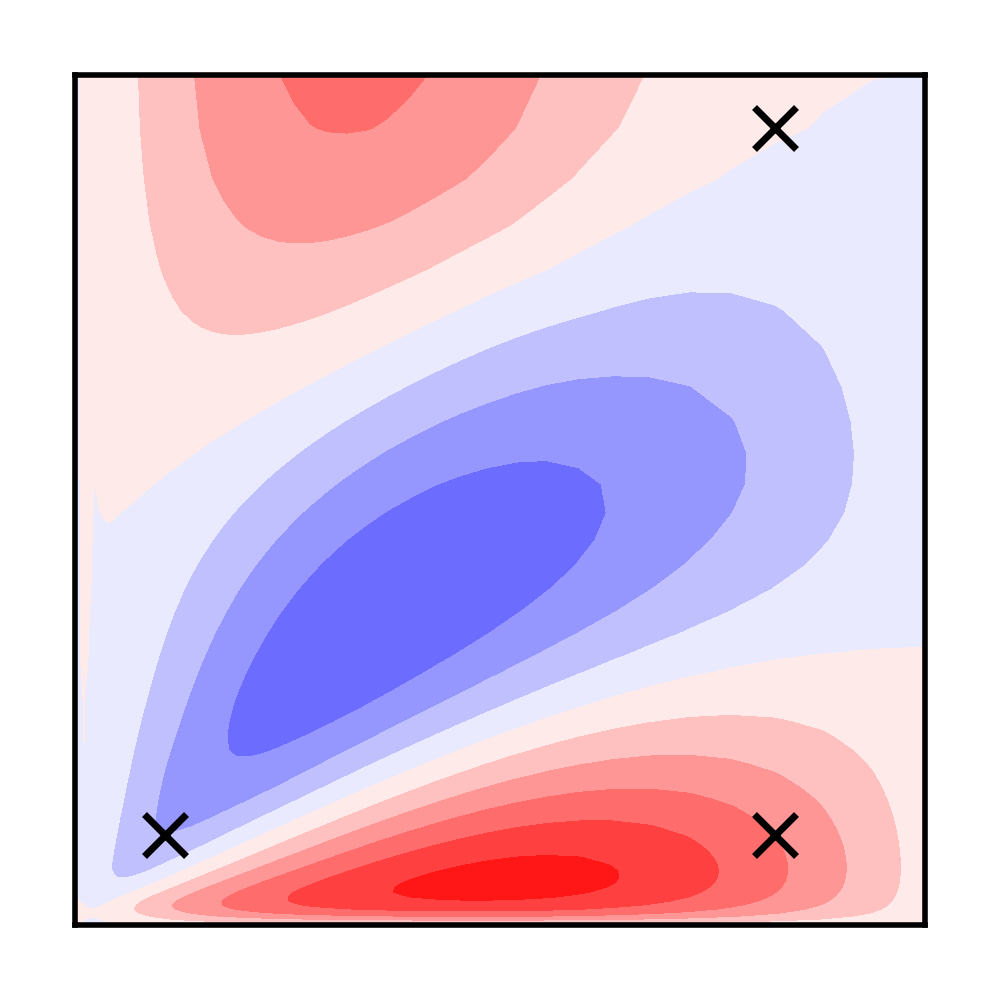}} &
        \raisebox{-.5\height}{\includegraphics[scale=0.5]{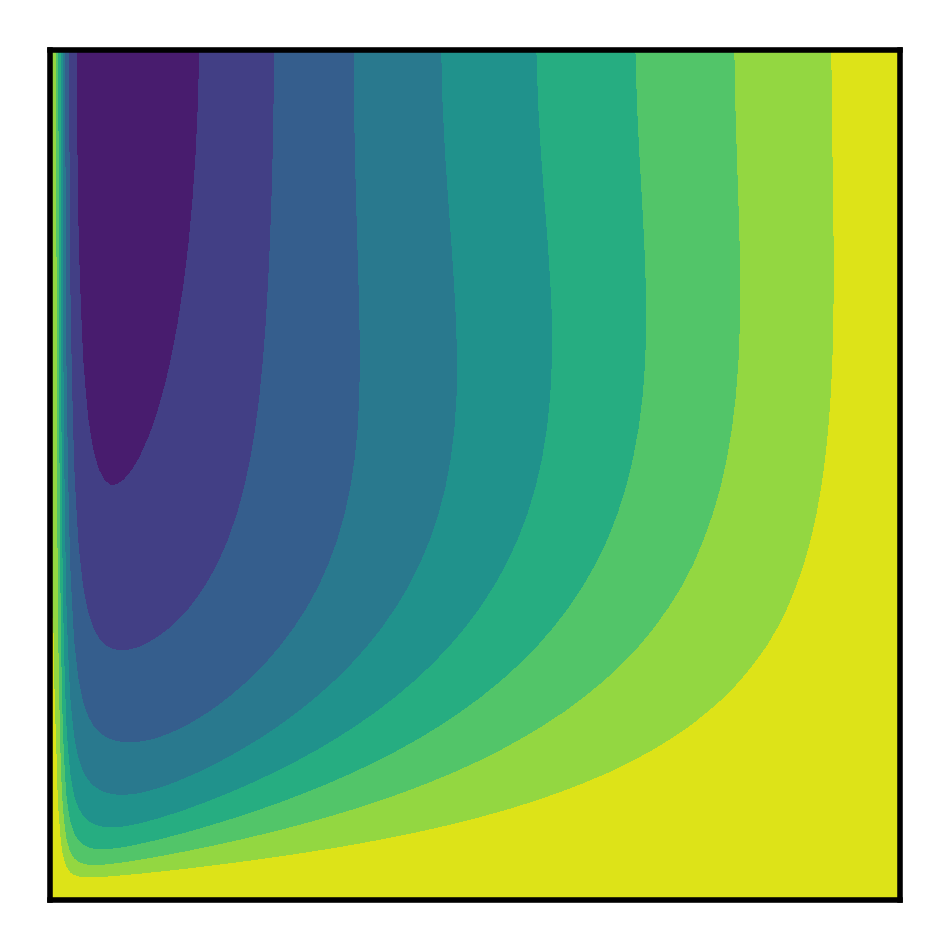}} &
        \raisebox{-.5\height}{\includegraphics[scale=0.5]{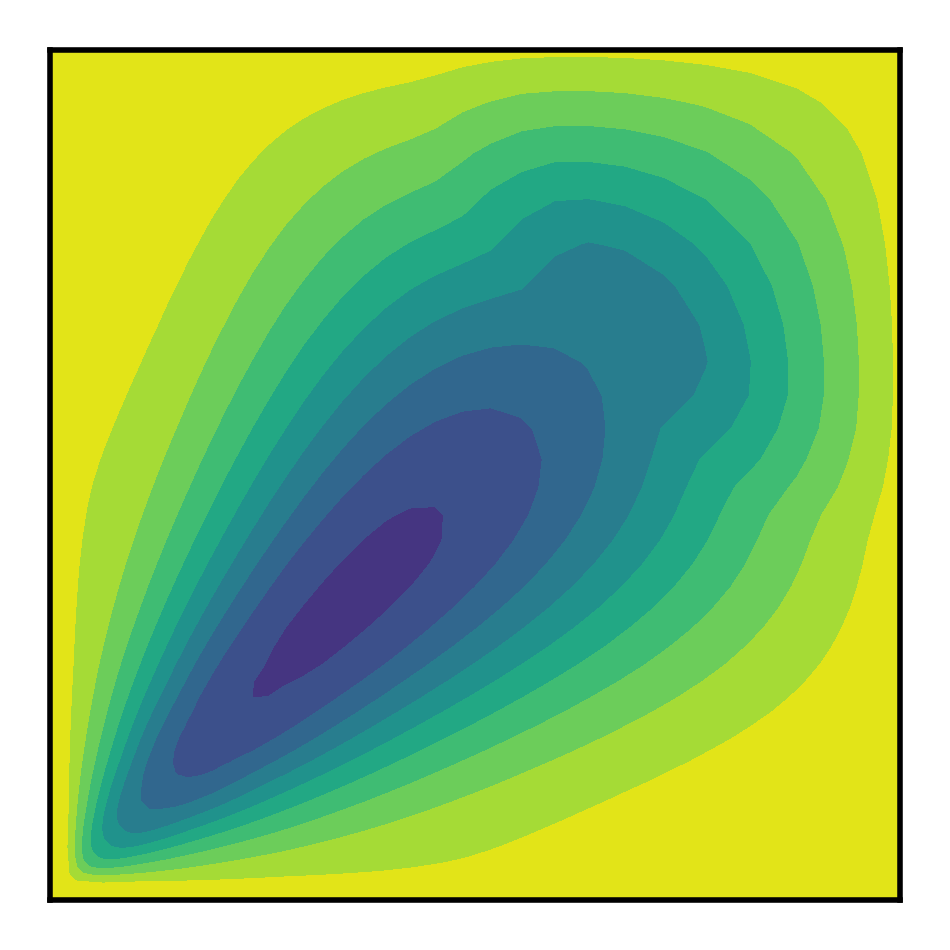}} &
        \raisebox{-.5\height}{\includegraphics[scale=0.5]{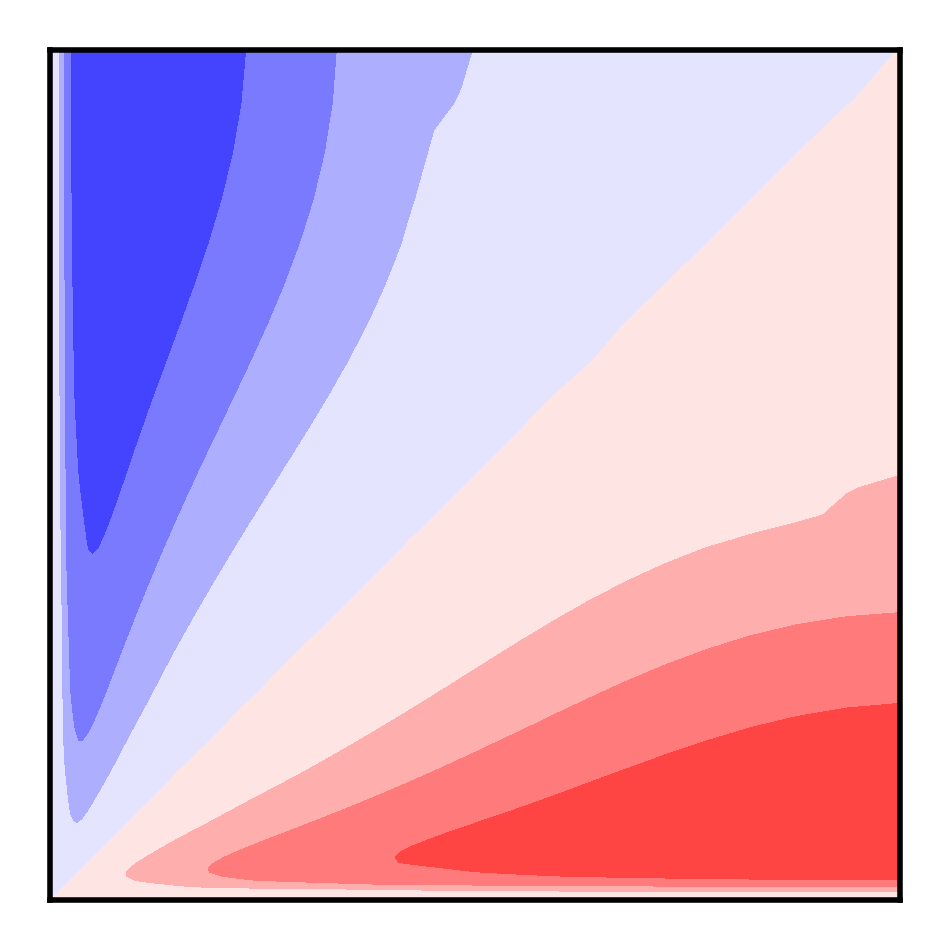}}
        \\
        & \raisebox{-.5\height}{\includegraphics[scale=0.7]{duct_Uy_cb.png}} 
        & \raisebox{-.5\height}{\includegraphics[scale=0.7]{duct_tau12_cb.png}} 
        & \raisebox{-.5\height}{\includegraphics[scale=0.7]{duct_tau23_cb.png}}
        &\raisebox{-.5\height}{\includegraphics[scale=0.7]{duct_tau22-tau33_cb.png}}
    \end{tabular}
    \caption{ 
     Plots of velocity~$u_y$, Reynolds shear stresses $\tau_{xy}$ and $\tau_{yz}$, and normal stresses imbalance $\tau_{yy}-\tau_{zz}$ case with different methods in the square duct. The models are learned from only indirect data with the ensemble Kalman method (center row) and hybrid gradient method (bottom row), compared with the ground truth (top row).
     }
    \label{fig:hybrid_results}
\end{figure}

We test the hybrid gradient method with comparison to the ensemble Kalman method in the square duct case.
Only indirect data are used to learn the neural network-based model.
The results are shown in Figure~\ref{fig:hybrid_results}.
It can be seen that the hybrid gradient method can also achieve good velocity reconstruction compared to ground truth.
The reconstructed Reynolds stress have slightly better than the ensemble method in the Reynolds shear stress.
The balance between the $\tau_{yy}$ and $\tau_{zz}$ is not well recovered for both the ensemble method and hybrid method.
The prediction errors in velocity are reduced to $11.9\%$, while the Reynolds stress is reduced to $9.8\%$.
For the ensemble method, the velocity errors are reduced to $8.88\%$, and the Reynolds stress can be reduced to $6.46\%$.
Hence, both the velocity and Reynolds stress with the hybrid gradient method are less accurate than that with the ensemble method. 

\section{Sensitivity study of the algorithmic hyperparameters and observation data}
\label{sec:sensitivity}

In this section we present the sensitivity study of the proposed ensemble-based learning algorithm in terms of the observation data error, data positions, data amounts, relative standard deviation of prior sampling, and the neural network architecture.

Our numerical results indicate that the prescribed data noise has a significant impact on the training performance with direct and indirect data.
Specifically, too small data noise, e.g., the relative standard deviation of $10^{-5}$, results in the divergence of the CFD solver during the training.
That is because the ensemble Kalman method approximates the gradient and Hessian information based on the linearization assumption.
The small data noise leads to extremely large updating steps that can result in nonphysical model output, such as negative eddy viscosity and further solver divergence.
Conversely, too large observation errors can reduce the convergence speed and lead to inferior training results in velocity and Reynolds stress.
Besides, it is observed that the relative standard deviation of~$10^{-2}$ for the direct data can provide slightly better predictions in both velocity and Reynolds stress than the baseline value of~$10^{-3}$.
Note that the inflation technique~\cite{luo2015iterative,zhang_ensemble-based_2022} can be introduced in the Kalman updating scheme to adjust the data error covariance adaptively.

We also investigate various positioning strategies of three data points, including near the wall, at the centerline, along the diagonal line, and at each corner of the computational domain (baseline).
The data positioning near the wall places the data points at $z/h=0.1$ and $y/H=0.1, 0.5, 0.9$, while the case at the centerline places the observation at $z/h=0.9$ and $y/H=0.1, 0.5, 0.9$.
The data positioning along the diagonal line places the three data points at $y/H=z/h=0.1$, $y/H=z/h=0.5$, and $y/H=z/h=0.9$, respectively.
Our results show the data positioning does not significantly affect the model accuracy among the four strategies. 
Additionally, the positioning at each corner of the domain achieves the best accuracy compared to other positioning strategies for the square duct case.
This is likely because the three points are positioned at the corner region, the duct center, and near the wall, respectively, which can cover the main characteristics in the input feature space.

Further, we conduct the sensitivity study in the data amount ranging from very sparse data to the full field data.
Specifically, we test the number of data points, including $3$ (baseline), $100$, $500$, and $2500$.
In the case with $100$ data points, we place the observation data along the line of $y/H=0.1$ and $z/H=0.9$.
For the case with $500$ data points, we select training data along the $y$ direction with an interval of $4$ grids, while the case with $2500$ data points uses the full field observation.
The results demonstrate that using more data can slightly improve the model prediction.
Specifically, when using only $3$ data points, the prediction errors in velocity and the Reynolds stress are around~$1.17\%$ and $5.24\%$, respectively.
Increasing the data number to $500$ data points, the predictive error for velocity and the Reynolds stress becomes $0.798\%$ and $6.55\%$, respectively, which is slightly improved in velocity predictions compared to the baseline case.
Further increasing the data number to $2500$, the predictive error in velocity is reduced to $0.46\%$, and the Reynolds stress discrepancy becomes similar to the baseline at $5.58\%$.
We note that the model is learned from the Reynolds stress data based on the first-order optimization in this work, which may have a relatively slow convergence speed.
One can improve the Reynolds stress prediction by running multiple regularization steps during one training step.

Besides, we test different values of the standard deviation of the prior weights, including $0.05, 0.1, 0.5$, and the baseline value $0.3$.
Our results suggest that using the large values of $0.5$ can lead to solver divergence. 
That is because a large standard deviation means that the initial guess is presumed to be far from the truth.
As a result, random samples can deviate significantly from the truth, leading to non-physical Reynolds stress and further the solver divergence.
In contrast, using relatively small prior errors, such as $0.05$ and $0.1$, can produce slightly inferior results to the baseline of $0.3$, likely due to the overconfidence in the prior weights from the Bayesian perspective~\cite{zhang2020evaluation}.

Finally, different neural network architectures are utilized to demonstrate the sensitivity of the proposed framework.
Specifically, three network architectures are tested: (1) two hidden layers with 5 neurons per layer (baseline); (2) two hidden layers with 10 neurons per layer; and (3) ten hidden layers with 10 neurons per layer. 
The results indicate that the neural network architectures have negligible effects on the velocity prediction but significantly affect the Reynolds stress prediction for the square duct case.
That is likely because the velocity prediction is improved with the ensemble Kalman updating, which has been shown to be insensitive to the neural network architecture in the square duct case based on our previous study~\cite{zhang_ensemble-based_2022}.
In contrast, the Reynolds stress prediction is enhanced with the back-propagation algorithm, which can be sensitive to the neural network architecture~\cite{uzair2020effects}.
The deep neural network has complex functional forms and a large set of model parameters to train and hence is difficult to converge with the backpropagation algorithm.
Therefore, the deep neural network with ten hidden layers provides relatively poor prediction in the Reynolds stress compared to the neural network with two hidden layers.

\begin{table}[!htb]
    \centering
    \begin{tabular}{c c c c}
    \hline
        parameter           &  value  & error($\bm{u}$) & error($\tau$) \\
        indirect data noise &  $0.00001$  &  diverge        &  diverge    \\  
                            &  $0.0001$   &  $0.624\%$ & $7.14\%$          \\ 
                            &  $\bm{0.001}$   &  $1.17\%$ & $5.24\%$          \\
                            &  $0.01$    &     $1.36\%$     &  $12.3\%$ \\ \\
        direct data noise & $0.00001$  &  diverge        &  diverge    \\  
                            & $0.0001$  &  $1.99\%$        & $9.09\%$ \\                  
                            &  $\bm{0.001}$   & $1.17\%$ & $5.24\%$  \\
                            &  0.01    &   $0.60\%$       & $4.78\%$  \\ \\
        prior std. dev.           &  0.05  & $2.10\%$         & $8.63\%$ \\
                            &  0.1   &  $2.09\%$        &  $8.66\%$ \\
                            &  $\bm{0.3}$   &  $1.17\%$        &  $5.24\%$ \\
                            &  0.5    &  diverge       & diverge  \\ \\
        data numbers      &  full   &   $0.46\%$      & $5.58\%$ \\
                            & 500   &  $0.798\%$       & $6.55\%$ \\
                            &  100    & $2.19\%$      &  $8.51\%$ \\
                            &  $\bm{3}$     &  $1.17\%$      &  $5.24\%$ \\ \\ 
        data positions      &  near wall   &  $2.16\%$        & $8.75\%$ \\
                            &  \textbf{corner}    & $1.17\%$         &  $5.24\%$ \\
                            &  along diagonal line    &   $1.76\%$       & $10.5 \%$ \\
                            &  at centerline   & $1.76 \%$         &  $10.6 \%$ \\ \\
        \shortstack{network architecture \\ (neurons/layer × layers)}      & $\bm{2} \times \bm{5}$ & $1.17\%$         &  $5.24\%$ \\
                            & $2 \times 10$ & $0.86\%$         &  $9.99\%$ \\
                            & $10 \times 10$ & $1.2\%$         &  $14.9\%$ \\
    \hline
    \end{tabular}
    \caption{Sensitivity of the training performance to the data noise, data amount, data positions, and network architecture for the square duct case. The values in bold indicate the baseline values.}
    \label{tab:sensitivity}
\end{table}

\section*{Acknowledgment}
The authors would like to thank the reviewers for their constructive and valuable comments, which helped improve the quality and clarity of this manuscript.
XLZ and GH are supported by the NSFC Basic Science Center Program for ``Multiscale Problems in Nonlinear Mechanics'' (No. 11988102). XLZ also acknowledges support from the National Natural Science Foundation
of China (No.~12102435) and the China Postdoctoral Science Foundation (No.~2021M690154). XL acknowledges partial financial support from the National Centre for Sustainable Subsurface Utilization of the Norwegian Continental Shelf (NCS2030), Norway, which is funded by the Research Council of Norway (project number: 331644) and industry partners.

\end{document}